\documentclass[11pt,a4paper]{article}
\usepackage{jheppub}

\usepackage{booktabs}
\usepackage{graphicx}
\usepackage{graphics}
\usepackage{subfig}
\usepackage{amssymb}
\usepackage{amsmath}
\usepackage{booktabs,multirow,tabularx}
\usepackage{slashed}
\usepackage{caption}
\usepackage{float}
\usepackage{placeins}
\usepackage{rotating}
\usepackage{lscape}
\usepackage{listings}
\usepackage{xspace}
\usepackage[Q=yes,pverb-linebreak=no]{examplep}

\usepackage{morefloats}
\usepackage{pgffor}
\usepackage{array}
\usepackage{ulem}

 \definecolor{fsblue}{rgb}{0.,.0,1.}
 \newcommand {\fsblue} {\color{fsblue}}
 
 \newcommand{\FSadd}[1]{{\fsblue #1}}

 \definecolor{isblue}{rgb}{0.3,0.3,1.}

\DeclareGraphicsExtensions{.pdf}

\newcommand{\code}[1]{{{\tt #1}}}

\newcommand{\pyrate}{\code{PyR@TE}\xspace}

\newcommand{\sarah}{\code{SARAH}\xspace}

\newcommand{\DR}{\ensuremath{\overline{\text{DR}}}\xspace}
\newcommand{\MS}{\ensuremath{\overline{\text{MS}}}\xspace}

\newcommand{\Tr}{\ensuremath{\operatorname{Tr}}\xspace}

\author[a]{Ingo Schienbein}
\affiliation[a]{
Laboratoire de Physique Subatomique et de Cosmologie, Universit\'e Grenoble Alpes, CNRS/IN2P3, 
 53 Avenue des Martyrs, F-38026 Grenoble, France
}
\emailAdd{ingo.schienbein@lpsc.in2p3.fr}

\author[b,c]{Florian Staub}
\affiliation[b]{
Institute for Theoretical Physics (ITP), Karlsruhe Institute of Technology, Engesserstra{\ss}e 7, D-76128 Karlsruhe, Germany}
\affiliation[c]{
Institute for Nuclear Physics (IKP), Karlsruhe Institute of Technology, Hermann-von-Helmholtz-Platz 1, D-76344 Eggenstein-Leopoldshafen,
Germany}
\emailAdd{florian.staub@kit.edu}

\author[d]{Tom Steudtner}
\affiliation[d]{
Department of Physics and Astronomy, U Sussex, Brighton, BN1 9QH, U.~K.
}
\emailAdd{T.Steudtner@sussex.ac.uk}

\author[e]{Kseniia Svirina}
\affiliation[e]{
Laboratoire de Physique Subatomique et de Cosmologie, Universit\'e Grenoble Alpes, CNRS/IN2P3, 
 53 Avenue des Martyrs, F-38026 Grenoble, France
}
\emailAdd{kseniia.svirina@lpsc.in2p3.fr}

\abstract{We revisit the renormalisation group equations (RGE) for general renormalisable gauge theories at one- and two-loop
accuracy. We identify and correct various mistakes in the literature for the $\beta$-functions of the dimensionful Lagrangian parameters (the fermion mass, the bilinear and trilinear scalar couplings) as well as the dimensionless
quartic scalar couplings. There are two sources for these discrepancies. 
Firstly, the known expressions for the scalar couplings assume a diagonal wave-function renormalisation
which is not appropriate for models with mixing in the scalar sector.
Secondly, the dimensionful parameters have been derived in the literature
using a dummy field method which we critically re-examine, obtaining revised expressions for the $\beta$-function of the fermion mass.
We perform an independent cross-check using well-tested supersymmetric RGEs
which confirms our results. The numerical impact of the changes in the $\beta$-function for the
fermion mass terms is illustrated using a toy model with a heavy vector-like fermion pair coupled to a scalar gauge singlet.
Unsurprisingly, the correction to the running of the fermion mass becomes sizeable for large Yukawa couplings
of the order of ${\cal O}(1)$. Furthermore, we demonstrate the importance of the correction to the $\beta$-functions of the scalar quartic couplings using a general type-III Two-Higgs-Doublet-Model.
All the corrected expressions have been implemented in updated versions of the Mathematica package 
\sarah and the Python package \pyrate.
}

\title{Revisiting RGEs for general gauge theories}

\preprint{KA-TP-27-2018}
\begin{document}

\maketitle

\newpage

\thispagestyle{empty}
 
 
 
\newpage

\section{Introduction}
\label{sec:intro}

Renormalisation Group Equations (RGEs) 
are important 
as they provide the necessary link between the physics at different energy scales.
The two-loop RGEs for all dimensionless parameters in general gauge theories have been
derived already more than 30 years ago \cite{Machacek:1983tz,Machacek:1983fi,Machacek:1984zw,Jack:1984vj,Jack:1982sr,Jack:1982hf}.
More recently, these results have been re-derived by Luo {\it et al.} \cite{Luo:2002ti} including the $\beta$-functions
for dimensionful parameters. The latter results are based on the $\beta$-functions of dimensionless
couplings by applying a so called ``dummy field'' method \cite{Martin:1993zk}. 
However, no independent direct calculation of the two-loop $\beta$-functions for scalar and fermion masses and scalar trilinear couplings exists so far 
in the literature. One of the aims of this paper is to provide a more detailed (pedagogical) discussion of the dummy field method
and to critically examine the $\beta$-functions for the dimensionful parameters. As a result we will correct the
$\beta$-functions for the fermion masses. 
We also find differences for the purely scalar couplings in certain models with respect to the literature.
These differences arise from not always justified assumption about the properties of the wave-function renormalisation.
We provide an independent cross-check using well tested supersymmetric RGEs which confirms our results. 
We believe that these corrections and validations are non-trivial and important in view of the wide use of the RGEs.
Still, an independent direct calculation of the dimensionful $\beta$-functions would be useful.

The general equations have been implemented in the Mathematica package \sarah 
\cite{Staub:2008uz,Staub:2009bi,Staub:2010jh,Staub:2012pb,Staub:2013tta}
and in the Python
package \pyrate \cite{Lyonnet:2013dna,Lyonnet:2016xiz}. 
More recent results which are (partially) included in these packages
such as kinetic mixing \cite{Fonseca:2013bua} or running VEVs \cite{Sperling:2013eva,Sperling:2013xqa} will not be discussed in this paper.
The overarching purpose is to present the current state-of-the art of the two-loop $\beta$-functions and to collect the
corrected expressions such that all the relevant information is at hand in one place.

\section{The Lagrangian for a general gauge theory}
\label{sec:Lagrangian}

In this section we review the Lagrangian for a general renormalisable field theory following \cite{Luo:2002ti}. 
The following particle content is considered:
\begin{itemize}
\item $V_\mu^A(x)$ ($A=1,\ldots, d$) are gauge fields of a compact simple group $G$ where $d$ is the dimension of $G$.
\item $\phi_a(x)$ ($a=1,\ldots,N_\phi$) denote real scalar fields transforming under a (in general) reducible representation of $G$.
The Hermitian generators of $G$ in this representation will be denoted $\Theta^A_{ab}$ ($A=1,\ldots, d$; $a,b = 1,\ldots, N_\phi$). 
Since the scalar fields are real, the generators $\Theta^A$ are purely imaginary and antisymmetric.
\item $\psi_j(x)$ ($j=1,\ldots, N_\psi$) are left-handed complex two-component fermion fields transforming under a representation of $G$ which is in general reducible as well. The Hermitian generators are denoted by $t^A_{jk}$ ($A=1,\ldots, d$; $j,k = 1,\ldots, N_\psi$).
\end{itemize}
The most general renormalisable Lagrangian can be decomposed into three parts,
\begin{equation}
{\cal L} = {\cal L}_0 + {\cal L}_1 + (\rm{gauge\ fixing + ghost\ terms})\, ,
\end{equation}
where ${\cal L}_0$ is free of dimensional parameters and ${\cal L}_1$ contains all terms
with  dimensional parameters. Here, ${\cal L}_0$ reads
\begin{align}
{\cal L}_0 & = -\frac{1}{4} F_A^{\mu\nu} F^A_{\mu \nu} + \frac{1}{2} D^\mu \phi_a D_\mu \phi_a + i \psi_j^\dagger \sigma^\mu D_\mu \psi_j
\nonumber\\
&\phantom{=}
- \frac{1}{2} \left(Y^a_{jk} \psi_j \zeta \psi_k \phi_a + Y^{a*}_{jk} \psi_j^\dagger \zeta \psi_k^\dagger \phi_a\right) - \frac{1}{4!} \lambda_{abcd} \phi_a \phi_b \phi_c \phi_d\, ,
\label{eq:L0}
\end{align}
where $F^A_{\mu \nu}(x)$ is the gauge field strength tensor defined in the usual way in terms of the structure constants $f^{ABC}$ of the
gauge group and the gauge coupling constant $g$:
\begin{equation}
F^A_{\mu \nu} = \partial_\mu V_\nu^A - \partial_\nu V_\mu^A + g f^{ABC} V_\mu^B V_\nu^C\, .
\end{equation}
The covariant derivatives of the scalar and fermion fields are given by
\begin{align}
D_\mu \phi_a & = \partial_\mu \phi_a - i g \Theta^A_{ab} V^A_\mu \phi_b,
\\
D_\mu \psi_j & = \partial_\mu \psi_j - i g t^A_{jk} V^A_\mu \psi_k\, .
\end{align}
Furthermore, $Y^a_{jk}$ ($a=1,\ldots,N_\phi; j,k=1,\ldots,N_\psi$) are complex Yukawa couplings and
$\zeta = i \sigma_2$ is the two-component spinor metric ($\sigma_2$ is the second Pauli matrix).
Finally, $\lambda_{abcd}$ denotes quartic scalar couplings which are real and invariant under permutations of the
set of indices $\{a,b,c,d\}$.

The Lagrangian containing the dimensionful parameters is given by
\begin{align}
{\cal L}_1 & = -\frac{1}{2} \left[(m_f)_{jk} \psi_j \zeta \psi_k + (m_f)^*_{jk} \psi^\dagger_j \zeta \psi^\dagger_k \right]
-\frac{m^2_{ab}}{2!} \phi_a \phi_b - \frac{h_{abc}}{3!} \phi_a \phi_b \phi_c\, .
\label{eq:L1}
\end{align}
Here $m_f$ is a complex matrix of fermion masses, $m^2$ is a real matrix of scalar masses squared, and
$h_{abc}$ are real cubic scalar couplings.
Our goal is to revisit the one- and two-loop $\beta$-functions for these dimensionful couplings which have been derived in Ref.~\cite{Luo:2002ti},
employing the so-called ``dummy field'' method which has been initially proposed in Ref.~\cite{Martin:1993zk}.

\section{Renormalisation Group Equations}

We are interested in the scale dependence of the Lagrangian parameters which, in general, 
is governed by RGEs. The RGEs can 
be calculated in different schemes. We are going to consider only dimensional regularisation with modified minimal subtraction, usually called \MS, for four dimensional field theories. In this scheme
the $\beta$-functions, which describe the renormalisation group running 
of the model parameters $(\Theta_i)$, 
are defined as
\begin{equation}
\beta_i = \mu \frac{d \Theta_i}{d\mu} \,,
\end{equation}
where $\mu$ is an arbitrary renormalisation scale. $\beta_i$ can be expanded in a perturbative series:
\begin{equation}
\beta_i = \sum_n \frac{1}{(16 \pi^2)^n} \beta_i^{(n)}\, ,
\end{equation}
where
$\beta_i^{(1)}$ and $\beta_i^{(2)}$ are the one- and two-loop contributions  to the running which we are interested in.
Generic expressions of the one- and two-loop $\beta$-functions for dimensionless parameters in a general quantum field theory were 
derived in Refs.~\cite{Machacek:1983tz,Machacek:1983fi,Machacek:1984zw}.

\section{The dummy field method}
\label{sec:dummy}

In principle, one could calculate the renormalisation constants for the dimensionful couplings (the fermion masses $(m_f)_{jk}$, 
the squared scalar masses $m^2_{ab}$, and the cubic scalar couplings $h_{abc}$)
and derive the $\beta$-functions directly from them. However, this is tedious and has not been attempted so far in the literature.
Instead, a ``dummy field'' method has been employed in \FSadd{Ref.}~\cite{Luo:2002ti} applying an idea, to our knowledge, 
first mentioned in Ref.~\cite{Martin:1993zk}.
Since a detailed description of this method is lacking in the literature we provide a careful discussion of it in this section.

The idea is to introduce a scalar ``dummy field'', i.e. a non-propagating real scalar field with no gauge interactions.
The dummy field will be denoted by an index with a hat, $\phi_{\hat d}$, and satisfies the condition $D_\mu \phi_{\hat d} = 0$.
As a consequence, expressions with two identical internal dummy indices (corresponding to a propagating dummy field) have to vanish.
Furthermore, since $D_\mu \phi_{\hat d} = 0$, all gauge boson - dummy scalar vertices vanish as well:
$\Gamma_{V\phi_a \phi_{\hat d} } = \Gamma_{V \phi_{\hat d} \phi_{\hat d}} = \Gamma_{VV\phi_a \phi_{\hat d} }=\Gamma_{VV\phi_{\hat d} \phi_{\hat d}}=0$. 

Let us now consider the Lagrangian ${\cal L}_0$ \eqref{eq:L0} in the presence of the same particle content plus one extra
scalar dummy field ($\phi_{\hat d}$) and separate the terms with the dummy field.
Using $D_\mu \phi_{\hat d}=0$,\; $\lambda_{ab\hat{d}\hat{d}} + \lambda_{a\hat{d} b \hat{d}} + \lambda_{\hat{d}  a b \hat{d}} +
\lambda_{a\hat{d}\hat{d}b}+ \lambda_{\hat{d}a\hat{d}b}+\lambda_{\hat{d}\hat{d}a b}=6 \lambda_{ab\hat{d}\hat{d}}$,\;
$\lambda_{abc\hat{d}} + \lambda_{ab\hat{d}c} + \lambda_{a\hat{d}bc} + \lambda_{\hat{d}abc} = 4\lambda_{abc\hat{d}}$, and
$\lambda_{a\hat{d} \hat{d}\hat{d}} + \lambda_{\hat{d} a\hat{d}\hat{d}} + \lambda_{\hat{d} \hat{d}a \hat{d}} + \lambda_{\hat{d} \hat{d}\hat{d}a} =
4 \lambda_{a\hat{d} \hat{d}\hat{d}}$ one easily finds (writing the sums over the scalar indices explicitly):
\begin{align}
{\cal L}_0 & = -\frac{1}{4} F_A^{\mu\nu} F^A_{\mu \nu} + \sum_{a=1}^{N_\phi}
\frac{1}{2} D^\mu \phi_a D_\mu \phi_a + i \psi_j^\dagger \sigma^\mu D_\mu \psi_j
\nonumber\\
&\phantom{=}
- \frac{1}{2} (\sum_{a=1}^{N_\phi}Y^a_{jk} \psi_j \zeta \psi_k \phi_a + \text{h.c.}) - 
\sum_{a,b,c,d=1}^{N_\phi} \frac{1}{4!} \lambda_{abcd} \phi_a \phi_b \phi_c \phi_d\, 
\nonumber\\
&\phantom{=}
- \frac{1}{2} (Y^{\hat d}_{jk} \psi_j \zeta \psi_k \phi_{\hat d} + \text{h.c.})
- 6 \sum_{a,b=1}^{N_\phi}\frac{1}{4!} \lambda_{ab\hat{d}\hat{d}} \phi_a \phi_b \phi_{\hat d} \phi_{\hat d}\,
- 4 \sum_{a,b,c=1}^{N_\phi}\frac{1}{4!} \lambda_{abc\hat{d}} \phi_a \phi_b \phi_c \phi_{\hat d}\,
\nonumber\\
&\phantom{=}
- 4 \sum_{a=1}^{N_\phi}\frac{1}{4!} \lambda_{a\hat{d} \hat{d}\hat{d}} \phi_a \phi_{\hat d} \phi_{\hat d} \phi_{\hat d}\,
- \frac{1}{4!} \lambda_{\hat{d} \hat{d} \hat{d}\hat{d}} \phi_{\hat d} \phi_{\hat d} \phi_{\hat d} \phi_{\hat d}\, .
\label{eq:dummyL}
\end{align}
A few comments are in order:
\begin{itemize}
\item The first two lines reproduce the Lagrangian ${\cal L}_0$ \eqref{eq:L0} 
with the original particle content without the dummy field.
\item The terms in the third line reproduce the Lagrangian ${\cal L}_1$ \eqref{eq:L1} if one makes the following
identifications:
\begin{align}
\label{eq:identification1}
Y^{\hat d}_{jk} \phi_{\hat d} & = (m_f)_{jk}\, , \quad
\lambda_{ab\hat{d}\hat{d}} \phi_{\hat d} \phi_{\hat d}  = 2 m^2_{ab}\, , \quad
\lambda_{abc\hat{d}} \phi_{\hat d}  = h_{abc}\, .
\end{align}
Note that we believe these are the correct relations while the notation below Eq.\ (21) in \cite{Luo:2002ti} 
is rather sloppy:
\begin{align}
\label{eq:identification2}
Y^{\hat d}_{jk} & = (m_f)_{jk}\, , \quad
\lambda_{ab\hat{d}\hat{d}} = 2 m^2_{ab}\, , \quad
\lambda_{abc\hat{d}}  = h_{abc}\, .
\end{align}
\item The terms in the fourth line of Eq.~\eqref{eq:dummyL} do not spoil the relations in Eq.\ \eqref{eq:identification1} or \eqref{eq:identification2}.
First of all, the second last term is only gauge invariant if $\phi_a$ is a gauge singlet.
Furthermore, it is an effective tadpole term which can be removed by a shift of the field $\phi$.\footnote{For the same reason such a term is not included in $\mathcal{L}_1$ in Eq.~\eqref{eq:L1}.}
The last term is just a constant.
In any case, contributions from the interactions in the fourth line
to the $\beta$-functions of the other dimensionful parameters would involve at least one internal dummy line
which gives a vanishing result.
\end{itemize}
The relations  \eqref{eq:identification2} have been used in Ref.~\cite{Luo:2002ti} to derive
the $\beta$-functions for the fermion masses from the known ones for the Yukawa interactions. Likewise, the $\beta$-functions
for the scalar masses and the trilinear scalar couplings were obtained from the scalar quartic $\beta$-functions.
This was achieved by removing contributions with a summation of $\hat d$-type indices
and terms with $\hat d$ indices appearing on the generators $\Theta$.
However, a subtlety arises
due to the wave-function renormalisation of external dummy scalar lines which leads to effective tadpole contributions.
Such contributions should be removed from the $\beta$-functions for the Yukawa interactions and quartic couplings
but are not necessarily eliminated by just suppressing the summation over $\hat d$-indices and associated gauge couplings.
For this reason, we re-examine in the following sections all the $\beta$-functions for the dimensionful parameters by verifying 
the dummy method on a diagram by diagram basis.

\section{$\beta$-functions for dimensionful parameters}
We now
apply the dummy method to obtain the $\beta$-functions 
of the dimensionful parameters using the generic results 
for the dimensionless parameters given in Refs.~\cite{Machacek:1983tz,Machacek:1983fi,Machacek:1984zw,Luo:2002ti}.
In Sec.\ \ref{sec:fermion_mass}, we start with the fermion mass term.
The trilinear scalar couplings will be discussed in Sec.\ \ref{sec:trilinear} before we turn to the scalar
mass terms in Sec.\ \ref{sec:scalar_mass}.
First of all,
it is necessary to introduce a number of group invariants and 
definitions for certain combinations of coupling constants. These definitions will be used to write the
expressions for the $\beta$-functions in a more compact form.

\paragraph{Group invariants}
$C_2(F)$ is the quadratic Casimir operator for the (in general) reducible fermion representation:
\begin{align}
C_2(F) :=  & \sum_{A=1}^d t^A t^A\, \quad \text{, i.e.} \,\, [C_2(F)]_{ij} \equiv C_2^{ij}(F)
= \sum_{A=1}^d \sum_{k=1}^{N_\psi} t^A_{ik} t^A_{kj}\, ,
\end{align}
where $i,j=1, \ldots, N_\psi$. Due to Schur's lemma, $C_2(F)$ is a diagonal $N_\psi \times N_\psi$ matrix with
the same eigenvalues for each irreducible representation.
Similarly, $C_2(S)$ is the quadratic Casimir operator for the (in general) reducible scalar representation:
\begin{align}
C_2(S) :=  & \sum_{A=1}^d \theta^A \theta^A\, \quad \text{, i.e.} \,\, [C_2(S)]_{ab} \equiv C_2^{ab}(S)
= \sum_{A=1}^d \sum_{c=1}^{N_\phi} \theta^A_{ac} \theta^A_{cb}\, ,
\end{align}
where $a,b=1, \ldots, N_\phi$. Again due to Schur's lemma, $C_2(S)$ is a diagonal $N_\phi \times N_\phi$ matrix.
Furthermore, $S_2(S)$ and $S_2(F)$ denote the Dynkin index of the scalar and fermion representations, respectively,
\begin{align}
\Tr[\theta^A \theta^B] =:  S_2(S) \delta^{AB}\, , \quad \Tr[t^A t^B] =: S_2(F) \delta^{AB}\, ,
\end{align}
and $C_2(G)$ is the quadratic Casimir operator of the (irreducible) adjoint representation
\begin{align}
C_2(G) \delta^{AB} := \sum_{C,D=1}^d f^{ACD}f^{BCD}\, .
\end{align}

\paragraph{Coupling combinations}
We start with two $N_\psi \times N_\psi$ matrices formed out of the Yukawa matrices $Y^a_{ij}$:
\begin{align}
Y_2(F) := & \sum_{a=1}^{N_\phi} Y^{\dagger a} Y^a\, , \quad  Y_2^\dagger(F) := \sum_{a=1}^{N_\phi} Y^a Y^{\dagger a}\, ,
\end{align}
where the sum includes all `active' (propagating) scalar indices but not the dummy index.
It should be noted that $Y_2^\dagger(F) \ne [Y_2(F)]^\dagger$; instead it represents the quantity $Y_2(F)$ where the Yukawa coupling
$Y^a$ has been replaced by its conjugate $Y^{\dagger a}$.
Furthermore, the following $N_\phi \times N_\phi$ matrices are needed below:
\begin{align}
Y_2^{ab}(S) := & \frac{1}{2} \Tr[Y^{\dagger a} Y^b + Y^{\dagger b} Y^a]\, ,
\\
H^2_{ab}(S) := & \frac{1}{2} \sum_{c=1}^{N_\phi} \Tr[Y^a Y^{\dagger b} Y^c Y^{\dagger c} + Y^{\dagger a} Y^b Y^{\dagger c} Y^c]\, ,
\\
\overline{H}_{ab}^2(S) := & \frac{1}{2}  \sum_{c=1}^{N_\phi} \Tr[Y^a Y^{\dagger c} Y^b Y^{\dagger c} + Y^{\dagger a} Y^c Y^{\dagger b} Y^c]\, ,
\\
\Lambda^2_{ab}(S) := & \frac{1}{6} \sum_{c,d,e=1}^{N_\phi}\lambda_{acde}\lambda_{bcde}\, ,
\\
Y^{2F}_{ab}(S) := & \frac{1}{2} \Tr[C_2(F) (Y^a Y^{\dagger b} + Y^b Y^{\dagger a})]\, .
\end{align}

There is one crucial comment in order concerning the properties of these objects: in previous works it is assumed that $Y_2^{ab}(S) = Y_2(S) \delta_{ab}$ and $\Lambda^2_{ab}(S) = \Lambda^2(S) \delta_{ab}$ holds. These properties are derived from group theoretical arguments. We agree with them as long as the considered model does not contain several scalar particles with identical quantum numbers. However, if this is the case than these relations are no longer valid. Or, in other words, the matrices $Y_2^{ab}$ and $\Lambda^2_{ab}$ are diagonal in the space of irreducible representations but not necessarily in the space of particles in the considered model. The consequence is that contributions from off-diagonal wave-function corrections may arise which are not included in Refs.~\cite{Machacek:1983tz,Machacek:1983fi,Machacek:1984zw,Luo:2002ti}. 
This is one source for the discrepancies between our results and previous ones. This does not only affect the dimensionful parameters but also the quartic scalar couplings.

\paragraph{RGEs for dimensionless parameters} The $\beta$-function for the dimensionful parameters are obtained from those of the dimensionless parameters using the 
dummy field method. The one- and two-loop expressions for the running of a Yukawa coupling are given by 
\begin{align}
 \label{eq:Yuk1}
 \beta_a^{I}=& \frac12 \left[Y_2^+(F) Y^a + Y^a Y_2(F) \right] + 2 Y^b Y^{+a} Y^b +  2 \kappa Y^b Y_2^{ab}(S) - 3 g^2 \{ C_2(F), Y^a \}\, , \\
\beta_a^{II} =& 2 Y^c Y^{+b} Y^a (Y^{+c} Y^b - Y^{+b} Y^c) - Y^b \left[Y_2(F) Y^{+a} + Y^{+a} Y_2^+(F) \right] Y^b \nonumber \\
&- \frac18 \left[Y^b Y_2(F) Y^{+b} Y^a + Y^a Y^{+b} Y_2^+(F) Y^b \right] - \underline{4\kappa Y_2^{bc}(S) Y^b Y^{+a} Y^c} - 2 \kappa Y^b \bar{H}^2_{ab}(S) \nonumber \\
&- \frac32\kappa Y_2^{bc}(S) (Y^b Y^{+c} Y^a + Y^a Y^{+c} Y^b ) - 3\kappa Y^b H_{ab}^2(S)- 2 \lambda_{abcd} Y^b Y^{+c} Y^d  \nonumber \\ 
&+ \frac12 \Lambda_{ab}^2(S) Y^b + 3 g^2 \{ C_2(F), Y^b Y^{+a} Y^b \} + 5 g^2 Y^b \{ C_2(F), Y^{+a} \} Y^b \nonumber \\ 
&- \frac74 g^2 [C_2(F) Y_2^+(F) Y^a + Y^a Y_2(F) C_2(F)] \nonumber \\
&- \frac14 g^2 [Y^b C_2(F) Y^{+b} Y^a + Y^a Y^{+b} C_2(F) Y^b] + 6 g^2 H_{2t}^a + 10 \kappa g^2 Y^b Y_{ab}^{2F}(S) \nonumber \\
&+ 6 g^2 [C_2^{bc}(S) Y^b Y^{+a} Y^c - 2 C_2^{ac}(S) Y^b Y^{+c} Y^b] + \frac92 g^2 C_2^{bc}(S) (Y^b Y^{+c} Y^a + Y^a Y^{+c} Y^b) \nonumber \\ 
&- \frac32 g^4 \{ \left[ C_2(F) \right]^2, Y^a \} + 6g^4 C_2^{ab}(S) \{ C_2(F), Y^b \} \nonumber \\
& + g^4 \left[ - \frac{97}{6}C_2(G) + \frac{10}{3}\kappa S_2(F)  +\frac{11}{12}S_2(S) \right] \{ C_2(F), Y^a \} - \frac{21}{2}g^4 C_2^{ab}(S) C_2^{bc}(S) Y^c \nonumber \\
& + g^4 C_2^{ab}(S) \left[ \frac{49}{4} C_2(G) - 2\kappa S_2(F) - \frac14 S_2(S) \right] Y^b\, ,
 \label{eq:Yuk2}
\end{align}
where the definition of $H^a_{2t}$ can be found in App.\ \ref{app:fermion} and the factor
$\kappa = 1/2$ for 2-component fermions and $\kappa = 1$ for 4-component fermions.
The underlined term differs from Refs.~\cite{Machacek:1983fi,Luo:2002ti} by a swapped index.  

For the quartic coupling, we are going to use the following expressions:
\begin{align}
\label{eq:quartic1}
 \beta_{abcd}^{I}  = & \Lambda_{abcd}^2 - 8\kappa H_{abcd} + 2\kappa \underline{\Lambda_{abcd}^Y} - 3g^2 \Lambda_{abcd}^S + 3g^4 A_{abcd}\, , \\
 \beta_{abcd}^{II} =& 
 \underline{\frac{1}{12} \sum_{per} \Lambda^2_{af} \lambda_{fbcd}} - \bar{\Lambda}_{abcd}^3 
 - 4\kappa \bar{\Lambda}_{abcd}^{2Y} + \kappa \left[ 8\bar{H}_{abcd}^\lambda - \underline{\frac16 \sum_{per} \left[ 3H^2_{af} + 2\bar{H}^2_{af} \right]  \lambda_{fbcd}} \right] \nonumber \\
&+ 4\kappa(H_{abcd}^Y + 2\bar{H}_{abcd}^Y + 2H_{abcd}^3) \nonumber \\
&+ g^2 \left[ 2\bar{\Lambda}_{abcd}^{2S} - 6\Lambda_{abcd}^{2g} + 4\kappa(H_{abcd}^S - H_{abcd}^F)   + \underline{\frac53 \kappa \sum_{per} Y^{2F}_{af} \lambda_{fbcd}} \right] \nonumber \\
&- g^4 \Big\{ \left[ \frac{35}{3}C_2(G) - \frac{10}{3}\kappa S_2(F) - \frac{11}{12}S_2(S) \right]
    \Lambda_{abcd}^S  -\frac32\Lambda_{abcd}^{SS} - \frac{5}{2}A_{abcd}^\lambda
   - \frac12\bar{A}_{abcd}^\lambda   \nonumber \\ 
& \hspace{2cm} + 4\kappa (B_{abcd}^Y - 10\bar{B}_{abcd}^Y) \Big\} \nonumber \\
&+ g^6 \left\{ \left[ \frac{161}{6}C_2(G) - \frac{32}{3}\kappa S_2(F) - \frac{7}{3}S_2(S) \right] A_{abcd}  - \frac{15}{2}A_{abcd}^S + 27A_{abcd}^g \right\}\, ,
\label{eq:quartic2}
\end{align}
where the quantities $\Lambda^2_{abcd}$, $H_{abcd}$, $\Lambda^Y_{abcd}$, 
$\Lambda^S_{abcd}$, and $A_{abcd}$ in Eq.\ \eqref{eq:quartic1} are described in Sec.\  \ref{sec:trilinear},
while the definitions for the quantities $\bar \Lambda^3_{abcd}$, \ldots, $A^g_{abcd}$ in Eq.\ \eqref{eq:quartic2} can be found
in App.\ \ref{app:cubic}.
Here, {${\phantom\,}{\sum\limits_{per}}$} denotes a sum over all permutations of uncontracted scalar indices.
Our equations \eqref{eq:quartic1} and \eqref{eq:quartic2} differ
from the results in Refs.~\cite{Machacek:1984zw,Luo:2002ti} in the terms which are underlined. The reason is that only the possibility of diagonal wave-function 
renormalisation is included Refs.~\cite{Machacek:1984zw,Luo:2002ti} as discussed above. 
\\

Finally, to have all RGEs at one place, we give here also the $\beta$-functions for the gauge coupling although we will not use
 them in the following:
\begin{align}
\beta_g^I =& -g^3 \left[ \frac{11}{3} C_2(G) - \frac{4}{3}\kappa S_2(F) - \frac{1}{6} S_2(S)\right] \, ,
\\
\beta_g^{II}=& - 2\kappa g^3 Y_4(F) - g^5 \left[ \frac{34}{3} C_2(G)^2 - \kappa \left( 4 C_2(F) + \frac{20}{3} C_2(G) \right) S_2(F) \right. \nonumber \\
         &\ \ \ \  - \left. \left( 2 C_2(S) + \frac{1}{3} C_2(G) \right) S_2(S) \right]\, .
\end{align}

\subsection{Fermion mass}
\label{sec:fermion_mass}

\allowdisplaybreaks
The $\beta$-function of the fermion mass term can be obtained from the expressions of the Yukawa coupling by considering 
the external scalar as dummy field. We follow a diagrammatic approach; for each class of diagrams we provide the coupling
structure and show the resulting diagram together with its expression after applying the dummy field method.
In accord with the discussion in Sec.\ \ref{sec:dummy}, the following mappings are performed:
\begin{align*}
a \to \hat d\, , \ Y^a \to & Y^{\hat d} \to m_f\, , \ Y^{\dagger a} \to Y^{\dagger \hat d}  \to m_f^{\dagger}\, , \ \lambda_{abcd} \to \lambda_{\hat d b c d} \to h_{bcd}\, .
\end{align*}
The fermion mass insertions will be represented by black dots in the Feynman diagrams. We recall that dummy scalars do neither couple to gauge bosons nor propagate.
There are two generically different wave function correction diagrams contributing to the running
of the Yukawa couplings: those stemming from either external fermions or scalars. For external fermions, the transition between the Yukawa coupling 
and fermion mass term looks as follows, where the grey blob depicts all loop corrections to the external line:
\begin{eqnarray}
\begin{aligned}
\includegraphics[width=0.3\linewidth]{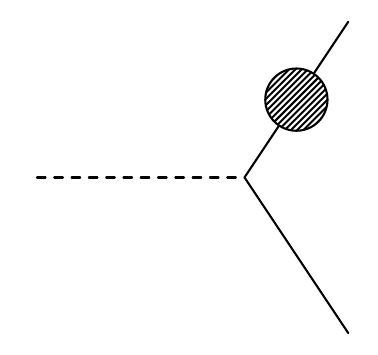}
\end{aligned}
&\quad \to \quad& 
\begin{aligned}
\includegraphics[width=0.3\linewidth]{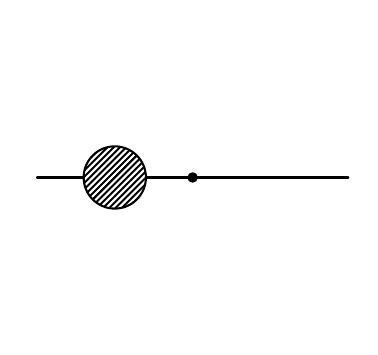}
\end{aligned}
\nonumber \\
Y_2^\dagger(F) Y^a + Y^a Y_2(F)
& \to &
Y_2^\dagger(F) m_f + m_f Y_2(F) \\
\{C_2(F),Y^a\}
& \to &
\{C_2(F),m_f\} \\
Y^b Y_2(F) Y^{\dagger b} Y^a + Y^a Y^{\dagger b} Y_2^\dagger(F) Y^b & \to &  
Y^b Y_2(F) Y^{\dagger b} m_f + m_f Y^{\dagger b} Y_2^\dagger(F) Y^b\\
Y_2^{bc}(S)(Y^b Y^{\dagger c} Y^a + Y^a Y^{\dagger c} Y^b) &\to & 
Y_2^{bc}(S)(Y^b Y^{\dagger c} m_f + m_f Y^{\dagger c} Y^b) \\
g_2^2 (C_2(F) Y_2^\dagger(F) Y^a + Y^a Y_2(F) C_2(F)) & \to &
g_2^2 (C_2(F) Y_2^\dagger(F) m_f + m_f Y_2(F) C_2(F)) \qquad \quad \\
g_2^2 (Y^b C_2(F) Y^{\dagger b} Y^a + Y^a Y^{\dagger b} C_2(F) Y^b) & \to &
g_2^2 (Y^b C_2(F) Y^{\dagger b} m_f + m_f Y^{\dagger b} C_2(F) Y^b) \\
g^2 C^{bc}_2(S)(Y^b Y^{\dagger c} Y^a + Y^a Y^{\dagger c} Y^b) &\to &
g^2 C^{bc}_2(S)(Y^b Y^{\dagger c} m_f + m_f Y^{\dagger c} Y^b) \\
g^4 \{|C_2(F)|^2,Y^a\} &\to& g^4 \{|C_2(F)|^2,m_f\} \\
g^4 C_2(G) \{C_2(F),Y^a\} &\to & g^4 C_2(G) \{C_2(F),m_f\} \\
g^4 ( x_1 S_2(F) + x_2 S_2(S)) \{C_2(F),Y^a\} &\to & g^4 ( x_1 S_2(F) + x_2 S_2(S)) \{C_2(F),m_f\}\,. 
\end{eqnarray}
Here, $x_1$ and $x_2$ are real numbers (cf.\ Eq.\ \eqref{eq:Yuk2}).

Thus, we find counterparts for all contributions in both cases. The wave-function renormalisation part stemming from the external scalar is completely different:
after applying the replacement with dummy fields, we find only tadpole contributions. However, those are usually absorbed into a re-definition of the vacuum, i.e., 
they don't contribute to the $\beta$-function of the fermion mass term, and the correct replacements are
\begin{eqnarray}
\begin{aligned}
\includegraphics[width=0.3\linewidth]{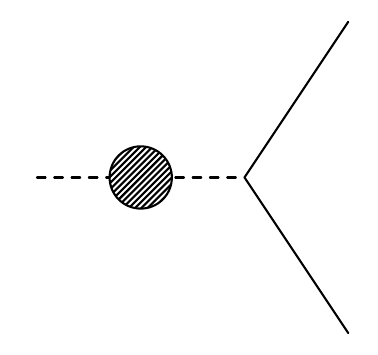}
\end{aligned}
&\quad \to \quad& 
\begin{aligned}
\includegraphics[width=0.3\linewidth]{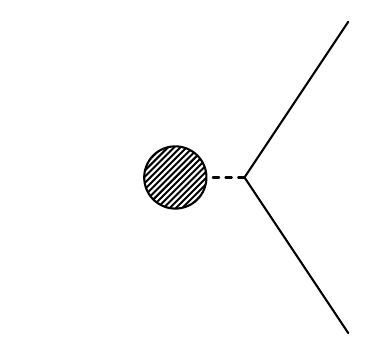}
\end{aligned}
\nonumber \\
\label{eq:rep1}
Y^b Y_2^{ab}(S) & \to & 0 \\
\label{eq:rep2}
Y^b \overline{H}^2_{ab}(S) & \to & 0 \\
\label{eq:rep3}
Y^b H^2_{ab}(S) & \to & 0 \\
\label{eq:rep4}
\Lambda^2_{ab}(S) Y^b & \to & 0 \\
\label{eq:rep5}
g^2 Y^b Y^{2F}_{ab}(S) & \to & 0 \\
\label{eq:rep6}
g^4 C_2^{ab}(S) \{C_2(F),Y^b\} & \to & 0 \\
\label{eq:rep7}
g^4 C_2^{ab}(S) C_2^{bc}(S) Y^c & \to & 0 \\
\label{eq:rep8}
g^4 C_2^{ab}(S) [x_1 C_2(G) + x_2 S_2(F) + x_3 S_2(S)] Y^b & \to & 0 \, .
\end{eqnarray}
However, we find differences compared to the results of Ref.~\cite{Luo:2002ti}, where the following replacements have been made:
\begin{align}
\label{eq:luorep1}
Y^b Y_2^{ab}(S) & \to \frac{1}{2} Y^b \Tr[m_f^\dagger Y^b+Y^{\dagger b} m_f]  \\
\label{eq:luorep2}
Y^b \overline{H}^2_{ab}(S) & \to \frac{1}{2} Y^b \Tr[m_f Y^{\dagger c} Y^b Y^{\dagger c} 
+ m_f^\dagger Y^c Y^{\dagger b} Y^c]\\
\label{eq:luorep3}
 Y^b H^2_{ab}(S)  & \to  \frac{1}{2} Y^b \Tr[m_f Y^{\dagger b} Y^b Y_2^{\dagger}(F)+ m_f^\dagger Y^b Y_2(F)]\\
\label{eq:luorep4}
 \Lambda^2_{ab}(S) Y^b & \to \frac{1}{6} h_{cde} \lambda_{bcde} Y^b\\
\label{eq:luorep5}
g^2 Y^b Y^{2F}_{ab}(S)  & \to \frac{1}{2} g^2 Y^b \Tr[C_2(F)(m_f Y^{\dagger b} + Y^b m_f^\dagger)]\\
\label{eq:luorep6}
g^4 C_2^{ab}(S) \{C_2(F),Y^b\} & \to  0 \\
\label{eq:luorep7}
g^4 C_2^{ab}(S) C_2^{bc}(S) Y^c & \to  0 \\
\label{eq:luorep8}
g^4 C_2^{ab}(S) [\ldots] Y^b & \to  0 \, .
\end{align}
Thus, there is a disagreement between Eqs.\ \eqref{eq:rep1} and \eqref{eq:luorep1} entering the one-loop beta-function for $m_f$.
Furthermore, there are differences between Eqs.\ \eqref{eq:rep2}--\eqref{eq:rep5} and Eqs.\ \eqref{eq:luorep2}--\eqref{eq:luorep5}
affecting the two-loop beta-function.

\begin{figure}[tb]
\begin{eqnarray*}
\begin{aligned}
\includegraphics[width=0.3\linewidth]{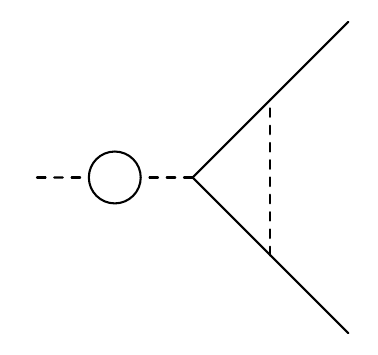}
\end{aligned}
&\quad \to \quad& 
\begin{aligned}
\includegraphics[width=0.3\linewidth]{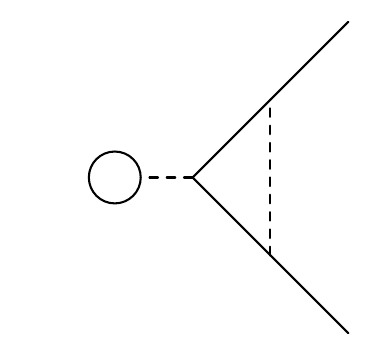}
\end{aligned}
\end{eqnarray*}
\caption{Two-loop diagram which does not contribute to the $\beta$-function of the fermion mass when replacing the external scalar by a dummy field as indicated here. The contribution depicted on the right 
hand side was included in Ref.~\cite{Luo:2002ti}. 
}
\label{fig:Mu2L}
\end{figure}

We now turn to the vertex corrections. At one-loop level, there is only one diagram which needs to be considered: 
\allowdisplaybreaks
\begin{eqnarray}
\begin{aligned}
\includegraphics[width=0.3\linewidth]{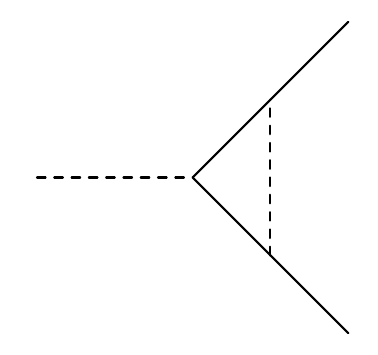}
\end{aligned}
&\quad \to \quad&  \hspace{-2cm}
\begin{aligned}
\includegraphics[width=0.3\linewidth]{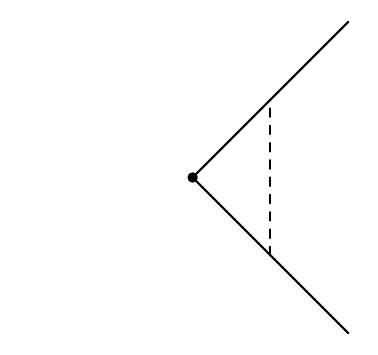}
\end{aligned}
\nonumber \\
Y^b Y^{\dagger a} Y^b
&& Y^b m_f^\dagger Y^b
\end{eqnarray}
At the two-loop level, there are many more contributions. The explicit diagrams are given in Appendix~\ref{app:fermion}. While we completely agree with Ref.~\cite{Luo:2002ti} for the one-loop vertex corrections, we also 
found differences at the two-loop level. Those stem from diagrams involving both, wave-function corrections of scalars as well as vertex corrections, as depicted in Fig.~\ref{fig:Mu2L}. According to our reasoning, these diagrams are also converted into 
tadpole diagrams which drop out. 

Summarising our results, we find that the one-loop $\beta$-functions of fermion masses 
have one term less than the expression given in Ref.~\cite{Luo:2002ti} and are given by the following form:
\begin{align}
{\beta}^I_{m_f} & = \frac{1}{2} \left[ Y_2^{\dagger}(F) m_f + m_f Y_2(F) \right] + 2Y^b m^{\dagger}_f Y^b
-3 g^2 \{C_2(F), m_f\} .
\label{eq:bm1}
\end{align}
At the two-loop level, we obtain
\begin{align}
{\beta}^{II}_{m_f} & = 2 Y^c Y^{\dagger b} m_f (Y^{\dagger c} Y^b - Y^{\dagger b} Y^c)
- Y^b \left[ Y_2(F) m^{\dagger}_f + m^{\dagger}_f Y_2^{\dagger}(F) \right] Y^b
\nonumber\\
&\phantom{=}
- \frac{1}{8} \left[ Y^b Y_2(F) Y^{\dagger b} m_f + m_f Y^{\dagger b} Y_2^{\dagger}(F) Y^b  \right]
- 4\kappa Y_2^{bc}(S) Y^b m_f^{\dagger} Y^{c}
\nonumber\\
&\phantom{=}
- \frac{3}{2} \kappa Y_2^{bc}(S) (Y^b Y^{\dagger c} m_f + m_f Y^{\dagger c} Y^b)
-2 h_{bcd} Y^b Y^{\dagger c} Y^d
\nonumber\\
&\phantom{=}
+ 3 g^2 \{C_2(F), Y^b m^{\dagger}_f Y^b \} + 5 g^2 Y^b \{C_2(F), m^{\dagger}_f \} Y^b
\nonumber\\
&\phantom{=}
- \frac{7}{4} g^2 \left[ C_2(F) Y_2^{\dagger}(F) m_f + m_f Y_2(F) C_2(F) \right]
\nonumber\\
&\phantom{=}
- \frac{1}{4} g^2 \left[Y^b C_2(F) Y^{\dagger b} m_f + m_f Y^{\dagger b} C_2(F) Y^b\right]
\nonumber\\
&\phantom{=}
+ 6 g^2 \left[t^{A *} m_f Y^{\dagger b} t^{A *} Y^b + Y^b t^A Y^{\dagger b} m_f t^A \right] +6 g^2 C_2^{bc}(S) Y^b m^{\dagger}_f Y^c
\nonumber\\
&\phantom{=}
 - \frac{3}{2} g^4 \{\left[C_2(F)\right]^2, m_f \} + \frac{9}{2} g^2 C_2^{bc}(S) (Y^b Y^{\dagger c} m_f + m_f Y^{\dagger c} Y^b)
\nonumber\\
&\phantom{=}
+ g^4 \left[ -\frac{97}{6} C_2(G) + \frac{10}{3} \kappa S_2(F)
+ \frac{11}{12} S_2(S) \right] 
\{C_2(F), m_f \}\,.
\label{eq:bm2}
\end{align}
Here, we disagree in several terms as discussed above. 
The numerical impact of these differences compared to earlier results is briefly discussed at the example 
of a specific model in Sec.~\ref{sec:numerics}.

\subsection{Trilinear coupling}
\label{sec:trilinear}

We now turn to the purely scalar interactions. The $\beta$-functions of the cubic interactions are obtained from the expressions for the quartic couplings by 
replacing one external scalar by a dummy field. The translation of the wave-function contributions between both cases is straightforward and can be summarized as follows:
\begin{eqnarray}
\begin{aligned}
\includegraphics[width=0.3\linewidth]{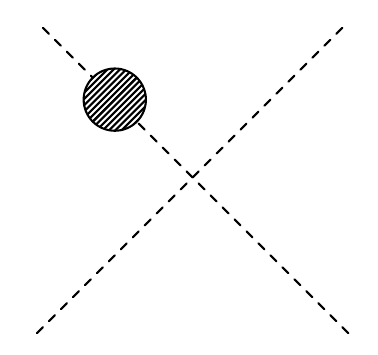}
\end{aligned}
&\quad \to \quad& 
\begin{aligned}
\includegraphics[width=0.3\linewidth]{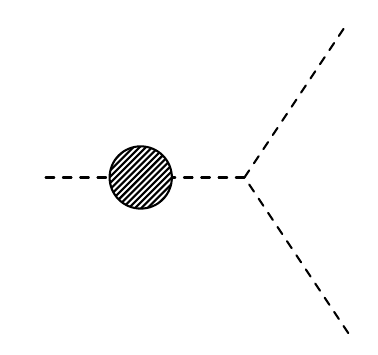}
\end{aligned}
\nonumber \\
\Lambda^Y_{abcd} = \frac16 \sum_{per} Y_2^{af}(S) \lambda_{fbcd} 
& \to &
\Lambda^Y_{abc} = \frac12 \sum_{per} Y_2^{af}(S) h_{fbc} \\
\Lambda^S_{abcd} = \sum_i C_2(i) \lambda_{abcd} 
& \to &
\Lambda^S_{abc} = \sum_i C_2(i) h_{abc} \\
\frac16 \sum_{per} \Lambda^2_{af}(S) \lambda_{fbcd} &\to &
\frac12 \sum_{per} \Lambda^2_{af}(S) h_{fbc} \\
\frac16 \sum_{per} (3 H^2_{af}(S) + 2 \overline{H}^2_{af}(S)) \lambda_{fbcd} &\to &
\frac12 \sum_{per} (3 H^2_{af}(S) + 2 \overline{H}^2_{af}(S)) h_{fbc}  \\
\frac16 \sum_{per} Y^{2F}_{af}(S) \lambda_{fbcd} &\to& 
\frac12 \sum_{per} Y^{2F}_{af}(S) h_{fbc} \\
X \Lambda^S_{abcd} & \to & X\Lambda^S_{abc} \\
\Lambda^{SS}_{abcd} = \sum_i |C_2(i)|^2 \lambda_{abcd} &\to & 
\Lambda^{SS}_{abc} = \sum_i |C_2(i)|^2 h_{abc}  
\end{eqnarray}
In this notation, the index $i$ is summed over all uncontracted scalar indices.
Furthermore, '$X$' denotes the combination of group invariants multiplying $\Lambda^S_{abcd}$ in
Eq.\ \eqref{eq:quartic2}.
As discussed above,
we have modi\-fied the parts which involve Yukawa or quartic couplings compared to Ref.~\cite{Luo:2002ti}. The reason is that in these cases new contributions can be present due to off-diagonal wave-function renormalisation corrections. 
There are three generically different vertex corrections which contribute to the RGE of the quartic interaction. However, since the dummy field does not interact with the gauge sector, those kind of contributions 
do not appear in the case of the cubic interaction. Therefore, the translation at the one-loop level becomes:
\begin{eqnarray}
\label{eq:cubic1start}
\begin{aligned}
\includegraphics[width=0.3\linewidth]{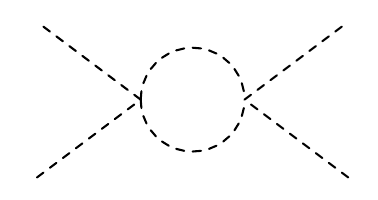}
\end{aligned}
&\quad \to \quad& 
\begin{aligned}
\includegraphics[width=0.3\linewidth]{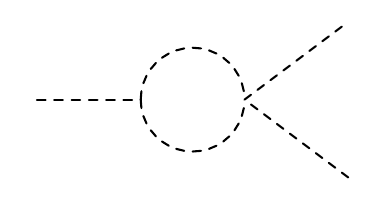}
\end{aligned}
\nonumber \\
\Lambda^2_{abcd}=\frac18 \sum_{per} \lambda_{abef} \lambda_{efcd} 
& &\Lambda^2_{abc}= \frac12 \sum_{per} \lambda_{abef} h_{efc}\\[4mm]
\midrule
\label{eq:cubic1end}
\begin{aligned}
\includegraphics[width=0.3\linewidth]{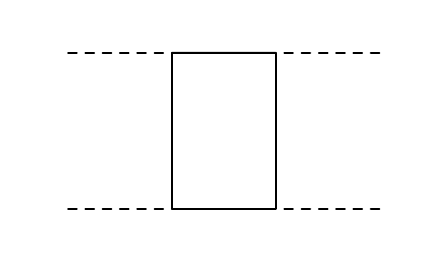}
\end{aligned}
&\quad \to \quad& 
\begin{aligned}
\includegraphics[width=0.3\linewidth]{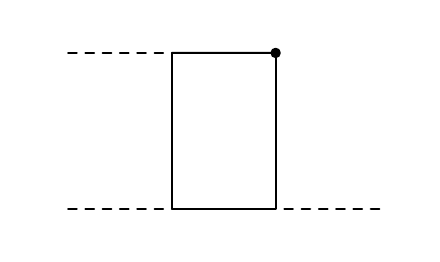} \hspace{-1cm}
\includegraphics[width=0.3\linewidth]{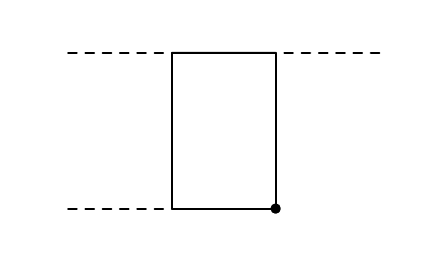}
\end{aligned}
\nonumber \\
H_{abcd}=\frac14 \sum_{per} \text{Tr}(Y^a Y^{\dagger b} Y^c Y^{\dagger d} )
& & 
\begin{array}{c}
H_{abc}= \frac12 \sum_{per} \text{Tr}(m_f Y^{\dagger a} Y^{ b} Y^{\dagger c} \\ + m_f^\dagger Y^a  Y^{\dagger b} Y^c) 
\end{array}\nonumber \\
\\
\midrule
\begin{aligned}
\includegraphics[width=0.3\linewidth]{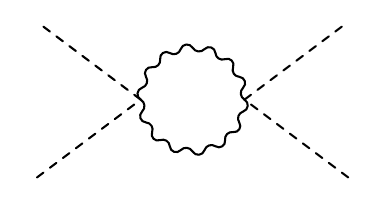}
\end{aligned}
&\quad \to \quad& \hspace{3cm}
\begin{aligned}
\includegraphics[width=0.1\linewidth]{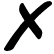}
\end{aligned}
\nonumber \\
A_{abcd}=\frac18 \sum_{per} \{\theta^A,\theta^B\}_{ab} \{\theta^A,\theta^B\}_{cd}
& & \hspace{3cm} 0 \\
\midrule
\nonumber 
\end{eqnarray}
The explicit form of the  two-loop diagrams as well as their expressions in both cases are given in Appendix~\ref{app:cubic}. We find agreement between our results and those of 
Ref.~\cite{Luo:2002ti} at the one- and two-loop level up to the differences from off-diagonal wave-function renormalisations. Thus, the $\beta$-functions at the one- and two-loop levels are
\begin{align}
\beta_{h_{abc}}^{I} =&
\Lambda_{abc}^2 - 8\kappa H_{abc} + 2\kappa \Lambda_{abc}^Y - 3g^2 \Lambda_{abc}^S \,,\\
\beta_{h_{abc}}^{II} =& {\frac14} \sum_{per} \Lambda^2_{af}(S) h_{fbc} 
 -  \bar{\Lambda}_{abc}^3 - 4\kappa \bar{\Lambda}_{abc}^{2Y}  \nonumber \\ 
&+ \kappa \left[ 8\bar{H}_{abc}^{\lambda m} + 8\bar{H}_{abc}^h - \frac12 \sum_{per} \left[ 3H^2_{af}(S) + 2\bar{H}^2_{af}(S) \right]  h_{fbc} \right] \nonumber \\
&+ 4\kappa(H_{abc}^Y + 2\bar{H}_{abc}^Y + 2H_{abc}^3) \nonumber \\
&+ g^2 \left[ 2\bar{\Lambda}_{abc}^{2S} - 6\Lambda_{abc}^{2g} + 4\kappa(H_{abc}^S - H_{abc}^F)
   + 5 \kappa \sum_{per} Y^{2F}_{af}(S) h_{fbc} \right] \nonumber \\
&- g^4 \left\{ \left[ \frac{35}{3}C_2(G) - \frac{10}{3}\kappa S_2(F) - \frac{11}{12}S_2(S) \right]
    \Lambda_{abc}^S \right. \nonumber \\
& \ \ \ \ \ \ \ \left. -\frac{3}{2}\Lambda_{abc}^{SS} - \frac{5}{2}A_{abc}^\lambda
   - {\frac12}\bar{A}_{abc}^\lambda  + 4\kappa (B_{abc}^Y - 10\bar{B}_{abc}^Y) \right\}\,,
\end{align}
where the invariants are defined in Eqs.~(\ref{eq:cubic1start})--(\ref{eq:cubic1end}) and (\ref{eq:cubic2start})--(\ref{eq:cubic2end}).

\subsection{Scalar mass}
\label{sec:scalar_mass}

Finally, we turn to the terms involving two scalar couplings. The procedure is very similar to the case 
of the cubic scalar coupling, and we find the following relations for the wave-function corrections to the terms appearing for the quartic scalar coupling:
\begin{eqnarray}
\begin{aligned}
\includegraphics[width=0.3\linewidth]{wave_quartic.pdf}
\end{aligned}
&\quad \to \quad& 
\begin{aligned}
\includegraphics[width=0.3\linewidth]{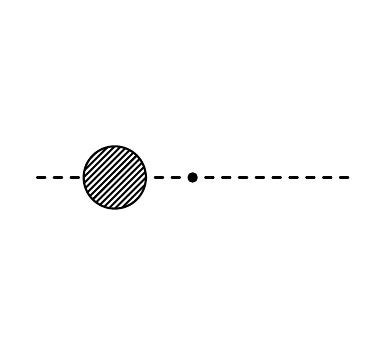}
\end{aligned}
\nonumber \\
\Lambda^Y_{abcd} = \frac16 \sum_{per} Y_2^{ae}(S) \lambda_{ebcd} & \to &
\Lambda^Y_{ab} = 2 \sum_{per} Y_2^{ae}(S) m^2_{eb} \\
\Lambda^S_{abcd} = \sum_i C_2(i) \lambda_{abcd} 
& \to &
\Lambda^S_{ab} = 2 \sum_i C_2(i) m^2_{ab} \\
\frac16 \sum_{per} \Lambda^2_{ae}(S) \lambda_{ebcd} &\to &
2 \sum_{per} \Lambda^2_{ae}(S) m^2_{eb} \\
\frac16 \sum_{per} (3 H^2_{af}(S) + 2 \overline{H}^2_{af}(S)) \lambda_{fbcd} &\to &
2 \sum_{per} (3 H^2_{af}(S) + 2 \overline{H}^2_{af}(S)) m^2_{fb}  \\
\frac16 \sum_{per} Y^{2F}_{af}(S) \lambda_{fbcd} &\to& 
2 \sum_{per} Y^{2F}_{af}(S) m^2_{fb} \\
X \Lambda^S_{abcd} & \to & X \Lambda^S_{ab} \\
\Lambda^{SS}_{abcd} = \sum_i |C_2(i)|^2 \lambda_{abcd} &\to & 
\Lambda^{SS}_{ab} = 2 \sum_i |C_2(i)|^2 m^2_{ab}\,.  
\end{eqnarray}
Again, '$X$' denotes the combination of group invariants multiplying $\Lambda^S_{abcd}$ in
Eq.\ \eqref{eq:quartic2}.

Again, we need to consider the three generically different diagrams which contribute to the running of the quartic functions. The one with vector bosons in the loop vanishes due to inserting dummy fields, 
while for the other two diagrams additional terms arise. 
\begin{eqnarray}
\label{eq:bi1start}
\begin{aligned}
\includegraphics[width=0.3\linewidth]{Lambda2S_1L.pdf}
\end{aligned}
&\quad \to \quad& 
\begin{aligned}
\includegraphics[width=0.3\linewidth]{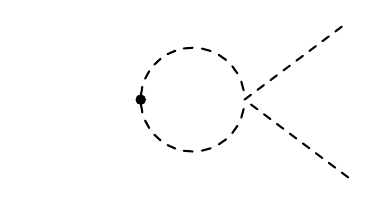} \\
\includegraphics[width=0.3\linewidth]{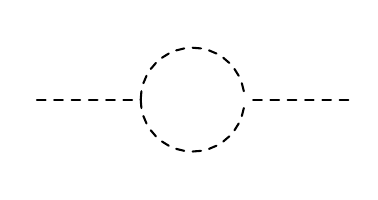}
\end{aligned}
\nonumber \\
\Lambda^2_{abcd}=\frac18 \sum_{per} \lambda_{abef} \lambda_{efcd} 
&&  2 m^2_{ef} \lambda_{abef} + 2 h_{aef} h_{bef}
\\[4mm]
\midrule
\label{eq:bi1end}
\begin{aligned}
\includegraphics[width=0.3\linewidth]{HY_1L.pdf}
\end{aligned}
&\quad \to \quad& 
\begin{aligned}
\includegraphics[width=0.3\linewidth]{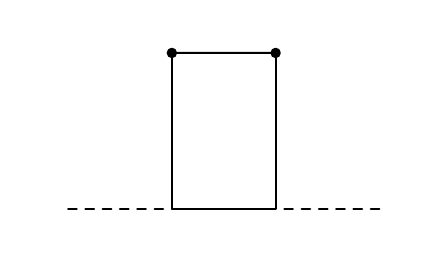} \hspace{-1cm}
\includegraphics[width=0.3\linewidth]{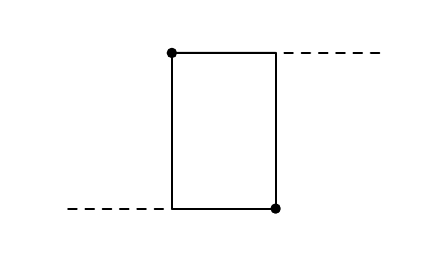}
\end{aligned}
\nonumber \\
H_{abcd}=\frac14 \sum_{per} \text{Tr}(Y^a Y^{\dagger b} Y^c Y^{\dagger d} )
&& 
\begin{array}{c}
H_{ab}=  \sum_{per} \text{Tr}(Y^a Y^{\dagger b} m_f m_f^\dagger + Y^{\dagger a} Y^b m_f^\dagger m_f\\ + \frac12 Y^{\dagger a} m_f Y^{b \dagger } m_f + \frac12 Y^a m_f^\dagger Y^b m_f^\dagger) 
\end{array}\nonumber \\
\\
\midrule
\begin{aligned}
\includegraphics[width=0.3\linewidth]{Lambda2g_1L.pdf}
\end{aligned}
&\quad \to \quad&  \hspace{3cm}
\begin{aligned}
\includegraphics[width=0.1\linewidth]{cross.pdf}
\end{aligned}
\nonumber \\
A_{abcd}=\frac18 \sum_{per} \{\theta^A,\theta^B\}_{ab} \{\theta^A,\theta^B\}_{cd}
&& \hspace{3cm}0 \\
\midrule
\nonumber
\end{eqnarray}
The two-loop diagrams are given in Appendix~\ref{app:bilinear}. We also find  agreement between our results here and the ones given in Ref.~\cite{Luo:2002ti} up to the wave-function renormalisation. One needs to be careful about 
some factor of $\frac12$ due to $\beta_{m^2_{ab}} = \frac12 \beta_{\lambda_{ab\hat{d}\hat{d}}}$, which we have included here explicitly into the definition of the $\beta$-function for $m^2_{ab}$, while it has been partially absorbed into other definitions in Ref.~\cite{Luo:2002ti}. Thus, with our conventions the one- and two-loop $\beta$-functions 
read
\begin{align}
\beta_{m^2_{ab}}^{I} =&
m^2_{ef} \lambda_{abef} + h_{aef} h_{bef} - 4\kappa H_{ab} + \kappa \Lambda_{ab}^Y - \frac32 g^2 \Lambda_{ab}^S \,,\\
\beta_{m^2_{ab}}^{II} =& \frac12 \sum_{per} \Lambda^2_{af}(S) m^2_{fb} 
 -  \frac12 \bar{\Lambda}_{ab}^3 - 2\kappa \bar{\Lambda}_{ab}^{2Y}  \nonumber \\ 
&+ \kappa \left[ 4\bar{H}_{ab}^{\lambda} -  \sum_{per} \left[ 3H^2_{af}(S) + 2\bar{H}^2_{af}(S) \right] m^2_{fb} \right] \nonumber \\
&+ 2\kappa(H_{ab}^Y + 2 \bar{H}_{ab}^Y + 2 H_{ab}^3) \nonumber \\
&+ g^2 \left[ \bar{\Lambda}_{ab}^{2S} - 3\Lambda_{ab}^{2g} + 2\kappa(H_{ab}^S - H_{ab}^F)
   + 10 \kappa \sum_{per} Y^{2F}_{af}(S) m^2_{fb} \right] \nonumber \\
&- g^4 \left\{ \left[ \frac{35}{6}C_2(G) - \frac{5}{3}\kappa S_2(F) - \frac{11}{24}S_2(S) \right]
    \Lambda_{ab}^S \right. \nonumber \\
& \ \ \ \ \ \ \ \left. -\frac{3}{4}\Lambda_{ab}^{SS} - \frac{5}{4}A_{ab}^\lambda
   - {\frac14}\bar{A}_{ab}^\lambda  + 2\kappa (B_{ab}^Y - 10\bar{B}_{ab}^Y) \right\}\,,
\end{align}
where we used the objects defined in Eqs.~(\ref{eq:bi1start})--(\ref{eq:bi1end}) and (\ref{eq:bi2start})--(\ref{eq:bi2end}).

\section{Comparison with supersymmetric RGEs}
\label{sec:susy}
We have now re-derived the full one- and two-loop RGEs for the dimensionful parameters. 
While we agree with Ref.~\cite{Luo:2002ti} concerning the bilinear and cubic scalar interactions (up to wave-function renormalisation), we find differences in the fermion mass terms. 
Therefore, we want to double-check our results by comparing to those obtained using supersymmetric (SUSY) RGEs. 
The general RGEs for a softly broken SUSY model have been independently calculated in Refs.~\cite{Martin:1993zk,Yamada:1994id,Jack:1997eh} 
and the general agreement between all results has been discussed in Ref.~\cite{Jack:1994rk}. Thus, there is hardly any doubt that these RGEs are absolutely correct. Therefore, we want to test
our results with a model in which we enforce SUSY relations among parameters. 
After a translation from the \MS to the \DR scheme one should recover the SUSY results. \\

Since a supersymmetric extension of the SM yields many couplings which are generically all of the same form, we opt for a more compact theory. 
We consider a toy model with one vector superfield $\hat{B}$ and three chiral superfields
\begin{align}
\hat{H}_d: \ \ & Q=-\frac12\, , \\
\hat{H}_u: \ \ & Q=\frac12\, , \\
\hat{S}: \ \ & Q=0 \, ,
\end{align}
where $Q$ denotes the electric charge.
The superpotential consists of two terms\footnote{We neglect terms $\sim \hat{S}^2,\,\hat{S}^3$ which are not essential for our argument.}
\begin{equation}
W = \lambda \hat{H}_u \hat{H}_d \hat{S} + \mu  \hat{H}_u \hat{H}_d 
\end{equation}
and the soft-breaking terms are 
\begin{align} 
- L_{SB} = \, & \left( B_{\mu} H_d H_u  +  T_{\lambda} H_d H_u S + \frac{1}{2} M_B \tilde{B}^{2}  + \mbox{h.c.}\right) + \nonumber  \\ 
& m_{H_d}^2 |H_d|^2  + m_{H_u}^2 |H_u|^2  + m_S^2 |S|^2\,.
\end{align} 
This model contains all of the relevant generic structure we need to test. Making use of the results of Ref.~\cite{Martin:1993zk}, which are also implemented in the package \sarah, we find 
the following expressions for the one- and two-loop RGEs for the different parts of the model:
\begin{enumerate}
\item {Gauge Couplings}
{\allowdisplaybreaks  \begin{align} 
\beta_{g}^{(1)} & =  
\frac{1}{2} g^{3} \\ 
\beta_{g}^{(2)} & =  
\frac{1}{2} g^{3} \Big(-2 |\lambda|^2  + g^{2}\Big)
\end{align}} 
\item {Gaugino Mass Parameters}
{\allowdisplaybreaks  \begin{align} 
\beta_{M_B}^{(1)} & =  
g^{2} M_B \\ 
\beta_{M_B}^{(2)} & =  
2 g^{2} \Big(g^{2} M_B  + \lambda^* \Big(- M_B \lambda  + T_{\lambda}\Big)\Big)
\end{align}} 
\item {Trilinear Superpotential Parameters}
{\allowdisplaybreaks  \begin{align} 
\beta_{\lambda}^{(1)} & =  
\lambda \Big(3 |\lambda|^2  - g^{2} \Big)\\ 
\beta_{\lambda}^{(2)} & =  
\lambda \Big(-6 |\lambda|^4  + g^{2} |\lambda|^2  + g^{4}\Big)
\end{align}} 
\item {Bilinear Superpotential Parameters}
{\allowdisplaybreaks  \begin{align} 
\beta_{\mu}^{(1)} & =  
- \mu \Big(-2 |\lambda|^2  + g^{2}\Big)\\ 
\beta_{\mu}^{(2)} & =  
\mu \Big(-4 |\lambda|^4  + g^{4}\Big)
\end{align}} 
\item {Trilinear Soft-Breaking Parameters}
{\allowdisplaybreaks  \begin{align} 
\beta_{T_{\lambda}}^{(1)} & =  
2 g^{2} M_B \lambda  - \Big(-9 |\lambda|^2  + g^{2}\Big)T_{\lambda} \\ 
\beta_{T_{\lambda}}^{(2)} & =  
-30 |\lambda|^4 T_{\lambda}  + g^{2} |\lambda|^2 \Big(-2 M_B \lambda  + 3 T_{\lambda} \Big) + g^{4} \Big(-4 M_B \lambda  + T_{\lambda}\Big)
\end{align}} 
\item {Bilinear Soft-Breaking Parameters}
{\allowdisplaybreaks  \begin{align} 
\beta_{B_{\mu}}^{(1)} & =  
2 g^{2} M_B \mu  + 4 |\lambda|^2 B_{\mu}  + 4 \mu \lambda^* T_{\lambda}  - g^{2} B_{\mu} \\ 
\beta_{B_{\mu}}^{(2)} & =  
\Big(2 g^{2} |\lambda|^2  -8 |\lambda|^4  + g^{4}\Big)B_{\mu}  -2 \mu \Big(10 |\lambda|^2 \lambda^* T_{\lambda}  + 2 g^{4} M_B  + g^{2} M_B |\lambda|^2 \Big)
\end{align}} 
\item {Soft-Breaking Scalar Masses}
{\allowdisplaybreaks  \begin{align} 
\beta_{m_{H_d}^2}^{(1)} & =  
-2 g^{2} |M_B|^2  + 2 \Big(m_{H_d}^2 + m_{H_u}^2 + m_S^2\Big)|\lambda|^2  + 2 |T_{\lambda}|^2  - \frac{1}{2} g^2 \Big(- m_{H_d}^2  + m_{H_u}^2\Big) \\ 
\beta_{m_{H_d}^2}^{(2)} & =  
6 g^{4} |M_B|^2  -8 \Big(m_{H_d}^2 + m_{H_u}^2 + m_S^2\Big)|\lambda|^4  -16 |\lambda|^2 |T_{\lambda}|^2 \nonumber \\ 
& + g^{4} m_{H_d}^2  + g^2 |\lambda|^2 \Big(m_{H_u}^2  - m_{H_d}^2\Big) \\[4mm] 
\beta_{m_{H_u}^2}^{(1)} & =  
-2 g^{2} |M_B|^2  + 2 \Big(m_{H_d}^2 + m_{H_u}^2 + m_S^2\Big)|\lambda|^2  + 2 |T_{\lambda}|^2  + \frac{1}{2} g^2 \Big(- m_{H_d}^2  + m_{H_u}^2\Big) \\ 
\beta_{m_{H_u}^2}^{(2)} & =  
6 g^{4} |M_B|^2  -8 \Big(m_{H_d}^2 + m_{H_u}^2 + m_S^2\Big)|\lambda|^4  -16 |\lambda|^2 |T_{\lambda}|^2 \nonumber \\
& + g^{4} m_{H_u}^2  +g^2 |\lambda|^2 \Big(-m_{H_u}^2  + m_{H_d}^2\Big)\\[4mm] 
\beta_{m_S^2}^{(1)} & =  
2 \Big(\Big(m_{H_d}^2 + m_{H_u}^2 + m_S^2\Big)|\lambda|^2  + |T_{\lambda}|^2\Big)\\ 
\beta_{m_S^2}^{(2)} & =  
2 \Big(-4 \Big(m_{H_d}^2 + m_{H_u}^2 + m_S^2\Big)|\lambda|^4 +\lambda^* \Big(g^{2} M_B^* \Big(2 M_B \lambda  - T_{\lambda} \Big) + \nonumber \\
& \lambda \Big(-8 |T_{\lambda}|^2  + g^{2} \Big(m_{H_d}^2 + m_{H_u}^2 + m_S^2\Big)\Big)\Big) +g^{2} T_{\lambda}^* \Big(- M_B \lambda  + T_{\lambda}\Big)\Big)
\end{align}} 
\end{enumerate}
As before, we have suppressed the pre-factors $\frac{1}{16\pi^2}$ and $\frac{1}{(16\pi^2)^2}$ for the one- and two-loop 
$\beta$-functions. 
With these functions, the running of all parameters at the one- and two-loop level is fixed. However, for later comparison, it will be convenient to know the $\beta$-functions for some products
of parameters as well. That is done by applying the chain rule:
\begin{align}
\beta^{(1)}_{\frac18 g^2} =& \frac14 g \beta^{(1)}_g =  \frac{1}{8} g^{4}\,, \\
\beta^{(2)}_{\frac18 g^2} =& \frac14 g \beta^{(2)}_g = - \frac14 |\lambda|^2 g^{4}  + \frac18 g^{6}\,, \\[4mm]
\beta^{(1)}_{|\lambda|^2} =& \lambda (\beta^{(1)}_\lambda)^* + \lambda^* \beta^{(1)}_\lambda =  2 |\lambda|^2 \Big(3 |\lambda|^2  - g^{2} \Big)\,,\\
\beta^{(2)}_{|\lambda|^2} =& \lambda (\beta^{(2)}_\lambda)^* + \lambda^* \beta^{(2)}_\lambda =  2 |\lambda|^2 \Big(-6 |\lambda|^4  + g^{2} |\lambda|^2  + 
g^{4}\Big)\,,\\[4mm]
\beta^{(1)}_{|\lambda|^2-\frac14 g^2} =& 2 \lambda \beta^{(1)}_\lambda -\frac12 g \beta^{(1)}_g =  -2 |\lambda|^2 g^2 +  6 |\lambda|^4 -\frac{1}{4} g^{4}\,,\\
\beta^{(2)}_{|\lambda|^2-\frac14 g^2} =& 2 \lambda \beta^{(2)}_\lambda -\frac12 g \beta^{(2)}_g =  -12 |\lambda|^6 -\frac{1}{4} g^{6}  +  \frac{5}{2} g^{4} |\lambda|^2   + 2 g^{2} |\lambda|^4  \,,\\[4mm]
\beta^{(1)}_{\lambda \mu^*} =& \lambda (\beta^{(1)}_\mu)^* + \mu^* \beta^{(1)}_{\lambda} =  \mu^* \lambda (-2 g^{2}  + 5 |\lambda|^2 ) \,,\\
\beta^{(2)}_{\lambda \mu^*} =& \lambda (\beta^{(2)}_\mu)^* + \mu^* \beta^{(2)}_{\lambda} =  \mu^* \lambda (-10 |\lambda|^{4} + 2 g^{4}  + g^{2} |\lambda|^2 ) \,,\\[4mm]
\beta^{(1)}_{|\mu|^2} =& \mu (\beta^{(1)}_\mu)^* + \mu^* \beta^{(1)}_\mu=-2 |\mu|^2 \Big(-2 |\lambda|^2  + g^{2}\Big) \,,\\
\beta^{(2)}_{|\mu|^2} =& \mu (\beta^{(2)}_\mu)^* + \mu^* \beta^{(2)}_\mu=2 |\mu|^2 \Big(-4 |\lambda|^4  + g^{4}\Big) \,.
\end{align}

We now consider the same model written as non-supersymmetric version. In this case, we have one gauge boson $B$, four fermions
\begin{align}
\tilde{H}_d:  \ \ & Q=-\frac12\, , \\ 
\tilde{H}_u:  \ \ & Q=\frac12\, , \\
\tilde{S}:  \ \ & Q=0\, , \\
\tilde{B}:  \ \ & Q=0\, ,
\end{align}
and three scalars
\begin{align}
H_d:  \ \ & Q=-\frac12 \, ,\\ 
H_u:  \ \ & Q=\frac12\, , \\
S:  \ \ & Q=0 \,.
\end{align}
The full potential for this models involves a substantial amount of different couplings
\begin{align}  \label{eq:non-susy}
V = \, & \left( T_1 S  |H_d|^2 + T_2  S |H_u|^2 + {T}_3 H_d H_u S  + \mbox{h.c.}\right) \nonumber \\
  & + m^2_{1} |H_d|^2  + m^2_{2} |H_u|^2  +  m^2_{3} |S|^2  \nonumber \\
  & \hspace{1cm} + \lambda_1 |S|^2 |H_d|^2 + \lambda_2 |S|^2 |H_u|^2 + \lambda_3 |H_d|^2 |H_u|^2  + \lambda_4 |H_d|^4 + \lambda_5 |H_u|^4 \nonumber \\
  & + \left(M_1 \tilde{B} \tilde{B} + M_2 \tilde{H}_d \tilde{H}_u + B H_d H_u +  \mbox{h.c.}\right) \nonumber \\
  & +\left(Y_1 S \tilde{H}_d \tilde{H}_u + Y_2 \tilde{S} H_d \tilde{H}_u + Y_{3} \tilde{S} \tilde{H}_d H_u - \frac{1}{\sqrt{2}} g_d \tilde{B} \tilde{H}_d H_d^{*} +  \frac{1}{\sqrt{2}} g_u \tilde{B} \tilde{H}_u H_u^{*} + \mbox{h.c.}\right)\,.
\end{align}
We think that this rather lengthy form justifies our approach to consider only a toy model, but not a realistic SUSY theory. We have neglected couplings that would be allowed by the symmetry of this theory, but vanish as we match to the SUSY model. In particular, CP even and odd part of the complex field $S$ will run differently unless specific (SUSY) relations among the parameters exist. Therefore, one would need to decompose $S$ into its real components and write down all possible potential terms involving these fields.
However, we are only interested in the $\beta$ functions in the SUSY limit where no splitting between these fields is introduced. Therefore, we retain the more compact notation in \eqref{eq:non-susy}.
We can now make use of our revised expressions to calculate the RGEs up to two-loop. For this purpose, we modified the packages \sarah and \pyrate accordingly. 
The lengthy expressions in the general case are given in Appendix~\ref{app:nonsusy}. In order to make connection to the SUSY case, we can make the following associations between parameters of these models:
\begin{align}
g_d = g_u =&  \ \ g \,,\\
Y_1=Y_2 = Y_3 =&  \ \ \lambda \,,\\
\lambda_1 = \lambda_2 = &  \ \ |\lambda|^2 \,,\\
\lambda_3  = &  \ \ |\lambda|^2 - \frac14 g^2 \,,\\
\lambda_4 = \lambda_5 =&   \ \ \frac18 g^2 \,,\\
T_1 = T_2 = &  \ \ \mu^* \lambda \,,\\
T_3 = &  \ \ T_\lambda \,,\\
M_1 = &  \ \ \frac12 M_B \,,\\
M_2 = &  \ \ \mu  \,,\\
m_1^2= &  \ \ m_{H_d}^2 + |\mu|^2 \,,\\
m_2^2= &  \ \ m_{H_u}^2 + |\mu|^2 \,,\\
m_3^2= &  \ \ m_S^2 \,,\\
B = &   \ \ B_\mu \,.
\end{align}
By doing that, we obtain the following RGEs:
\begin{enumerate}
\item {Gauge Couplings}
{\allowdisplaybreaks  \begin{align} 
\beta_{g}^{(1)} & =  
\frac{1}{2} g^{3} \\ 
\beta_{g}^{(2)} & =  
\frac{1}{2} g^{3} \Big(-2 |\lambda|^2  + g^{2}\Big)
\end{align}} 
\item {Quartic scalar couplings}
{\allowdisplaybreaks  \begin{align} 
\beta_{\lambda_1}^{(1)} & =  \beta_{\lambda_2}^{(1)} =
2 |\lambda|^2 \Big(3 |\lambda|^2  - g^{2} \Big)\\ 
\beta_{\lambda_1}^{(2)} & =   \beta_{\lambda_2}^{(2)} =
2 |\lambda|^2 \Big( -6 |\lambda|^4+ \frac54 g^{2} |\lambda|^2   + \frac{17}{8} g^{4}  \Big)\\[4mm] 
\beta_{\lambda_3}^{(1)} & =  
-2 g^{2} |\lambda|^2  + 6 |\lambda|^4  -\frac{1}{4} g^{4} \\ 
\beta_{\lambda_3}^{(2)} & =  
-12 |\lambda|^{6}  -\frac{17}{8} g^{6}  + \frac{31}{4} g^{4} |\lambda|^2  + g^{2} |\lambda|^4 \\[4mm] 
\beta_{\lambda_4}^{(1)} & =   \beta_{\lambda_5}^{(1)} = 
\frac{1}{8} g^{4} \\ 
\beta_{\lambda_4}^{(2)} & =  \beta_{\lambda_5}^{(2)} = 
\frac78 g^{4} |\lambda|^2  - g^{2} |\lambda|^4  +\frac{1}{16} g^{6}\\ 
\end{align}} 
\item {Yukawa Couplings}
{\allowdisplaybreaks  \begin{align} 
\beta_{g_{d}}^{(1)} & =   \beta_{g_{u}}^{(1)} =
\frac{1}{2} g^{3} \\ 
\beta_{g_{d}}^{(2)} & =  \beta_{g_{u}}^{(2)} = 
\frac12 g^3 \Big(-\frac{22}{8}|\lambda|^2  + \frac{11}{8} g^2\Big)\\[4mm]
\beta_{Y_1}^{(1)} & =  
\lambda \Big(3 |\lambda|^2  - g^{2} \Big)\\ 
\beta_{Y_1}^{(2)} & =  
\lambda \Big(- 6 |\lambda|^4  + \frac14 g^{2} |\lambda|^2   + \frac{11}{8} g^{4}  \Big)\\[4mm]
\beta_{Y_2}^{(1)} & =  \beta_{Y_3}^{(1)} =
\lambda \Big(3 |\lambda|^2  - g^{2} \Big)\\ 
\beta_{Y_2}^{(2)} & =  \beta_{Y_3}^{(2)} = 
 \lambda \Big(-6 |\lambda|^4  + \frac{11}{8} g^{2} |\lambda|^2+\frac{13}{16} g^{4}    \Big)
\end{align}} 
\item {Fermion Mass Terms}
{\allowdisplaybreaks  \begin{align} 
\beta_{M_1}^{(1)} & =  
\frac{1}{2} g^{2} M_B \\ 
\beta_{M_1}^{(2)} & =  
g^{2} \Big(\frac98 g^{2} M_B  + \lambda^* \Big(-M_B \lambda  + T_\lambda \Big)\Big)\\[4mm]
\beta_{M_2}^{(1)} & =  
- \mu \Big(-2 |\lambda|^2  + g^{2}\Big)\\ 
\beta_{M_2}^{(2)} & =  
\mu \Big(-4 |\lambda|^4 + \frac{11}{8} g^{4}  -\frac14 g^{2} |\lambda|^2   \Big)
\end{align}}  
\item {Trilinear Scalar couplings}
{\allowdisplaybreaks  \begin{align} 
\beta_{T_1}^{(1)} & =  \beta_{T_2}^{(1)}= \lambda \mu^* \Big(-2 g^{2}  + 5 |\lambda|^2 \Big) \\ 
\beta_{T_1}^{(2)} & =  \beta_{T_2}^{(2)} = 
 \lambda \mu^* \Big(-10 |\lambda|^4 + \frac{17}{4} g^{4}    + 2 g^{2} |\lambda|^2 \Big) \\[4mm]
\beta_{{T}_3}^{(1)} & =  
2 g^{2} M_B \lambda  - \Big(-9 |\lambda|^2  + g^{2}\Big)T_\lambda \\ 
\beta_{{T}_3}^{(2)} & =  
-30 |\lambda|^4 T_\lambda   + g^{2} |\lambda|^2 \Big(-2 M_B \lambda  + 3 T_\lambda \Big) + g^{4}\Big(-4  M_B \lambda  + \frac{7}{4} T_\lambda \Big)
\end{align}} 
\item {Scalar Mass Terms}
{\allowdisplaybreaks  \begin{align} 
\beta_{{B}}^{(1)}  =&  2 g^{2} M_B \mu  + 4 B_\mu |\lambda|^2  + 4 \mu \lambda^* T_\lambda  - B_\mu g^{2} \\ 
\beta_{{B}}^{(2)}  =&  \Big(-8 |\lambda|^4  + \frac{5}{2} g^{2} |\lambda|^2  + \frac{7}{4} g^{4}\Big)B_\mu -2\mu \Big(10 |\lambda|^2 \lambda^* T_\lambda + 2 g^{4} M_B  + g^{2} |\lambda|^2 M_B\Big) \\[4mm]
\beta_{m^2_{1}}^{(1)}  =&  
-2 g^{2} |M_B|^2  + 2 |\lambda|^2 \Big(m_{H_d}^2 + m_{H_u}^2 + m^2_S\Big) + 2 |T_\lambda|^2  + \frac12 g^{2} \Big( m_{H_d}^2 - m_{H_u}^2 \Big) \nonumber \\
 & \hspace{2cm}+ \Big(4 |\lambda|^2 -2 g^{2}\Big) |\mu|^2  \\ 
\beta_{m^2_{1}}^{(2)}  =&  
\frac{11}{2} g^{4} |M_B|^2 -8 \Big(m_{H_d}^2 + m_{H_u}^2 + m^2_S\Big)|\lambda|^4 -16 |T_\lambda|^2  |\lambda|^2  \nonumber \\
&   + \frac12 |\lambda|^2  g^{2} \Big(2 m_{H_u}^2   - m_{H_d}^2 \Big) + \frac14 g^{4} \Big(+ 2 m_{H_u}^2  + 9 m_{H_d}^2 \Big) \nonumber \\
& + |\mu|^2 \Big(\frac32 |\lambda|^2 g^2  -8 |\lambda|^4 +  \frac{17}{4} g^4  \Big) \\[4mm] 
\beta_{m^2_{2}}^{(1)} =&  
-2 g^{2} |M_B|^2  + 2 |\lambda|^2 \Big(m_{H_d}^2 + m_{H_u}^2 + m^2_S\Big) + 2 |T_\lambda|^2  + \frac12 g^{2} \Big(m_{H_u}^2 - m_{H_d}^2  \Big) \nonumber \\
 & \hspace{2cm}+ \Big(4 |\lambda|^2 -2 g^{2}\Big) |\mu|^2 \\ 
\beta_{m^2_{2}}^{(2)} =&  
\frac{11}{2} g^{4} |M_B|^2 -8 \Big(m_{H_d}^2 + m_{H_u}^2 + m^2_S \Big)|\lambda|^4 -16 |T_\lambda|^2  |\lambda|^2  \nonumber \\
&   + \frac12 |\lambda|^2 g^{2} \Big(2 m_{H_d}^2   - m_{H_u}^2 \Big)  + \frac14g^{4} \Big( + 2 m_{H_d}^2  + 9 m_{H_u}^2 \Big)\nonumber \\
& + |\mu|^2 \Big(\frac32 |\lambda|^2 g^2  -8 |\lambda|^4 +  \frac{17}{4} g^4  \Big) \\[4mm] 
\beta_{m^2_{3}}^{(1)} =&  
2 \Big(\Big(m_{H_d}^2 + m_{H_u}^2 + m^2_S\Big)|\lambda|^2  + |T_\lambda|^2\Big)\\ 
\beta_{m^2_{3}}^{(2)} =&  
-2 \Big(4 \Big(m_{H_d}^2+ m_{H_u}^2+ m^2_S\Big)\lambda^{2}  + g^{2} \mu^{2} \Big)\lambda^{*\,2} -2 g^{2} \Big(\lambda^{2} \mu^{*\,2}  + T_\lambda^* \Big(M_B\lambda  - T_\lambda \Big)\Big)\nonumber \\ 
 &+\lambda^* \Big(g^{2} \lambda \Big(2 m_{H_d}^2+ 2 m_{H_u}^2 + 4 |M_B|^2  + 8 |\mu|^2  + m^2_S\Big) -2 \Big(8 \lambda T_\lambda^*  + g^{2} M_B^* \Big)T_\lambda\Big)
\end{align}} 
\end{enumerate}

We see that all one-loop expressions as well as the two-loop $\beta$-function of the gauge coupling agree with the SUSY expressions. The remaining discrepancies at two-loop are due to the differences between \MS and \DR scheme. In order to translate the non-SUSY expressions to the \DR-scheme, we need to apply the following shifts \cite{Martin:1993yx} 
\begin{align}
g_{d,u} \ \ \to\ \  & g_{d,u} \Big(1 - \frac{1}{16\pi^2} \cdot \frac18 g^2 \Big) \,,\\ 
Y_1 \ \ \to\ \  & Y_1 \Big(1 + \frac{1}{16\pi^2} \cdot  \frac14 g^2  \Big) \,,\\
Y_{2,3} \ \ \to\ \  & Y_{2,3} \Big(1 - \frac{1}{16\pi^2} \cdot  \frac18 g^2  \Big) \,,\\
\lambda_3 \ \ \to\ \  &\lambda_3 - \frac{1}{16\pi^2} \cdot  \frac14 g^4 \,,\\
\lambda_{4,5} \ \ \to\ \  & \lambda_{4,5} - \frac{1}{16\pi^2} \cdot  \frac18 g^4 \,,\\
M_2 \ \ \to \ \  &  M_2 \Big(1+ \frac{1}{16\pi^2} \cdot  \frac14 g^2 \Big) \,,
\end{align}
which have to be applied to the expressions of the one-loop $\beta$ functions to obtain the corresponding two-loop shifts. In addition, one must take into account that for the quartic couplings and the Yukawa couplings an additional shift appears `on the left hand side' of the expression, e.g. 
\begin{equation}
\beta^{\DR}_Y = \frac{d}{dt} Y^{\DR} =  \frac{d}{dt} \left(Y^{\MS} \Big(1 + \frac{c}{16\pi^2}  g^2 \Big) \right) = \beta^{\MS}_Y \left(1  + \frac{c}{16 \pi^2} g^2\right) + 2 g Y^{\MS} \frac{c}{16 \pi^2} \beta_g   
\end{equation}
with some coefficient $c$ depending on the charges of the involved fields. \\
We find the following shifts for the different couplings:
\begin{align}
\Delta \lambda_1 &=  -\frac{1}{2} g^{2} |\lambda|^4  -\frac{9}{4} g^{4} |\lambda|^2  \\
\Delta \lambda_3 &= \frac{15}{8} g^{6}  -\frac{21}{4} g^{4} |\lambda|^2  + g^{2} |\lambda|^4  \\
\Delta \lambda_4 &= \frac{1}{16} g^{6}  -\frac{9}{8} g^{4} |\lambda|^2  + g^{2} |\lambda|^4  \\[4mm]
\Delta g_d &= -\frac{3}{16} g^{5}  + \frac{3}{8} g^{3} |\lambda|^2  \\
\Delta Y_1 &= \frac{3}{4} g^{2} \lambda|\lambda|^{2}  -\frac{3}{8} g^{4} \lambda \\
\Delta Y_2 &= \frac{3}{16} g^{4} \lambda  -\frac{3}{8} g^{2} \lambda |\lambda|^{2} \\[4mm]
\Delta M_1 &= -\frac{1}{8} g^{4} M_B\\
\Delta M_2 &= \frac{1}{4} g^{2} \mu |\lambda|^2  -\frac{3}{8} g^{4} \mu \\[4mm]
\Delta T_1 &=   -\frac{1}{4} g^{2} \lambda \Big(4 |\lambda|^2  + 9 g^{2} \Big)\mu^* \\
\Delta T_3 &= -\frac{3}{4} g^{4} T_\lambda \\[4mm]
\Delta B &= -\frac{1}{4} {B} g^{2} \Big(2 |\lambda|^2  + 3 g^{2} \Big) \\
\Delta m_1^2 &= -\frac{1}{4} g^{2} \Big(-2 g^{2} |M_B|^2  + 2 |\lambda|^2 \Big(3 |\mu|^2 + m_{H_d}^2\Big) + g^{2} \Big(2 m_{H_u}^2  + 5 m_{H_d}^2  + 9 |\mu|^2
\Big)\Big) \\
\Delta m_2^2 &= -\frac{1}{4} g^{2} \Big(-2 g^{2} |M_B|^2  + 2 |\lambda|^2 \Big(3 |\mu|^2 + m_{H_u}^2\Big) + g^{2} \Big(2 m_{H_d}^2  + 5 m_{H_u}^2  + 9 |\mu|^2 \Big)\Big)  \\
\Delta m_3^2 &=  g^{2} \Big(2 \lambda^{2} {\mu^*}^{2}  + 2 \mu^{2} (\lambda^*)^2  + |\lambda|^2 \Big(-8 |\mu|^2 + m^2_S\Big)\Big)
\end{align}
This gives a complete agreement between the two-loop $\beta$-functions of both calculations. Thus, our revised results for the RGEs of a general quantum field theory are confirmed. 

\section{Numerical impact}
\label{sec:numerics}
\subsection{Running of fermion mass terms}
We briefly want to discuss the numerical impact on the changes in the $\beta$-function for the fermion mass term. Differences in the running will only appear in models in which
the Lagrangian contains fermionic terms
\begin{equation}
{\cal L} \supset Y S f_1 f_2 + \mu f_1 f_2  + \text{h.c.}
\end{equation}
with a Yukawa-like coupling $Y$ between two Weyl fermions $f_1$, $f_2$ and a scalar $S$ as well as a fermion mass term $\mu$. Both terms can only be present if $S$ is a gauge singlet and if $f_1$, $f_2$ form a vector-like fermion pair. As concrete example, we consider the case of heavy top-like states and a real singlet, i.e.
\begin{align}
T' : \quad & ({\bf 3}, {\bf 1})_{-\frac13}\, , \\
\bar{T}' : \quad & ({\bf \overline{3}}, {\bf 1})_{\frac13}\, , \\
S: \quad & ({\bf 1}, {\bf 1})_{0}\, ,
\end{align}
and the potential reads
\begin{align}
V = &  V_{SM} +  \frac14 \lambda_S S^4 + \frac12 \lambda_{SH} |H|^2 S^2  + \kappa_{SH} |H|^2 S + \frac13 \kappa_S S^3 + \frac12 m_S^2 S^2 \nonumber \\ 
& \hspace{1cm} +  \left( Y_T S \bar{T}' T' + \mu_T \bar{T}' T'  + \text{h.c.}\right)\,.
\end{align}
The one- and two-loop $\beta$-functions are computed using our corrected expression and read
\begin{align}
\beta_{\mu_{T}}^{(1)} & =  
2 Y_{T}^{2} \mu_{T}^*  -\frac{2}{5} \Big(20 g_{3}^{2}  + g_{1}^{2}\Big)\mu_{T}  + \mu_{T} |Y_{T}|^2 \,,\\ 
\beta_{\mu_{T}}^{(2)} & =  
\frac{1}{450} \Big(667 g_{1}^{4} -240 g_{1}^{2} g_{3}^{2}  -46600 g_{3}^{4} \Big)\mu_{T} +\frac{4}{15} \Big(  2 g_{1}^{2} \mu_{T}  + 40 g_{3}^{2} \mu_{T} -15 \kappa_S Y_{T} \Big)|Y_{T}|^2  \nonumber \\  & \quad \quad -\frac{37}{4} \mu_{T} |Y_{T}|^4+\frac{2}{15} Y_{T}^{2} \Big(-105 |Y_{T}|^2  + 8 \Big(20 g_{3}^{2}  + g_{1}^{2}\Big)\Big)\mu_{T}^* \,,
\end{align}
while the differences compared to the old results are 
\begin{align}
\label{eq:diffVLQ1}
\Delta \beta_{\mu_T}^{(1)} &=  -6 \mu_{T} Y_{T}^{2} \,,\\
\label{eq:diffVLQ2}
\Delta \beta_{\mu_T}^{(2)} &= Y_{T} (-2 \kappa_{HS} \lambda_{HS}  - \kappa_S \lambda_S  + \mu_{T} Y_{T} (27 Y_{T}^{2}  -2 g_{1}^{2}  -40 g_{3}^{2} ) - 12 \mu_T^* Y_T |Y_T|^2 )\,.
\end{align}

\begin{figure}
\centering
\includegraphics[width=0.75\linewidth]{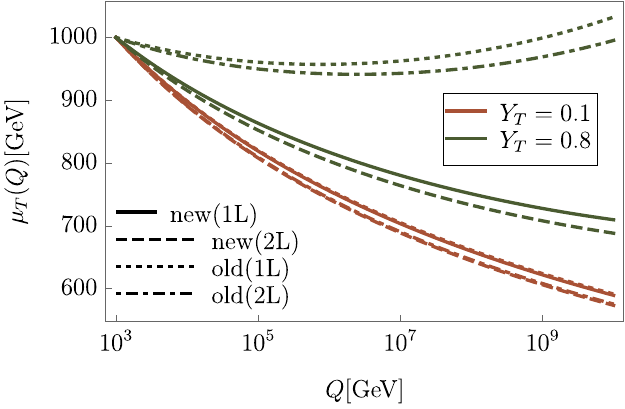} 
\caption{The running mass $\mu_T$ of the vector-like top partners at one- and two-loop level for two different choices of the Yukawa coupling $Y_T$. Here, we show the results using the incorrect (`old') expressions in literature as well as our derived expressions (`new'). The other parameters are set to $\lambda_{HS}=0$, $\lambda_S=1$, $\kappa_{HS}=\kappa=1$~TeV.}
\label{fig:VLQ}
\end{figure}

The numerical impact of this difference is depicted in Fig.~\ref{fig:VLQ} where we assumed a value of 1~TeV for $\mu_T$ at the scale $Q=1$~TeV and used different values $Y_T$.  As expected from 
Eq.~(\ref{eq:diffVLQ1}), the discrepancy between the old and new results rapidly grows with increasing $Y_T$. Thus, the correction in the RGEs is crucial for instance to study grand unified theories which also predict additional vector-like fermions with large Yukawa couplings to a gauge singlet. 

\subsection{Off-diagonal wave-function renormalisation}
\begin{figure}
\centering
\includegraphics[width=0.49\linewidth]{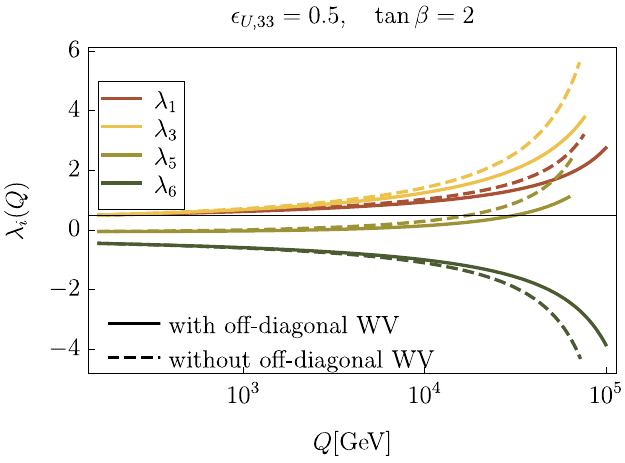}  \hfill
\includegraphics[width=0.49\linewidth]{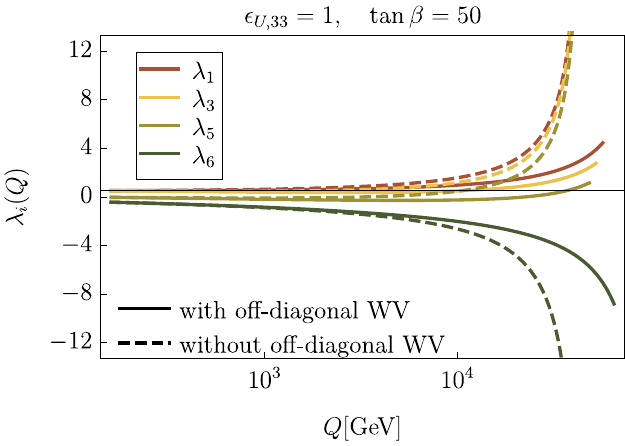}  \hfill
\caption{The running of different quartic couplings in the THDM-III with and without the contributions of off-diagonal wave-function renormalisation to the $\beta$-functions of the quartic couplings. Here, we have used the  input parameters $\lambda_1=\lambda_3=\lambda_4=0.5$, $\lambda_5=-0.05$, $\lambda_6=\lambda_7=-0.45$, $\tan\beta=2$ and $M_{12}=500^2\ \text{GeV}^2$ at $Q=m_t$. On the left, we have used $\epsilon_{U,33}=0.5$, $\lambda_2=0.5$, $\tan\beta=2$, and on the right $\epsilon_{U,33}=1$, $\lambda_2=0.15$, $\tan\beta=50$. All other $\epsilon_i$ are zero.}
\label{fig:THDMIII}
\end{figure}

We now turn to the numerical impact of the off-diagonal wave-function renormalisation which is not included in the previous works. For this purpose, we consider the general Two-Higgs-Doublet-Model type-III
with the following scalar potential:
\begin{align}
V =  & m_1^2 |H_1|^2 + m_2^2 |H_2|^2 + \lambda_1 |H_1|^4 + \lambda_2 |H_2|^4 + \lambda_3 |H_1|^2 |H_2|^2 + \lambda_4 |H_2^\dagger H_1|^2  \nonumber\\
& +\left( \frac12 \lambda_5 (H_2^\dagger H_1) +  \lambda_6 |H_1|^2 (H_1^\dagger H_2) + \lambda_7 |H_2|^2 (H_1^\dagger H_2) - M_{12} H_1^\dagger H_2 + \text{h.c.} \right) 
\end{align}
and the Yukawa interactions
\begin{align}
{\cal L}_Y =  -\left( Y_d H_1^\dagger d q + Y_e H_1^\dagger e l - Y_u H_2 u q + \epsilon_d H_2^\dagger d q + \epsilon_e H_2^\dagger e l - \epsilon_u H_1 u q + \text{h.c.} \right)\,.
\end{align}
Due to the presence of all Yukawa interactions allowed by gauge invariance, the anomalous dimensions of the Higgs doublets $H_1$ and $H_2$ are no longer diagonal, but a mixing is induced 
proportional to $\text{Tr}(Y_i \epsilon_i)$ with $i=e,d,u$. If we neglect for the moment all terms involving either the electroweak gauge couplings ($g_1$, $g_2$), a lepton or down-quark Yukawa coupling ($Y_d$, $Y_e$, $\epsilon_d$, $\epsilon_e$), 
the one-loop 
$\beta$-functions for the quartic coupling read
{\allowdisplaybreaks  
\begin{align} 
\beta_{\lambda_1}^{(1)} & =  
24 \lambda_{1}^{2} +2 \lambda_{3}^{2} +2 \lambda_3 \lambda_4 +\lambda_{4}^{2}+|\lambda_5|^2+12 |\lambda_6|^2 \nonumber \\ 
 &{\phantom =} +12 \lambda_1 \mbox{Tr}\Big({\epsilon_u  \epsilon_{u}^{\dagger}}\Big) + \underline{6 \text{Re}(\lambda_6) \mbox{Tr}\Big({\epsilon_u  Y_{u}^{\dagger}}\Big)} -6 \mbox{Tr}\Big({\epsilon_u  \epsilon_{u}^{\dagger}  \epsilon_u  \epsilon_{u}^{\dagger}}\Big) \,,\\ 
\beta_{\lambda_2}^{(1)} & =  
24 \lambda_{2}^{2} +2 \lambda_{3}^{2} +2 \lambda_3 \lambda_4 +\lambda_{4}^{2}+|\lambda_5|^2+12 |\lambda_7|^2 \nonumber \\ 
 &{\phantom =} +12 \lambda_2 \mbox{Tr}\Big({Y_u  Y_{u}^{\dagger}}\Big) + \underline{6 \text{Re}(\lambda_7) \mbox{Tr}\Big({\epsilon_u  Y_{u}^{\dagger}}\Big)} -6 \mbox{Tr}\Big({Y_u  Y_{u}^{\dagger}  Y_u  Y_{u}^{\dagger}}\Big) \,,\\ 
\beta_{\lambda_3}^{(1)} & =  
2 |\lambda_5|^2 + 2 \lambda_{4}^{2}  + 4 \lambda_{3}^{2} +\underline{6 \text{Re}\Big(\lambda_6 + \lambda_7\Big)\mbox{Tr}\Big({\epsilon_u  Y_{u}^{\dagger}}\Big)} + 4 |\lambda_7|^2 + 4 |\lambda_6|^2 + 16 \text{Re}(\lambda_6\lambda_7^*) \nonumber \\ 
 &{\phantom =} +6 \lambda_3 \mbox{Tr}\Big({\epsilon_u  \epsilon_{u}^{\dagger}}+{Y_u  Y_{u}^{\dagger}}\Big)+ 4 \Big(\lambda_1 + \lambda_2\Big)\Big(3 \lambda_3 + \lambda_4 \Big)  -12 \mbox{Tr}\Big({\epsilon_u  \epsilon_{u}^{\dagger}  Y_u  Y_{u}^{\dagger}}\Big)  \,,\\ 
\beta_{\lambda_4}^{(1)} & =  
4 \lambda_4 \Big(2 \lambda_3  + \lambda_1 + \lambda_2 + \lambda_4\Big)+8 |\lambda_5|^2 +\underline{6 \text{Re}\Big(\lambda_6 + \lambda_7\Big)\mbox{Tr}\Big({\epsilon_u  Y_{u}^{\dagger}}\Big)} + 2 \lambda_6^*  \Big(5 \lambda_6  + \lambda_7\Big)\nonumber \\ 
 &+2\lambda_7^*  \Big(5 \lambda_7  + \lambda_6\Big)+6 \lambda_4 \mbox{Tr}\Big({\epsilon_u  \epsilon_{u}^{\dagger}}+ {Y_u  Y_{u}^{\dagger}}\Big)-12 \mbox{Tr}\Big({\epsilon_u  Y_{u}^{\dagger}  Y_u  \epsilon_{u}^{\dagger}}\Big) \,,\\ 
\beta_{\lambda_5}^{(1)} & =  
2 \Big(2 \Big(2 \lambda_3  + 3 \lambda_4  + \lambda_1 + \lambda_2\Big)\lambda_5 +5 \lambda_{6}^{*\,2} +2 \lambda_6^* \lambda_7^* +5 \lambda_{7}^{*\,2} + \underline{3 \Big(\lambda_6^* + \lambda_7^*\Big)\mbox{Tr}\Big({\epsilon_u  Y_{u}^{\dagger}}\Big)} \nonumber \\ 
 & +3 \lambda_5 \Big(\mbox{Tr}\Big({\epsilon_u  \epsilon_{u}^{\dagger}}\Big) + \mbox{Tr}\Big({Y_u  Y_{u}^{\dagger}}\Big)\Big)-6 \mbox{Tr}\Big({\epsilon_u  Y_{u}^{\dagger}  \epsilon_u  Y_{u}^{\dagger}}\Big) \Big)\,,\\ 
\beta_{\lambda_6}^{(1)} & =  
24 \lambda_1 \lambda_6 +6 \lambda_3 \Big(\lambda_6 + \lambda_7\Big)+4 \lambda_4 \Big(2 \lambda_6  + \lambda_7\Big)+\lambda_5^* \Big(10 \lambda_6^*  + 2 \lambda_7^*\Big)  + \underline{3 \lambda_5^* \mbox{Tr}\Big({\epsilon_u  Y_{u}^{\dagger}}\Big)} \nonumber \\ 
 &+\underline{3 \Big(2 \lambda_1  + \lambda_3 + \lambda_4\Big)\mbox{Tr}\Big({Y_u  \epsilon_{u}^{\dagger}}\Big)} +3 \lambda_6 \mbox{Tr}\Big(3 {\epsilon_u  \epsilon_{u}^{\dagger}}  + {Y_u  Y_{u}^{\dagger}}\Big)-12 \mbox{Tr}\Big({\epsilon_u  \epsilon_{u}^{\dagger}  Y_u  \epsilon_{u}^{\dagger}}\Big) \,,\\ 
\beta_{\lambda_7}^{(1)} & =  
4 \lambda_4 \lambda_6 +8 \Big(3 \lambda_2  + \lambda_4\Big)\lambda_7 +6 \lambda_3 \Big(\lambda_6 + \lambda_7\Big)+\lambda_5^* \Big(10 \lambda_7^*  + 2 \lambda_6^*\Big)  + \underline{ 3  \lambda_5^*\mbox{Tr}\Big({\epsilon_u  Y_{u}^{\dagger}}\Big)}\nonumber \\ 
 &+\underline{3 \Big(2 \lambda_2  + \lambda_3 + \lambda_4\Big)\mbox{Tr}\Big({Y_u  \epsilon_{u}^{\dagger}}\Big)} +3 \lambda_7 \mbox{Tr}\Big(3 {Y_u  Y_{u}^{\dagger}} + {\epsilon_u  \epsilon_{u}^{\dagger}}\Big)-12 \mbox{Tr}\Big({Y_u  \epsilon_{u}^{\dagger}  Y_u  Y_{u}^{\dagger}}\Big) \,.
\end{align}}
The underlined terms stem from the off-diagonal wave-function renormalisation and are missing in the results of Refs.~\cite{Machacek:1983tz,Machacek:1983fi,Machacek:1984zw,Luo:2002ti}.
In Fig.~\ref{fig:THDMIII} we show the numerical impact of the additional one-loop contributions on the running of the quartic couplings for two different points. The chosen sets of the quartic couplings, $\tan\beta$ and $M_{12}$ result in a tree-level Higgs mass of 125~GeV \footnote{While it is in principle possible to renormalise the Higgs sector of the THDM-III on-shell, large radiative corrections can occur when extracting the \MS parameters which enter the RGEs \cite{Braathen:2017jvs}. Therefore, the given example is meant as an illustration on the difference in the running, but the input parameters in the running will change 
when including those corrections.}. 
We see that the additional terms can lead to sizeable differences already for $\epsilon_{u,33}=0.5$ and small $\tan\beta=2$. This is due to $\text{Tr}(\epsilon_u Y_u^\dagger)$. When increasing 
$\epsilon_{u,33}$ to 1 and $\tan\beta=50$, one obtains $\text{Tr}(\epsilon_u Y_u^\dagger) \simeq 1$ and the impact on the running couplings is tremendous. \\

Of course, there are also differences at the two-loop level. Those read within the same approximation:
\begin{align}
\Delta \beta^{(2) }_{\lambda_1} &=\frac{1}{4} (6 \lambda_5^* \lambda_{6}^{*\,2} +6 \lambda_6 ( (2 \lambda_2  + \lambda_3 + \lambda_4)\lambda_7^*  + \lambda_5 \
(\lambda_6 + \lambda_7)) \nonumber \\ 
 &+ \lambda_6 \epsilon_t Y_t (-27 \epsilon_{u}^{2}  -27 Y_{u}^{2}  + 80 g_{3}^{2} ) +\lambda_6^* (12 \lambda_2 \lambda_7  + 24 \lambda_1 \lambda_6  -27 \epsilon_{u}^{3} Y_t  -27 \epsilon_t Y_{u}^{3}  \nonumber \\ 
 &+ 6 (\lambda_3 + \lambda_4)(2 \lambda_6  + \lambda_7) + 6 \lambda_5^* \lambda_7^*  + 80 \epsilon_t g_{3}^{2} Y_t )) \,,\\
\Delta \beta^{(2) }_{\lambda_2} &= \frac{1}{4} (\lambda_7 (6 \lambda_5 (\lambda_6 + \lambda_7) + \epsilon_t Y_t (-27 \epsilon_{u}^{2}  -27 Y_{u}^{2}  + 80 g_{3}^{2} \
)) \nonumber \\ 
&+6 \lambda_6^* ((2 \lambda_1  + \lambda_3 + \lambda_4)\lambda_7  + \lambda_5^* \lambda_7^* )
 +\lambda_7^* (12 \lambda_1 \lambda_6  + 24 \lambda_2 \lambda_7  -27 \epsilon_{u}^{3} Y_t  -27 \epsilon_t Y_{u}^{3}  \nonumber \\ 
 &+ 6 (\lambda_3 + \lambda_4)(2 \lambda_7  + \lambda_6) + 6 \lambda_5^* \lambda_7^*  + 80 \epsilon_t g_{3}^{2} Y_t )) \,,\\
\Delta \beta^{(2) }_{\lambda_3} &=\frac{1}{4} ((\lambda_6 + \lambda_7)(6 \lambda_5 (\lambda_6 + \lambda_7) + \epsilon_t Y_t (-27 \epsilon_{u}^{2}  -27 Y_{u}^{2}  + 80 g_{3}^{2} ))+6 \lambda_5^* \lambda_{6}^{*\,2} +6 \lambda_5^* \lambda_{7}^{*\,2} \nonumber \\ 
 & +12 (\lambda_5^* \lambda_6^* \lambda_7^* + (2 \lambda_2  + \lambda_3 + \lambda_4)(|\lambda_7|^2+|\lambda_6|^2)  + 2 (\lambda_1 + \lambda_2 + \lambda_3 + \lambda_4)\text{Re}(\lambda_7^* \lambda_6) ) \nonumber \\
 &+ (\lambda_7^*+\lambda_6^*)(-27 \epsilon_{u}^{3} Y_t  -27 \epsilon_t Y_{u}^{3}   + 80 \epsilon_t g_{3}^{2} Y_t ) \,,\\
\Delta \beta^{(2) }_{\lambda_4} &=\frac{1}{4} ((\lambda_6 + \lambda_7)(6 \lambda_5 (\lambda_6 + \lambda_7) + \epsilon_t Y_t (-27 \epsilon_{u}^{2}  -27 Y_{u}^{2}  + 80 g_{3}^{2} ))+6 \lambda_5^* \lambda_{6}^{*\,2}  + 6 \lambda_5^* \lambda_7^{*2}  \nonumber \\ 
 &+12 (\lambda_5^* \lambda_6^* \lambda_7^* + (2 \lambda_2  + \lambda_3 + \lambda_4)(|\lambda_7|^2+|\lambda_6|^2)  +2 (\lambda_1 + \lambda_2 + \lambda_3 + \lambda_4) \text{Re}(\lambda_7^*\lambda_6 ))\nonumber \\ 
 & + (\lambda_7^*+\lambda_6^*) (-27 \epsilon_{u}^{3} Y_t  -27 \epsilon_t Y_{u}^{3}  +  + 80 \epsilon_t g_{3}^{2} Y_t ) \,, \\
\Delta \beta^{(2) }_{\lambda_5} &=\frac{1}{2} (\lambda_6^* + \lambda_7^*)(6 (2 \lambda_1  + \lambda_3 + \lambda_4)\lambda_6^*  + 6 (2 \lambda_2  + \lambda_3 + \lambda_4)\lambda_7^*  + 6 \lambda_5 (\lambda_6 + \lambda_7) \nonumber \\
& + \epsilon_t Y_t (-27 \epsilon_{u}^{2}  -27 Y_{u}^{2}  + 80 g_{3}^{2} )) \,, \\
\Delta \beta^{(2) }_{\lambda_6} &=\frac{1}{4} ((2 \lambda_1  + \lambda_3 + \lambda_4)(12 (\lambda_1 \lambda_6  + \lambda_2 \lambda_7)  -27 (\epsilon_{u}^{3} Y_t  + \epsilon_t Y_{u}^{3})  + 6 (\lambda_3 + \lambda_4)(\lambda_6 + \lambda_7)\nonumber \\ 
 &+ 80 \epsilon_t g_{3}^{2} Y_t ) +\lambda_5^* (12 (2 \lambda_1  + \lambda_3 + \lambda_4)\lambda_6^*  + 12 (\lambda_1 + \lambda_2 + \lambda_3 + \lambda_4)\lambda_7^*  \nonumber \\
 &+ 6 \lambda_5 (\lambda_6 + \lambda_7) + \epsilon_t Y_t (-27 (\epsilon_{u}^{2}  + Y_{u}^{2})  + 80 g_{3}^{2} ))) \,, \\
\Delta \beta^{(2) }_{\lambda_7} &=\frac{1}{4} (6 \lambda_5^* \lambda_{6}^{*,2} +\lambda_6 (6 (2 \lambda_2  + \lambda_3 + \lambda_4)\lambda_7^*  + 6 \lambda_5 (\lambda_6 + \lambda_7) + \epsilon_t Y_t (-27 ( \epsilon_{u}^{2}  + Y_{u}^{2} ) \nonumber \\ 
 & + 80 g_{3}^{2} )) +\lambda_6^* (12 \lambda_2 \lambda_7  + 24 \lambda_1 \lambda_6  -27 (\epsilon_{u}^{3} Y_t  + \epsilon_t Y_{u}^{3})  + 6 (\lambda_3 + \lambda_4)(2 \lambda_6  + \lambda_7)  \nonumber \\ 
 &+ 6 \lambda_5^* \lambda_7^* + 80 \epsilon_t g_{3}^{2} Y_t )) \,.
\end{align}

\section{Conclusions} 
\label{sec:conclusion}

In this paper, we have revisited the general RGEs with the goal to present the current state-of-the-art and
to correct some mistakes in the literature. 
In particular, the known expressions for the scalar quartic couplings \cite{Machacek:1984zw, Luo:2002ti} assume a diagonal wave-function renormalisation
which is not appropriate for models with mixing in the scalar sector. We therefore have corrected/generalized the expressions
for the $\beta$-functions of the quartic couplings in \eqref{eq:quartic1} and \eqref{eq:quartic2}.
While finalizing this work, a related paper appeared on the arxiv \cite{Bednyakov:2018cmx} 
which confirms our findings concerning the couplings in the scalar sector.
Furthermore, we have carefully re-examined the dummy field method and have provided a detailed description of
it, which has so far been missing in the literature. We then have used this method to re-derive the $\beta$-functions
for the dimensionful parameters (fermion masses, scalar masses, and the cubic scalar couplings).
For cubic scalar couplings and scalar masses, the only differences to Ref.\ \cite{Luo:2002ti} are due to the aforementioned off-diagonal wave-function renormalisation. However, discrepancies for the  fermion mass $\beta$-functions in \cite{Luo:2002ti} have been found and reconciled in \eqref{eq:bm1} and \eqref{eq:bm2}.
We have also performed an independent cross-check of our results using well-tested supersymmetric RGEs
and we find complete agreement. 

We have illustrated the numerical impact on the changes in the $\beta$-function for the
fermion mass terms using a toy model with a heavy vector-like fermion pair coupled to a scalar gauge singlet.
Unsurprisingly, the correction to the running of the fermion mass rapidly grows with increasing Yukawa coupling.
Thus it is crucial to use the corrected RGEs if one wants to study for instance grand unified theories which
predict additional vector-like fermions with large Yukawa couplings to a gauge singlet.
In addition, we have demonstrated the importance of the correction to the $\beta$-functions of the scalar quartic couplings using a general type-III Two-Higgs-Doublet-Model. As can be seen in Fig.\ \ref{fig:THDMIII} the corrections to the running couplings
are non-negligible and can become very large in certain regions of the parameter space.

All the corrected expressions have been implemented in updated versions of the Mathematica package 
\sarah and the Python package \pyrate.
We hope that this paper will be a useful resource in which all the relevant information on the two-loop
$\beta$-functions is at hand in one place.

\clearpage

\begin{acknowledgments}
We are grateful to Dominik St\"ockinger, Anders Eller Thomsen and Colin Poole who first pointed out mistakes in the literature. 
KS and IS would like to thank Steven Martin for very helpful discussions. 
FS is supported by the ERC Recognition Award ERC-RA-0008 of the Helmholtz Association.
\end{acknowledgments}

\begin{appendix}

\section{The dummy field method at two-loop}
In this appendix, we list all two-loop vertex corrections which are needed to obtain the $\beta$ functions for dimensionful parameters.
\subsection{Fermion mass}
\label{app:fermion}
\begin{eqnarray}
\begin{aligned}
\includegraphics[width=0.3\linewidth]{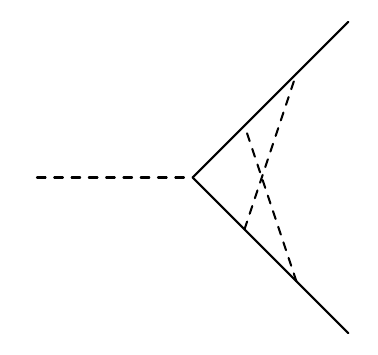}
\end{aligned}
&\quad \to \quad&  \hspace{-2cm}
\begin{aligned}
\includegraphics[width=0.3\linewidth]{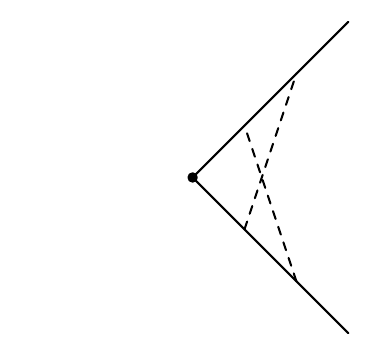}
\end{aligned}
\nonumber \\
Y^c Y^{\dagger b} Y^a (Y^{\dagger c} Y^b - Y^{\dagger b} Y^c) 
& &
Y^c Y^{\dagger b} m_f (Y^{\dagger c} Y^b - Y^{\dagger b} Y^c) 
\\[4mm]
\midrule
\begin{aligned}
\includegraphics[width=0.3\linewidth]{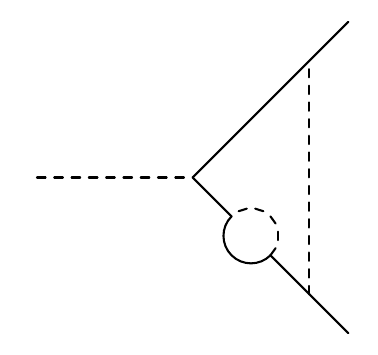}
\end{aligned}
&\quad \to \quad&  \hspace{-2cm}
\begin{aligned}
\includegraphics[width=0.3\linewidth]{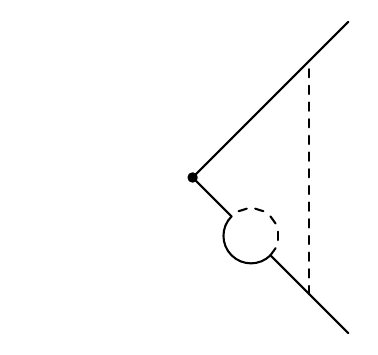}
\end{aligned}
\nonumber \\
Y^b (Y_2(F) Y^{\dagger a} + Y^{\dagger a} Y_2^{\dagger}(F)) Y^b
&&
Y^b (Y_2(F) m_f^{\dagger } + m_f^{\dagger } Y_2^{\dagger}(F)) Y^b \\[4mm]
\midrule
\begin{aligned}
\includegraphics[width=0.3\linewidth]{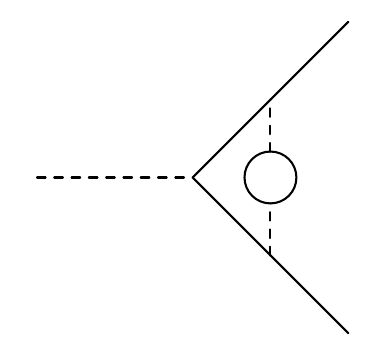}
\end{aligned}
&\quad \to \quad& 
\begin{aligned}
\includegraphics[width=0.3\linewidth]{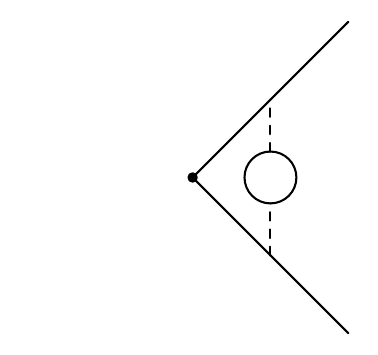}
\end{aligned}
\nonumber \\
Y_2^{bc}(S) Y^b Y^{\dagger a} Y^c
&&
Y_2^{bc}(S) Y^b m_f^{\dagger} Y^c \\[4mm]
\midrule
\begin{aligned}
\includegraphics[width=0.3\linewidth]{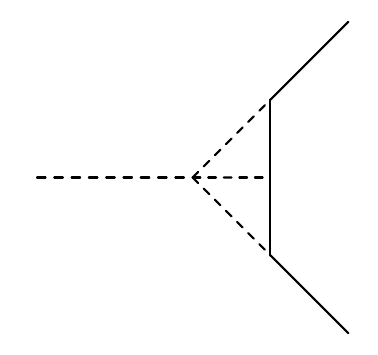}
\end{aligned}
&\quad \to \quad&  \hspace{-2cm}
\begin{aligned}
\includegraphics[width=0.3\linewidth]{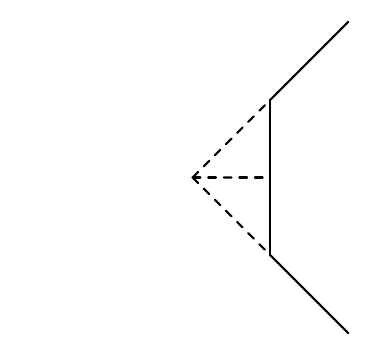}
\end{aligned}
\nonumber \\
\lambda_{abcd} Y^b Y^{\dagger c} Y^d 
 & & h_{abc} Y^a Y^{\dagger b} Y^c
\\[4mm]
\midrule
\begin{aligned}
\includegraphics[width=0.3\linewidth]{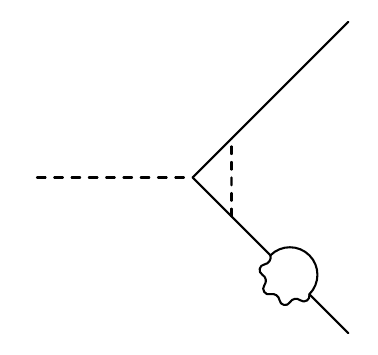}
\end{aligned}
&\quad \to \quad&  \hspace{-2cm}
\begin{aligned}
\includegraphics[width=0.3\linewidth]{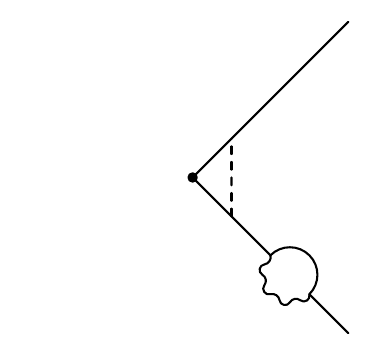}
\end{aligned}
\nonumber \\
g^2 \{C_2(F),Y^b Y^{\dagger a} Y^b\} 
& & g^2 \{C_2(F),Y^b m_f Y^b\} \\[4mm]
\midrule
\begin{aligned}
\includegraphics[width=0.3\linewidth]{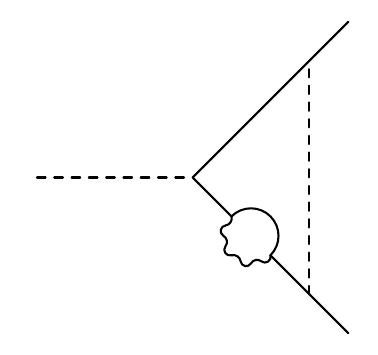}
\end{aligned}
&\quad \to \quad&  \hspace{-2cm}
\begin{aligned}
\includegraphics[width=0.3\linewidth]{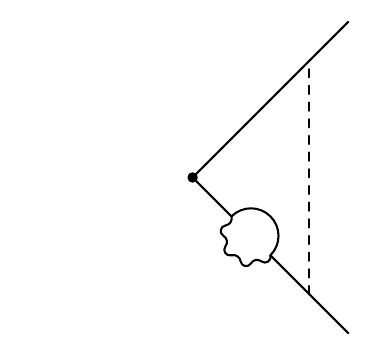}
\end{aligned}
\nonumber \\
\begin{array}{c}
g^2 Y^b \{C_2(F),Y^{\dagger a}\} Y^b 
\end{array}
& & 
\begin{array}{c}
g^2 Y^b \{C_2(F),m^\dagger_f\} Y^b 
\end{array}
\\[4mm]
\midrule
\begin{aligned}
\includegraphics[width=0.3\linewidth]{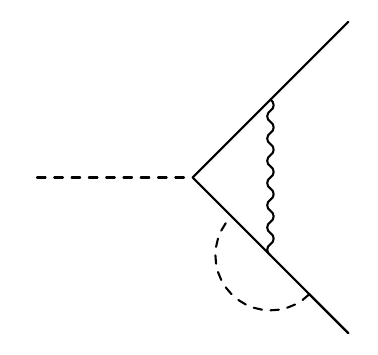}
\end{aligned}
&\quad \to \quad& \hspace{-2cm}
\begin{aligned}
\includegraphics[width=0.3\linewidth]{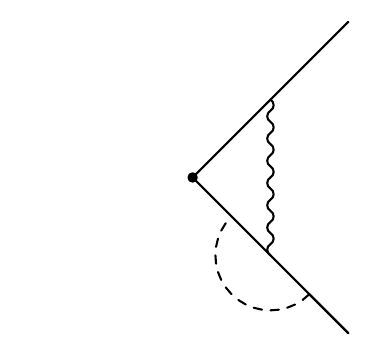}
\end{aligned}
\nonumber \\
\begin{array}{c}
g^2 H^a_{2t} = g^2(t^{A *} Y^a Y^{\dagger b} t^{A *} Y^b \\ + Y^b t^A Y^{\dagger b} Y^a t^A)
\end{array}
& &
\begin{array}{c}
 g^2(t^{A *} m_f Y^{\dagger b} t^{A *} Y^b + \\ Y^b t^A Y^{\dagger b} m_f t^A) 
 \end{array} \nonumber \\
\\[4mm]
\midrule
\begin{aligned}
\includegraphics[width=0.3\linewidth]{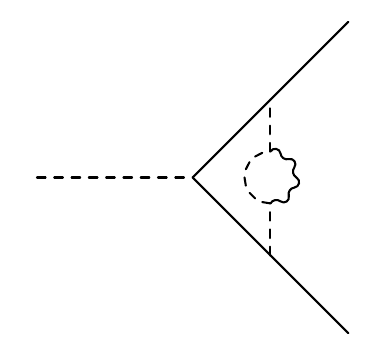}
\end{aligned}
&\quad \to \quad& \hspace{-2cm}
\begin{aligned}
\includegraphics[width=0.3\linewidth]{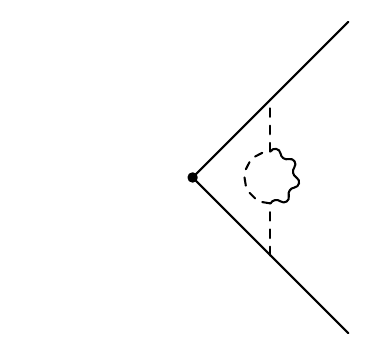}
\end{aligned}
\nonumber \\
g^2 C_2^{bc}(S) Y^b Y^{\dagger a} Y^c 
& & g^2 C_2^{bc}(S) Y^b m_f^\dagger Y^c \\[4mm]
\midrule
\begin{aligned}
\includegraphics[width=0.3\linewidth]{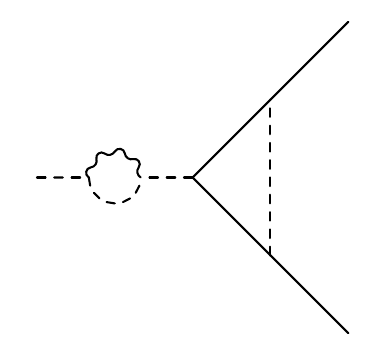}
\end{aligned} 
&\quad \to \quad&  \hspace{2cm}
\begin{aligned}
\includegraphics[width=0.1\linewidth]{cross.pdf}
\end{aligned}
\nonumber \\
g^2 C_2^{ac} Y^b Y^{\dagger c} Y^b
& &  \hspace{2cm} 0 \\
\midrule
\nonumber
\end{eqnarray}

\subsection{Cubic scalar coupling}
\label{app:cubic}
\begin{enumerate}
\item Scalar-only contributions:
\begin{eqnarray}
\label{eq:cubic2start}
\begin{aligned}
\includegraphics[width=0.3\linewidth]{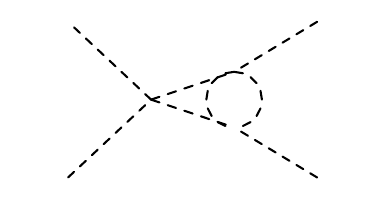}
\end{aligned}
&\quad \to \quad& 
\begin{aligned}
\includegraphics[width=0.3\linewidth]{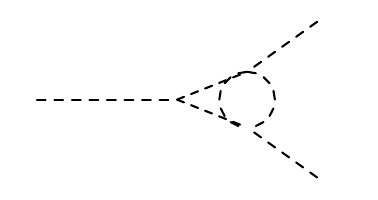} \\
\includegraphics[width=0.3\linewidth]{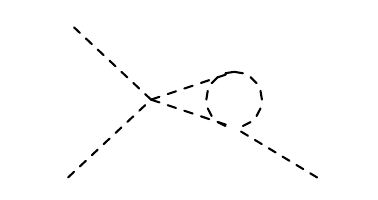} 
\end{aligned}
\nonumber \\
\overline{\Lambda}^3_{abcd}=\frac14 \sum_{per} \lambda_{abef} \lambda_{cegh} \lambda_{dfgh} 
&& \overline{\Lambda}^3_{abc}= \frac12  \sum_{per} [h_{aef} \lambda_{begh} \lambda_{cfgh} + \lambda_{abef} \lambda_{cegh} h_{fgh}]\nonumber \\
\\
\midrule
\nonumber
\end{eqnarray}
\item Scalar-Fermion contributions: 
\begin{eqnarray}
\begin{aligned}
\includegraphics[width=0.3\linewidth]{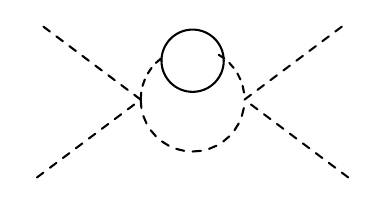}
\end{aligned}
&\quad \to \quad& 
\begin{aligned}
\includegraphics[width=0.3\linewidth]{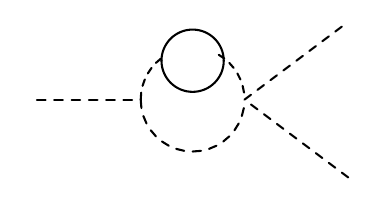}
\end{aligned}
\nonumber \\
\begin{array}{c}
\overline{\Lambda}^{2Y}_{abcd}=\\ \frac18 \sum_{per} Y_2^{fg}(S) \lambda_{abef} \lambda_{cdeg}
\end{array}
&& \overline{\Lambda}^{2Y}_{abc}= \frac12  \sum_{per} Y_2^{fg}(S) \lambda_{abef} h_{ceg}   \\[4mm]
\midrule
\begin{aligned}
\includegraphics[width=0.3\linewidth]{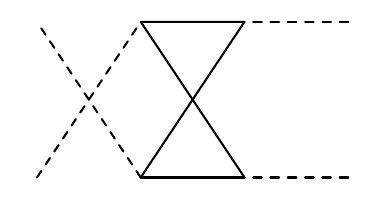}
\end{aligned}
&\quad \to \quad& 
\begin{aligned}
\begin{array}{c}
\includegraphics[width=0.2\linewidth]{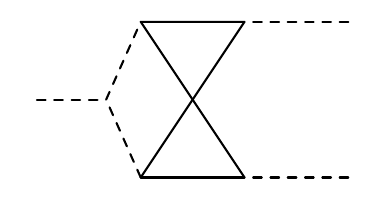} \\
\includegraphics[width=0.2\linewidth]{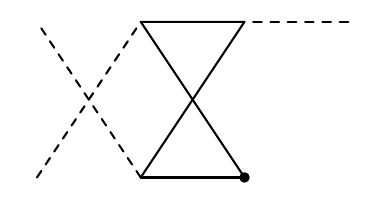}
\includegraphics[width=0.2\linewidth]{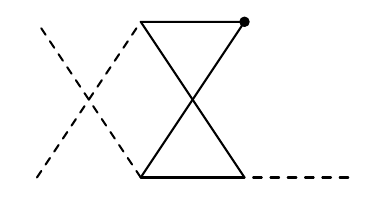}
\end{array}
\end{aligned}
\nonumber \\
\begin{array}{c}
\overline{H}^\lambda_{abcd}= \\ \frac18 \sum_{per} \lambda_{abef} \text{Tr}(Y^c Y^{\dagger e} Y^d Y^{\dagger f} \\ + Y^{\dagger c} Y^e Y^{\dagger d} Y^f)
\end{array}
&& 
\begin{array}{c}
\overline{H}^{h}_{abc} + \overline{H}^{\lambda m}_{abc} = \\ \frac14 \sum_{per} h_{aef} \text{Tr}(Y^b Y^{\dagger e} Y^c Y^{\dagger f} \\
+ Y^{\dagger b} Y^e Y^{\dagger c} Y^f) + \\
\frac14 \sum_{per} \lambda_{abef} \text{Tr}(m_f Y^{\dagger e} Y^c Y^{\dagger f}  \\ + Y^{\dagger c} Y^e m_f^\dagger Y^f) 
\end{array} \nonumber \\
\\[4mm]
\midrule
\begin{aligned}
\includegraphics[width=0.3\linewidth]{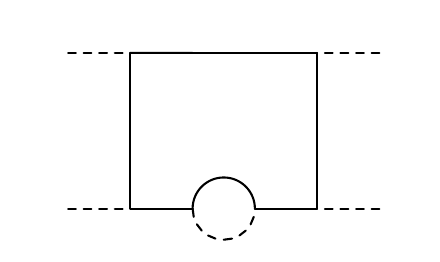}
\end{aligned}
&\quad \to \quad& 
\begin{aligned}
\includegraphics[width=0.2\linewidth]{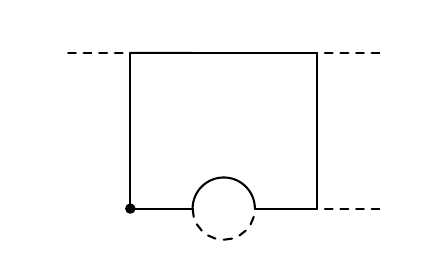} 
\includegraphics[width=0.2\linewidth]{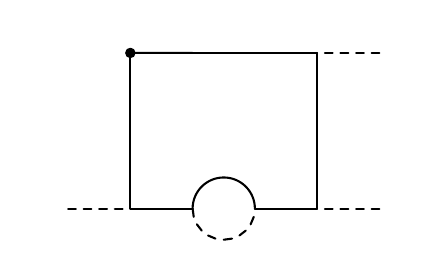} \\
\includegraphics[width=0.2\linewidth]{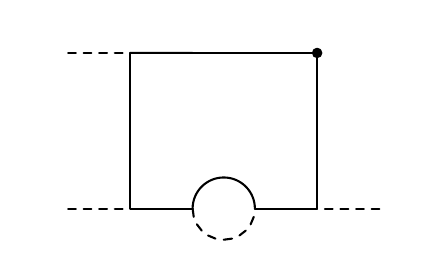}
\includegraphics[width=0.2\linewidth]{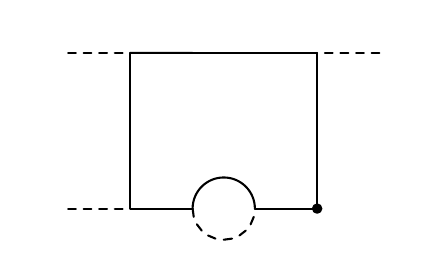}
\end{aligned}
\nonumber \\
\begin{array}{c}
H^Y_{abcd} =  \\  \sum_{per} \text{Tr}(Y_2(F) Y^{\dagger a} Y^{b} Y^{\dagger c} Y^{d})
\end{array}
&&
\begin{array}{c}
H^Y_{abc} =  \sum_{per} \text{Tr} \Big(Y_2(F) [ m_f^\dagger Y^a Y^{\dagger b} Y^c  \\ + Y^{\dagger a} m_f Y^{\dagger b} Y^{c} 
+ Y^{\dagger a} Y^b m_f^\dagger Y^c \\ + Y^{\dagger a} Y^b Y^{\dagger c} m_f]\Big) 
\end{array} \\[4mm]
\midrule
\begin{aligned}
\includegraphics[width=0.3\linewidth]{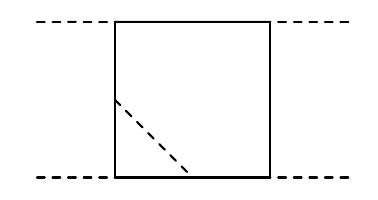}
\end{aligned}
&\quad \to \quad& 
\begin{aligned}
\includegraphics[width=0.2\linewidth]{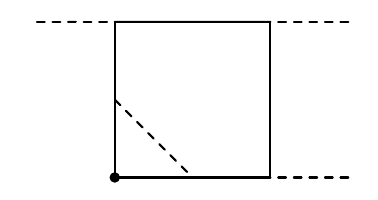} 
\includegraphics[width=0.2\linewidth]{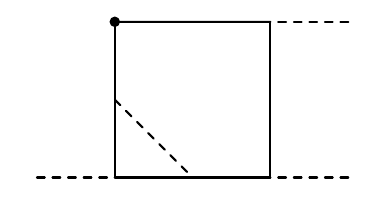} \\
\includegraphics[width=0.2\linewidth]{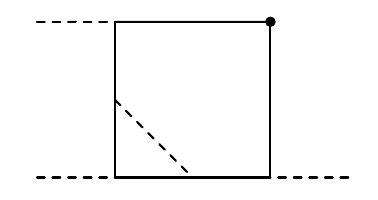} 
\includegraphics[width=0.2\linewidth]{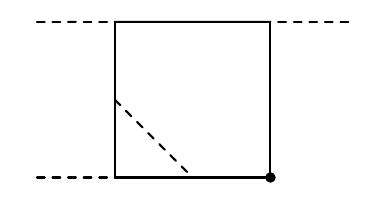} 
\end{aligned}
\nonumber \\
\begin{array}{c}
\overline{H}^Y_{abcd} = \\  \sum_{per} \frac12 \text{Tr}(Y^e Y^{\dagger a} Y^e Y^{\dagger b} Y^c Y^{\dagger d}  \\
+ Y^{\dagger e} Y^a Y^{\dagger e} Y^b Y^{\dagger c} Y^d) 
\end{array}
&&
\begin{array}{c}
\overline{H}^Y_{abc} = \\ \frac12  \sum_{per} \text{Tr}(Y^e m_f^\dagger Y^e Y^{\dagger a} Y^b Y^{\dagger c}  
+ \\ Y^e Y^{\dagger a} Y^e m_f^{\dagger} Y^b Y^{\dagger c} \\
+ Y^e Y^{\dagger a} Y^{e} Y^{\dagger b} m_f Y^{\dagger c} + \\ Y^e Y^{\dagger a} Y^{e} Y^{\dagger b} Y^{c} m_f^\dagger + \text{h.c.}) 
\end{array} \nonumber  \\
\\[4mm]
\midrule
\begin{aligned}
\includegraphics[width=0.3\linewidth]{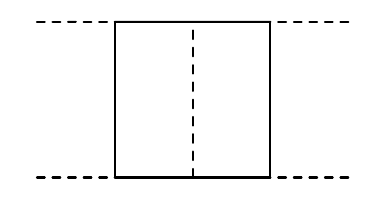}
\end{aligned}
&\quad \to \quad& 
\begin{aligned}
\includegraphics[width=0.2\linewidth]{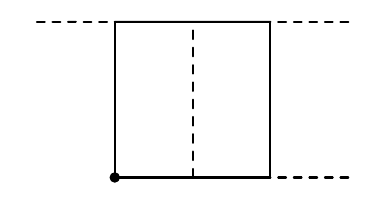}
\includegraphics[width=0.2\linewidth]{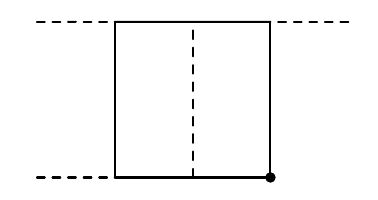} \\
\includegraphics[width=0.2\linewidth]{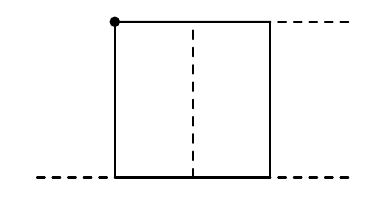}
\includegraphics[width=0.2\linewidth]{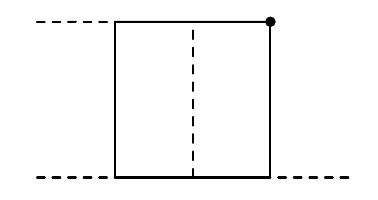}
\end{aligned}
\nonumber \\
\begin{array}{c}
H^3_{abcd}=\\ \frac12 \sum_{per} \text{Tr}(Y^a Y^{\dagger b} Y^e Y^{\dagger c} Y^d Y^{\dagger e})
\end{array}
&&
\begin{array}{c}
H^a_{abc} =  \frac12 \sum_{per} \text{Tr} (m_f Y^{a\dagger} Y^e Y^{\dagger b} Y^c Y^{\dagger e} \\+ Y^a m_f^\dagger Y^e Y^{\dagger b} Y^c Y^{\dagger e} 
\\ + Y^a Y^{\dagger b} Y^e m_f^\dagger Y^c Y^{\dagger e} \\+ Y^a Y^{\dagger b} Y^e Y^{\dagger c} m_f Y^{\dagger e})
\end{array} \nonumber \\
\\
\midrule
\begin{aligned}
\includegraphics[width=0.3\linewidth]{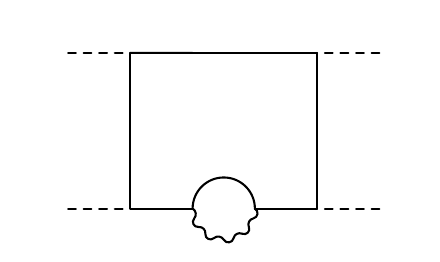}
\end{aligned}
&\quad \to \quad& 
\begin{aligned}
\includegraphics[width=0.2\linewidth]{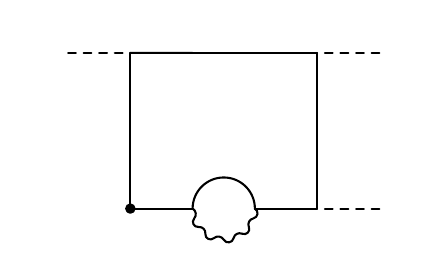}
\includegraphics[width=0.2\linewidth]{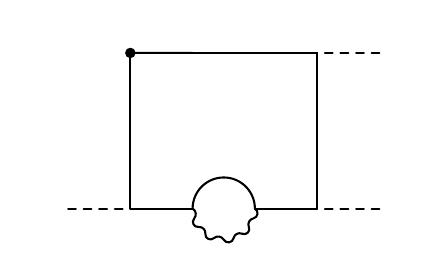} \\
\includegraphics[width=0.2\linewidth]{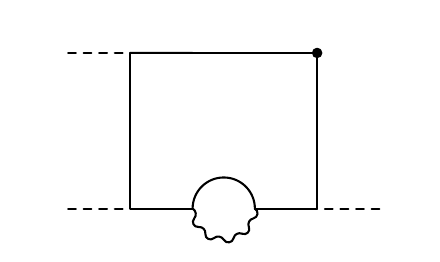}
\includegraphics[width=0.2\linewidth]{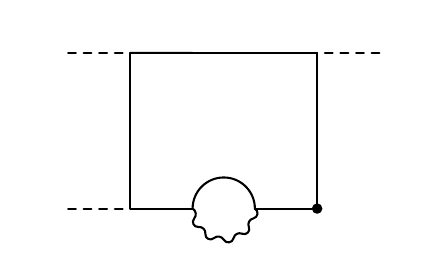}
\end{aligned}
\nonumber \\
\begin{array}{c}
H^F_{abcd}=  \\ \sum_{per} \text{Tr}(\{C_2(F),Y^a\} Y^{\dagger b} Y^{c} Y^{\dagger d})
\end{array}
&&
\begin{array}{c}
H^F_{abc} =  \sum_{per} \text{Tr} (\{C_2(F),m_f\} Y^{a\dagger} Y^b Y^{\dagger c} \\
+ \{C_2(F),Y^a\} m_f^\dagger Y^{b} Y^{\dagger c} 
\\+ \{C_2(F),Y^a\} Y^{\dagger b} m_f Y^{\dagger c} \\
+ \{C_2(F),Y^a\} Y^{\dagger b} Y^c m_f^\dagger)
\end{array} \nonumber \\
\\
\midrule
\begin{aligned}
\includegraphics[width=0.3\linewidth]{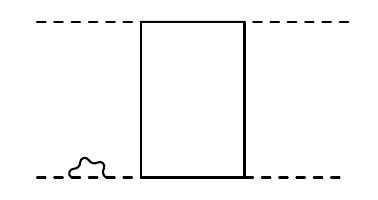}
\end{aligned}
&\quad \to \quad& 
\begin{aligned}
\includegraphics[width=0.3\linewidth]{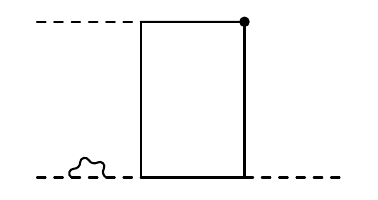}
\end{aligned}
\nonumber \\
H^S_{abcd}=  \sum_i C_2(i) H_{abcd}
&&
H^S_{abc} =  \sum_i C_2(i) H_{abc}\\
\midrule
\nonumber
\end{eqnarray}
\item Scalar-Vector contributions
\begin{eqnarray}
\begin{aligned}
\includegraphics[width=0.3\linewidth]{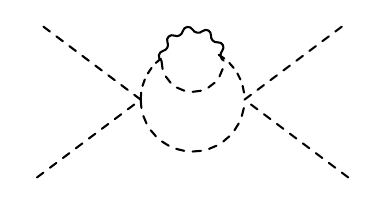}
\end{aligned}
&\quad \to \quad& 
\begin{aligned}
\includegraphics[width=0.3\linewidth]{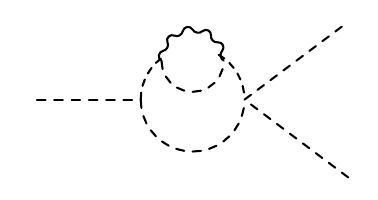}
\end{aligned} 
\nonumber \\
\overline{\Lambda}^{2S}_{abcd} = \frac18 \sum_{per} C_2^{fg}(S) \lambda_{abef} \lambda_{cdeg} 
& &
\overline{\Lambda}^{2S}_{abc} = \frac12 \sum_{per} C_2^{fg}(S) h_{aef} \lambda_{bceg} \nonumber \\ \\[4mm]
\midrule
\begin{aligned}
\includegraphics[width=0.3\linewidth]{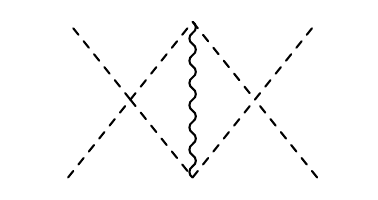}
\end{aligned}
&\quad \to \quad& 
\begin{aligned}
\includegraphics[width=0.3\linewidth]{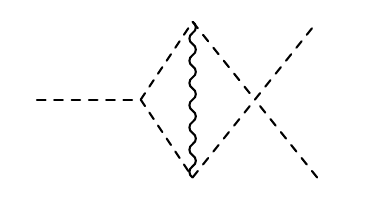}
\end{aligned} 
\nonumber \\
\overline{\Lambda}^{2g}_{abcd} = \frac18 \sum_{per} \lambda_{abef} \lambda_{cdgh} \theta^A_{eg} \theta^A_{fh} 
& &
\overline{\Lambda}^{2g}_{abc} = \frac12 \sum_{per} h_{aef} \lambda_{bcgh} \theta^A_{eg}\theta^A_{fh} \nonumber \\ \\[4mm]
\midrule
\begin{aligned}
\includegraphics[width=0.3\linewidth]{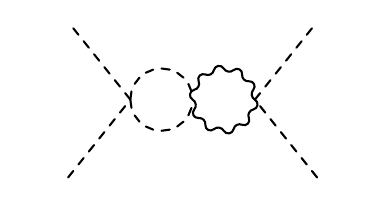}
\end{aligned}
&\quad \to \quad& 
\begin{aligned}
\includegraphics[width=0.3\linewidth]{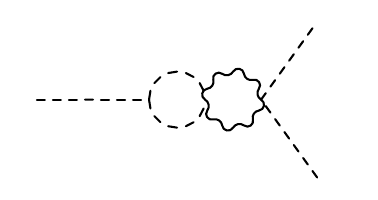}
\end{aligned} 
\nonumber \\
A^\lambda_{abcd}=\frac14 \sum_{per} \lambda_{abef} \{\theta^A,\theta^B\}_{ef} \{\theta^A,\theta^B\}_{cd} 
&&
\begin{array}{c}
A^\lambda_{abc}= \\ \frac12 \sum_{per} h_{aef} \{\theta^A,\theta^B\}_{ef} \{\theta^A,\theta^B\}_{bc}
\end{array} \nonumber \\
\\[4mm]
\midrule
\begin{aligned}
\includegraphics[width=0.3\linewidth]{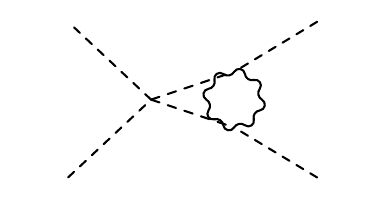}
\end{aligned}
&\quad \to \quad& 
\begin{aligned}
\includegraphics[width=0.3\linewidth]{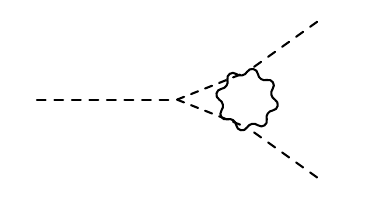}
\end{aligned} 
\nonumber \\
\overline{A}^\lambda_{abcd}=\frac14 \sum_{per} \lambda_{abef} \{\theta^A,\theta^B\}_{ce} \{\theta^A,\theta^B\}_{df} 
&&
\begin{array}{c}
\overline{A}^\lambda_{abc}=\\ \frac12 \sum_{per} h_{aef} \{\theta^A,\theta^B\}_{be} \{\theta^A,\theta^B\}_{cf}
\end{array} \nonumber \\
\\[4mm]
\midrule
\begin{aligned}
\includegraphics[width=0.3\linewidth]{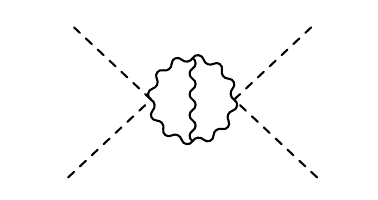}
\end{aligned}
&\quad \to \quad&  \hspace{2cm}
\begin{aligned}
\includegraphics[width=0.1\linewidth]{cross.pdf}
\end{aligned} 
\nonumber \\
\begin{array}{c}
A^g_{abcd}= \\ \frac18 f^{ACE} f^{BDE} \sum_{per} \{\theta^A,\theta^B\}_{ab} \{\theta^C,\theta^D\}_{cd}
\end{array}
&& \hspace{2cm}
0 \\[4mm]
\midrule
\begin{aligned}
\includegraphics[width=0.3\linewidth]{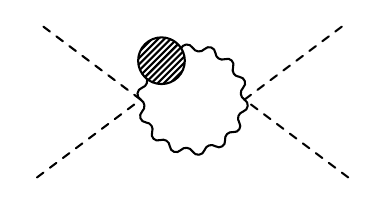}
\end{aligned}
&\quad \to \quad&  \hspace{2cm}
\begin{aligned}
\includegraphics[width=0.1\linewidth]{cross.pdf}
\end{aligned} 
\nonumber \\
\begin{array}{c}
X A_{abcd} = X \{\theta^A,\theta^B\}_{ab} \{\theta^A,\theta^B\}_{cd} 
\end{array}
&& \hspace{2cm}
0 \\[4mm]
\midrule
\begin{aligned}
\includegraphics[width=0.3\linewidth]{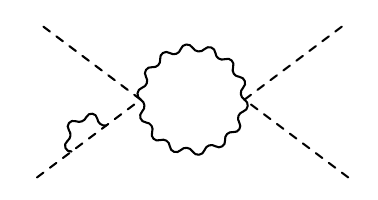}
\end{aligned}
&\quad \to \quad&  \hspace{2cm}
\begin{aligned}
\includegraphics[width=0.1\linewidth]{cross.pdf}
\end{aligned} 
\nonumber \\
\begin{array}{c}
A^S_{abcd} = \\  \sum_i C_2(i) \{\theta^A,\theta^B\}_{ab} \{\theta^A,\theta^B\}_{cd} 
\end{array}
&& \hspace{2cm}
0 \\
\midrule
\nonumber
\end{eqnarray}
\item Scalar-Fermion-Vector contributions
\begin{eqnarray}
\begin{aligned}
\includegraphics[width=0.3\linewidth]{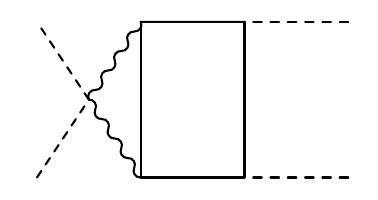}
\end{aligned}
&\quad \to \quad& 
\begin{aligned}
\includegraphics[width=0.2\linewidth]{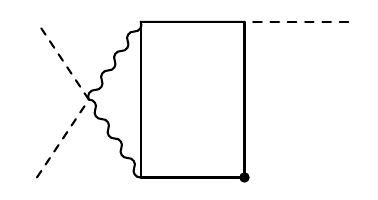} 
\includegraphics[width=0.2\linewidth]{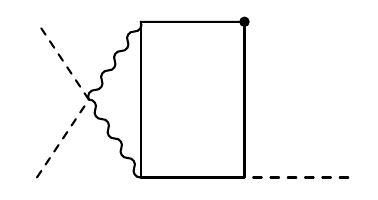}
\end{aligned} 
\nonumber \\
\begin{array}{c}
B^Y_{abcd}=\\ \frac14 \sum_{per} \{\theta^A,\theta^B\}_{ab} \text{Tr}(t^{A*} t^{B*} Y^c Y^{\dagger d} \\
+ Y^c t^{A} t^{B} Y^{\dagger d})
\end{array}
&&
\begin{array}{c}
B^Y_{abc}=\frac14 \sum_{per} \{\theta^A,\theta^B\}_{ab} \text{Tr}(t^{A*} t^{B*} m_f Y^{\dagger c} \\
+ m_f t^A t^B Y^{\dagger c} + t^{A*} t^{B*} Y^c m_f^\dagger \\
+ Y^c t^A t^B m_f^\dagger)
\end{array} \nonumber \\ 
\\[4mm]
\midrule
\label{eq:cubic2end}
\begin{aligned}
\includegraphics[width=0.3\linewidth]{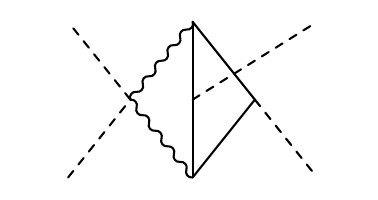}
\end{aligned}
&\quad \to \quad& 
\begin{aligned}
\includegraphics[width=0.2\linewidth]{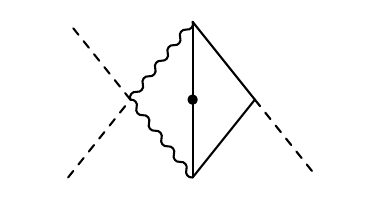}  
\includegraphics[width=0.2\linewidth]{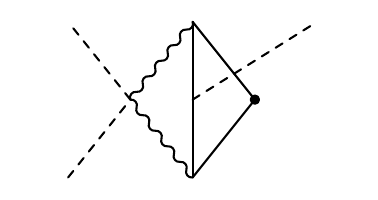}
\end{aligned} 
\nonumber \\
\begin{array}{c}
\overline{B}^Y_{abcd}=\\ \frac14 \sum_{per} \{\theta^A,\theta^B\}_{ab} \text{Tr}(t^{A*} Y^c t^B Y^{\dagger d})
\end{array}
&&
\begin{array}{c}
\overline{B}^Y_{abc}=\frac14 \sum_{per} \{\theta^A,\theta^B\}_{ab} \text{Tr}(t^{A*} m_f t^B Y^{\dagger c} \\
+ t^{A*} Y^c t^B m_f^{\dagger})
\end{array}\nonumber \\ \\
\midrule
\nonumber
\end{eqnarray} 
\end{enumerate}

\subsection{Bilinear scalar}
\label{app:bilinear}
\begin{enumerate}
\item Scalar-only contributions:
\begin{eqnarray}
\label{eq:bi2start}
\begin{aligned}
\includegraphics[width=0.3\linewidth]{Lambda3.pdf}
\end{aligned}
&\quad \to \quad& 
\begin{aligned}
\begin{array}{c}
\includegraphics[width=0.3\linewidth]{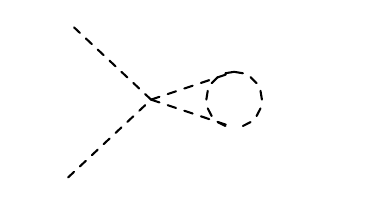} 
\mkern-40mu
\includegraphics[width=0.3\linewidth]{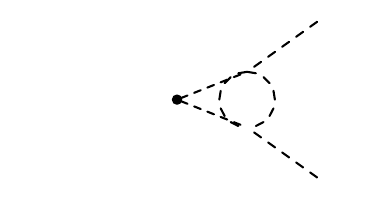} \\
\includegraphics[width=0.3\linewidth]{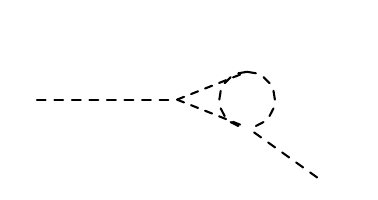} 
\end{array}
\end{aligned}
\nonumber \\
\begin{array}{c}
\overline{\Lambda}^3_{abcd}= \\ \frac14 \sum_{per} \lambda_{abef} \lambda_{cegh} \lambda_{dfgh} 
\end{array}
&& 
\begin{array}{c}
\overline{\Lambda}^3_{ab}=  \lambda_{abef} h_{egl} h_{fgl} + 2 m_{ef}^2 \lambda_{aegl} \lambda_{bfgl} \\
+2 \sum_{per} h_{aef} h_{fgl} \lambda_{begl}  
\end{array} \\
\midrule
\nonumber
\end{eqnarray}
\item Scalar-Fermion contributions:
\begin{eqnarray}
\begin{aligned}
\includegraphics[width=0.3\linewidth]{Lambda2Y.pdf}
\end{aligned}
&\quad \to \quad& \hspace{-2cm}
\begin{aligned}
\includegraphics[width=0.3\linewidth]{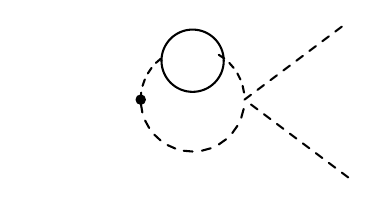}
\includegraphics[width=0.3\linewidth]{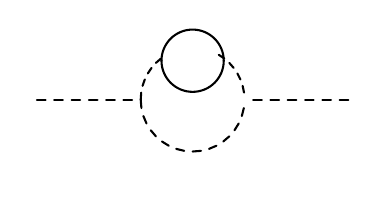}
\end{aligned}
\nonumber \\
\overline{\Lambda}^{2Y}_{abcd}=\frac18 \sum_{per} Y_2^{fg}(S) \lambda_{abef} \lambda_{cdeg}
&& \overline{\Lambda}^{2Y}_{ab}= 2 Y_2^{fg}(S) (m^2_{eg} \lambda_{abef} + h_{aef} h_{beg})   
\nonumber \\
\\[4mm]
\midrule
\begin{aligned}
\includegraphics[width=0.3\linewidth]{Hlambda.pdf}
\end{aligned}
&\quad \to \quad& 
\begin{aligned}
\begin{array}{c}
\includegraphics[width=0.3\linewidth]{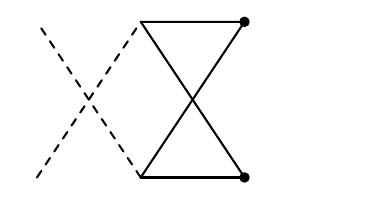} \hspace{-2cm}
\includegraphics[width=0.3\linewidth]{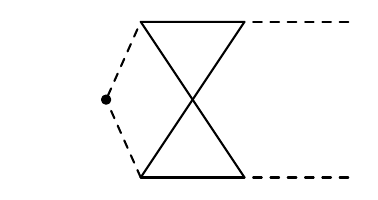} \\
\includegraphics[width=0.3\linewidth]{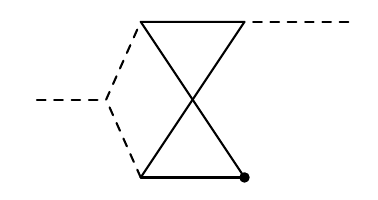}
\end{array}
\end{aligned}
\nonumber \\
\begin{array}{c}
\overline{H}^\lambda_{abcd}= \\ \frac18 \sum_{per} \lambda_{abef} \text{Tr}(Y^c Y^{\dagger e} Y^d Y^{\dagger f} \\ + Y^{\dagger c} Y^e Y^{\dagger d} Y^f)
\end{array}
&& 
\begin{array}{c}
\overline{H}^{\lambda}_{ab}= \frac12 \lambda_{abef} \text{Tr}(m_f Y^{\dagger e} m_f Y^{\dagger f} + \text{h.c.}) \\
+ m^2_{ef} \text{Tr}(Y^a Y^{\dagger e} Y^b Y^{\dagger f}  + \text{h.c.}) \\
+  \sum_{per} h_{aef} \text{Tr}(Y^b Y^{\dagger e} m_f Y^{\dagger f} + \text{h.c.})
\end{array} \nonumber \\
\\[4mm]
\midrule
\begin{aligned}
\includegraphics[width=0.3\linewidth]{HY4.pdf}
\end{aligned}
&\quad \to \quad& 
\begin{aligned}
\includegraphics[width=0.2\linewidth]{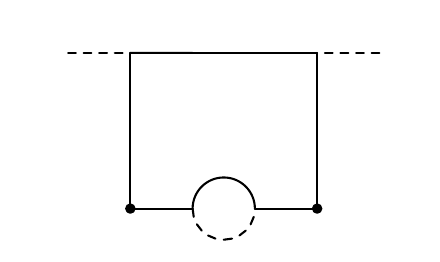} 
\includegraphics[width=0.2\linewidth]{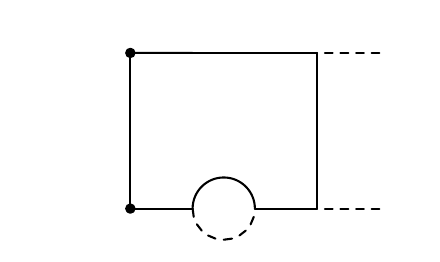} \\
\includegraphics[width=0.2\linewidth]{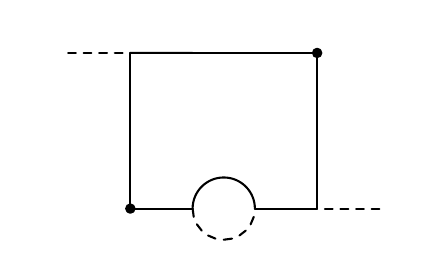} 
\includegraphics[width=0.2\linewidth]{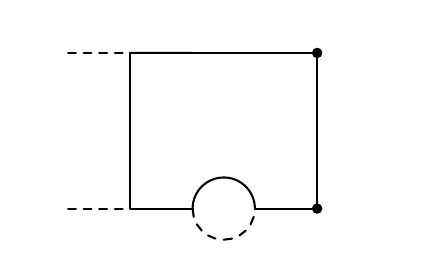} \\
\includegraphics[width=0.2\linewidth]{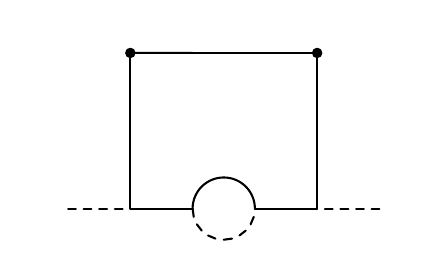} 
\includegraphics[width=0.2\linewidth]{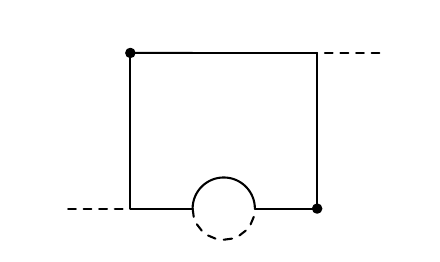} 
\end{aligned}
\nonumber \\
\begin{array}{c}
H^Y_{abcd} = \\ \sum_{per}  \text{Tr}(Y_2(F) Y^{\dagger a} Y^{b} Y^{\dagger c} Y^{d})
\end{array}
&&
\begin{array}{c}
H^Y_{ab} = 2 \sum_{per} \Big[\text{Tr}( \{Y_2(F),m_f^\dagger m_f\} Y^{\dagger a} Y^b) + \\
 \text{Tr}(Y_2(F) Y^{\dagger a} m_f (Y^{\dagger b} m_f + m_f^\dagger Y^b) + \\
Y_2(F) m_f^\dagger Y^a  (Y^{\dagger b} m_f + m_f^\dagger Y^b))\Big]
\end{array}\nonumber \\
\\[4mm]
\midrule
\begin{aligned}
\includegraphics[width=0.3\linewidth]{HYb4.pdf}
\end{aligned}
&\quad \to \quad& 
\begin{aligned}
\includegraphics[width=0.2\linewidth]{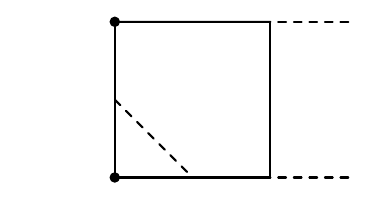} 
\includegraphics[width=0.2\linewidth]{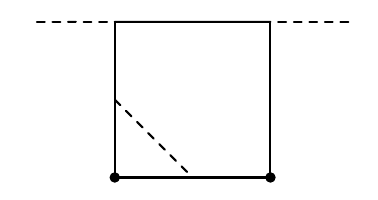} \\
\includegraphics[width=0.2\linewidth]{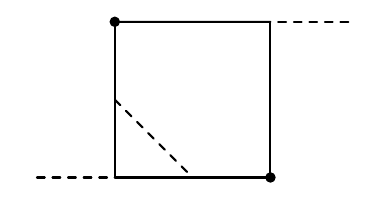} 
\includegraphics[width=0.2\linewidth]{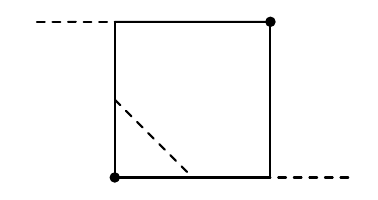} \\
\includegraphics[width=0.2\linewidth]{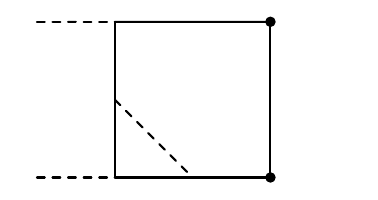} 
\includegraphics[width=0.2\linewidth]{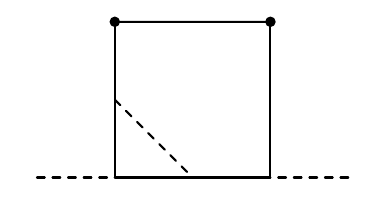} 
\end{aligned}
\nonumber \\
\begin{array}{c}
\overline{H}^Y_{abcd} = \\ \sum_{per} \frac12 \text{Tr}(Y^e Y^{\dagger a} Y^e Y^{\dagger b} Y^c Y^{\dagger d}  \\
+ Y^{\dagger e} Y^a Y^{\dagger e} Y^b Y^{\dagger c} Y^d) 
\end{array}
&& \hspace{0.5cm}
\begin{array}{c}
\overline{H}^Y_{ab} = \sum_{per}\Big[ \text{Tr}\Big(Y^e Y^{\dagger a} Y^e Y^{\dagger b} m_f m_f^\dagger + \\
Y^e m_f^\dagger Y^e m_f^\dagger Y^a Y^{\dagger b} + \\
(Y^e Y^{\dagger a} Y^e m_f^\dagger +Y^e m_f^\dagger Y^e Y^{\dagger a}) \times \\ (Y^b m_f^\dagger + m_f Y^{\dagger b}) + \text{h.c.}  \Big)\Big]
\end{array} \nonumber  \\
\\[4mm]
\midrule
\begin{aligned}
\includegraphics[width=0.3\linewidth]{H34.pdf}
\end{aligned}
&\quad \to \quad& 
\begin{aligned}
\includegraphics[width=0.2\linewidth]{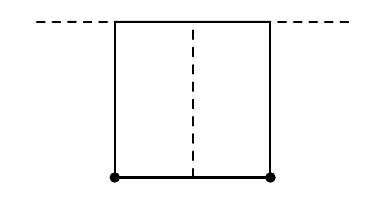}
\includegraphics[width=0.2\linewidth]{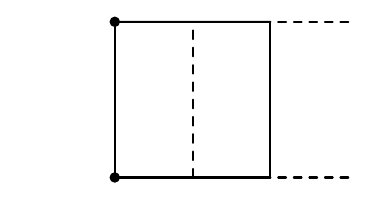} \\
\includegraphics[width=0.2\linewidth]{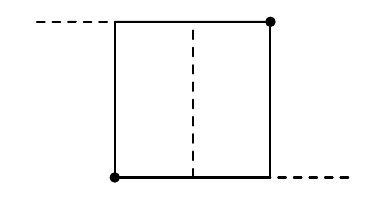}
\includegraphics[width=0.2\linewidth]{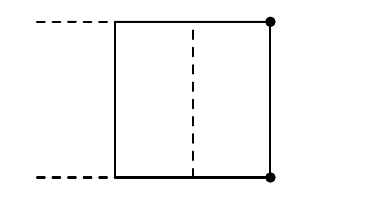} \\
\includegraphics[width=0.2\linewidth]{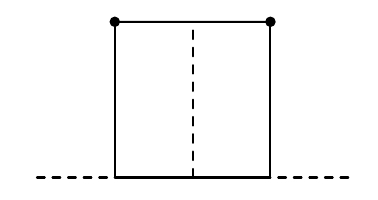}
\includegraphics[width=0.2\linewidth]{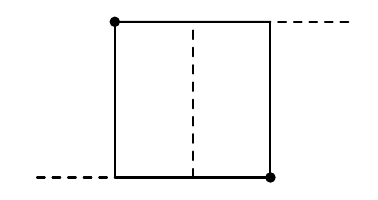}
\end{aligned}
\nonumber \\
\begin{array}{c}
H^3_{abcd}=  \frac12 \sum_{per}  \\ \text{Tr}(Y^a Y^{\dagger b} Y^e Y^{\dagger c} Y^d Y^{\dagger e})
\end{array}
&&
\begin{array}{c}
H^3_{ab} =  \sum_{per} \Big[\text{Tr}\Big(Y^a Y^{\dagger b} Y^e m_f^\dagger m_f Y^{\dagger e} \\
+ m_f m_f^\dagger Y^e Y^{\dagger a} Y^b Y^{\dagger e} \\
Y^a m_f^{\dagger} Y^e (Y^{\dagger b} m_f + m_f^\dagger Y^b) Y^{\dagger e} \\
+ m_f Y^{\dagger a} Y^e (Y^{\dagger b} m_f + m_f^\dagger Y^b) Y^{\dagger e} 
\Big)
\Big]
\end{array}\nonumber \\
\\
\midrule
\begin{aligned}
\includegraphics[width=0.3\linewidth]{HF.pdf}
\end{aligned}
&\quad \to \quad& 
\begin{aligned}
\includegraphics[width=0.2\linewidth]{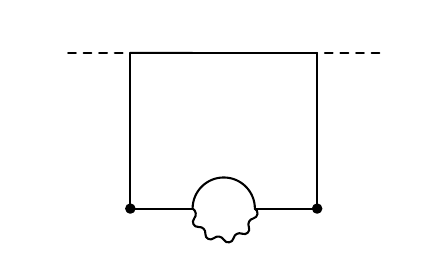}
\includegraphics[width=0.2\linewidth]{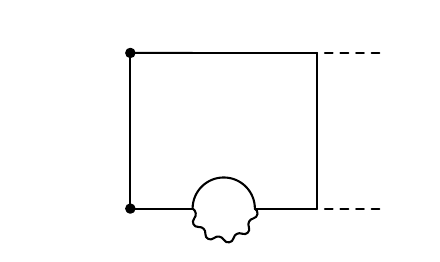} \\
\includegraphics[width=0.2\linewidth]{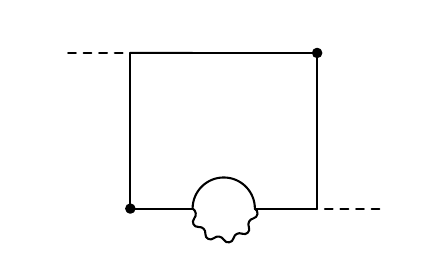}
\includegraphics[width=0.2\linewidth]{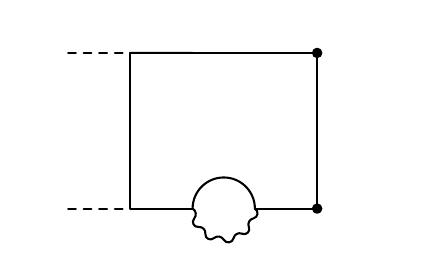}\\
\includegraphics[width=0.2\linewidth]{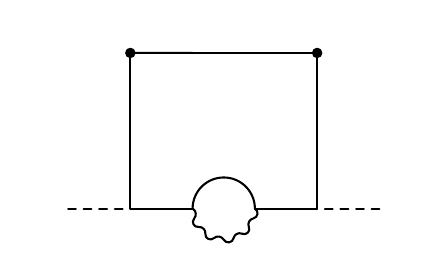}
\includegraphics[width=0.2\linewidth]{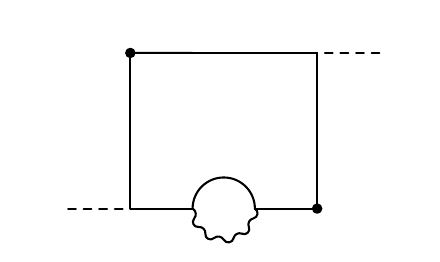}
\end{aligned}
\nonumber \\
\begin{array}{c}
H^F_{abcd}= \sum_{per}  \\ \text{Tr}(\{C_2(F),Y^a\} Y^{\dagger b} Y^{c} Y^{\dagger d})
\end{array}
&&
\begin{array}{c}
H^F_{ab} =  2 \sum_{per} \text{Tr} \big[\{C_2(F),m_f\} Y^{a\dagger} (Y^b m_f^\dagger +\text{h.c.}) \\
+ \{C_2(F),Y^a\} m_f^\dagger (Y^{b} m_f^{\dagger}  + \text{h.c.}) \\
+ \{C_2(F),Y^a\} Y^{\dagger b} m_f m_f^\dagger \\
+ \{C_2(F),m_f\} m_f^\dagger Y^a Y^{\dagger b}\big]
\end{array} \nonumber \\
\\
\midrule
\begin{aligned}
\includegraphics[width=0.3\linewidth]{HS4.pdf}
\end{aligned}
&\quad \to \quad& 
\begin{aligned}
\includegraphics[width=0.3\linewidth]{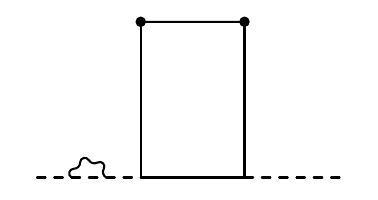}
\end{aligned}
\nonumber \\
H^S_{abcd}=  \sum_i C_2(i) H_{abcd}
&&
H^S_{ab} =  \sum_i C_2(i) H_{ab} \\
\midrule 
\nonumber
\end{eqnarray}
\item Scalar-Vector contributions
\begin{eqnarray}
\begin{aligned}
\includegraphics[width=0.3\linewidth]{Lambda2S.pdf}
\end{aligned}
&\quad \to \quad& 
\begin{aligned}
\includegraphics[width=0.3\linewidth]{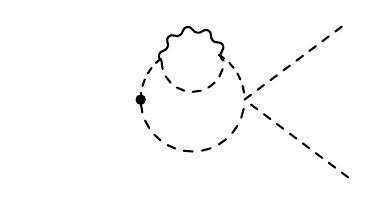} \\
\includegraphics[width=0.3\linewidth]{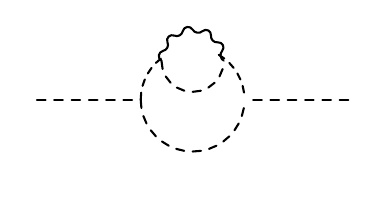}
\end{aligned} 
\nonumber \\
\overline{\Lambda}^{2S}_{abcd} = \frac18 \sum_{per} C_2^{fg}(S) \lambda_{abef} \lambda_{cdeg} 
& &
\overline{\Lambda}^{2S}_{ab} = 2 C_2^{fg}(S) (\lambda_{abef} m_{eg}^2 + h_{aef} h_{beg} )  \nonumber \\
\\[4mm]
\midrule
\begin{aligned}
\includegraphics[width=0.3\linewidth]{Lambda2g.pdf}
\end{aligned}
&\quad \to \quad& 
\begin{aligned}
\includegraphics[width=0.3\linewidth]{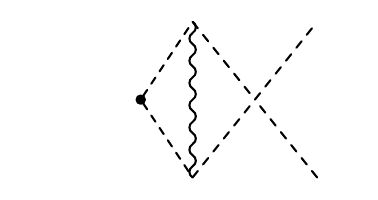} \\
\includegraphics[width=0.3\linewidth]{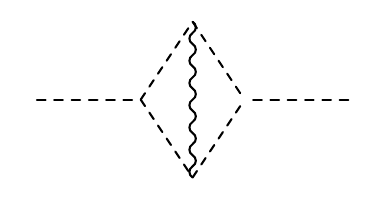}
\end{aligned} 
\nonumber \\
\overline{\Lambda}^{2g}_{abcd} = \frac18 \sum_{per} \lambda_{abef} \lambda_{cdgh} \theta^A_{eg} \theta^A_{fh} 
& &
\overline{\Lambda}^{2g}_{ab} = 2 (\lambda_{abef} m_{gh}^2 + h_{aef} h_{bgh} )\theta^A_{eg}\theta^A_{fh}  \nonumber \\
\\[4mm]
\midrule
\begin{aligned}
\includegraphics[width=0.3\linewidth]{Alambda.pdf}
\end{aligned}
&\quad \to \quad& 
\begin{aligned}
\includegraphics[width=0.3\linewidth]{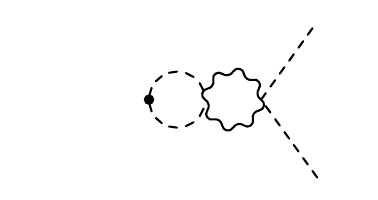}
\end{aligned} 
\nonumber \\
\begin{array}{c}
A^\lambda_{abcd}=\\ \frac14 \sum_{per} \lambda_{abef} \{\theta^A,\theta^B\}_{ef} \{\theta^A,\theta^B\}_{cd} 
\end{array}
&&
A^\lambda_{ab}= 2 m^2_{ef} \{\theta^A,\theta^B\}_{ef} \{\theta^A,\theta^B\}_{ab}\nonumber \\
\\[4mm]
\midrule
\begin{aligned}
\includegraphics[width=0.3\linewidth]{AlambdaBar.pdf}
\end{aligned}
&\quad \to \quad& 
\begin{aligned}
\includegraphics[width=0.3\linewidth]{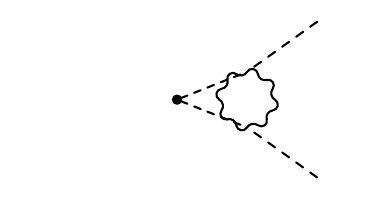}
\end{aligned} 
\nonumber \\
\begin{array}{c}
\overline{A}^\lambda_{abcd}= \\ \frac14 \sum_{per} \lambda_{abef} \{\theta^A,\theta^B\}_{ce} \{\theta^A,\theta^B\}_{df} 
\end{array}
&&
\overline{A}^\lambda_{ab}= 2 m^2_{ef} \{\theta^A,\theta^B\}_{ae} \{\theta^A,\theta^B\}_{bf}\nonumber \\
\\[4mm]
\midrule
\begin{aligned}
\includegraphics[width=0.3\linewidth]{Ag.pdf}
\end{aligned}
&\quad \to \quad&  \hspace{2cm}
\begin{aligned}
\includegraphics[width=0.1\linewidth]{cross.pdf}
\end{aligned} 
\nonumber \\
\begin{array}{c}
A^g_{abcd}=\\ \frac18 f^{ACE} f^{BDE} \sum_{per} \{\theta^A,\theta^B\}_{ab} \{\theta^C,\theta^D\}_{cd}
\end{array}
&& \hspace{2cm}
0 \\[4mm]
\midrule
\begin{aligned}
\includegraphics[width=0.3\linewidth]{XAabcd.pdf}
\end{aligned}
&\quad \to \quad&  \hspace{2cm}
\begin{aligned}
\includegraphics[width=0.1\linewidth]{cross.pdf}
\end{aligned} 
\nonumber \\
\begin{array}{c}
X A_{abcd} = X \{\theta^A,\theta^B\}_{ab} \{\theta^A,\theta^B\}_{cd} 
\end{array}
&& \hspace{2cm}
0 \\[4mm]
\midrule
\begin{aligned}
\includegraphics[width=0.3\linewidth]{ASabcd.pdf}
\end{aligned}
&\quad \to \quad&  \hspace{2cm}
\begin{aligned}
\includegraphics[width=0.1\linewidth]{cross.pdf}
\end{aligned} 
\nonumber \\
\begin{array}{c}
A^S_{abcd} = \\  \sum_i C_2(i) \{\theta^A,\theta^B\}_{ab} \{\theta^A,\theta^B\}_{cd} 
\end{array}
&& \hspace{2cm}
0 \\
\midrule
\nonumber 
\end{eqnarray}
\item Scalar-Fermion-Vector contributions
\begin{eqnarray}
\begin{aligned}
\includegraphics[width=0.3\linewidth]{BY4.pdf}
\end{aligned}
&\quad \to \quad& 
\begin{aligned}
\includegraphics[width=0.3\linewidth]{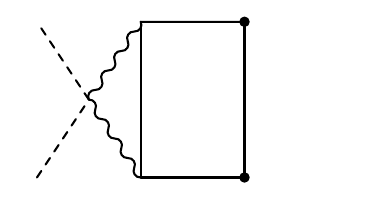}
\end{aligned} 
\nonumber \\
\begin{array}{c}
B^Y_{abcd}=\frac14 \sum_{per} \{\theta^A,\theta^B\}_{ab} \text{Tr}(t^{A*} t^{B*} Y^c Y^{\dagger d} \\
+ Y^c t^{A} t^{B} Y^{\dagger d})
\end{array}
&&
\begin{array}{c}
B^Y_{ab}= \{\theta^A,\theta^B\}_{ab} \text{Tr}(t^{A*} t^{B*} m_f m_f^{\dagger} \\ + m_f t^A t^B m_f^{\dagger} )
\end{array} \nonumber \\
\\[4mm]
\midrule
\label{eq:bi2end}
\begin{aligned}
\includegraphics[width=0.3\linewidth]{BYbar.pdf}
\end{aligned}
&\quad \to \quad& 
\begin{aligned}
\includegraphics[width=0.3\linewidth]{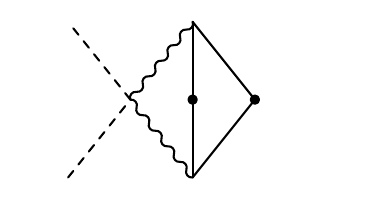}  
\end{aligned} 
\nonumber \\
\overline{B}^Y_{abcd}=\frac14 \sum_{per} \{\theta^A,\theta^B\}_{ab} \text{Tr}(t^{A*} Y^c t^B Y^{\dagger d})
&&
\overline{B}^Y_{abc}= \{\theta^A,\theta^B\}_{ab} \text{Tr}(t^{A*} m_f t^B m_f^\dagger)\nonumber \\
\\
\midrule
\nonumber
\end{eqnarray}
\end{enumerate}

\section{Full two-loop RGEs without SUSY relations}
\label{app:nonsusy}
In this appendix, the full $\beta$-functions for all parameters of the non-supersymmetric toy model in Sec.~\ref{sec:susy} are listed up to two-loop order.
\subsection{Gauge couplings}
{\allowdisplaybreaks  \begin{align} 
\beta_{g}^{(1)} & =  
\frac{1}{2} g^{3} \\ 
\beta_{g}^{(2)} & =  
\frac{1}{8} g^{3} \Big(-2 |Y_2|^2  -2 |Y_3|^2  -4 |Y_1|^2  + 6 g^{2}  - |g_{d}|^2  - |g_{u}|^2 \Big)
\end{align}} 
\subsection{Quartic scalar couplings}
{\allowdisplaybreaks  \begin{align} 
\beta_{\lambda_4}^{(1)} & =  
20 \lambda_{4}^{2}  + 2 \lambda_4 |g_{d}|^2  -2 |Y_2|^4  -3 g^{2} \lambda_4  + 4 \lambda_4 |Y_2|^2  -\frac{1}{2} |g_{d}|^4  + \frac{3}{8} g^{4}  + \lambda_{1}^{2} + \lambda_{3}^{2}\\ 
\beta_{\lambda_4}^{(2)} & =  
-\frac{25}{16} g^{6} -4 \lambda_{1}^{3} +\frac{5}{4} g^{4} \lambda_3 +2 g^{2} \lambda_{3}^{2} -4 \lambda_{3}^{3} +\frac{63}{8} g^{4} \lambda_4 -10 \lambda_{1}^{2} \lambda_4 -10 \lambda_{3}^{2} \lambda_4 +28 g^{2} \lambda_{4}^{2} \nonumber \\ 
 &-240 \lambda_{4}^{3} -2 \lambda_{1}^{2} |Y_1|^2 -\frac{1}{4} g^{4} |Y_2|^2 +\frac{5}{2} g^{2} \lambda_4 |Y_2|^2 -40 \lambda_{4}^{2} |Y_2|^2 -2 \lambda_{3}^{2} |Y_3|^2 +2 \lambda_4 |Y_2|^4 \nonumber \\ 
 &+2 |Y_1|^2 |Y_2|^4 +2 |Y_3|^2 |Y_2|^4 +g_{d}^{3} g_{d}^{*\,3} -3 \lambda_4 Y_2 |Y_1|^2 Y_2^* +8 Y_{2}^{3} Y_{2}^{*\,3} \nonumber \\ 
 &+\frac{1}{4} g_{d} g_{d}^{*\,2} \Big(2 \Big(-2 g_{u} Y_3 Y_2^*  + g_{d} \lambda_4  + g_{d} |Y_1|^2  + g_{d} |Y_3|^2 \Big) + g_{d} |g_{u}|^2 \Big)-3 \lambda_4 Y_3 |Y_2|^2 Y_3^* \nonumber \\ 
 &-\frac{1}{2} g_{u}^* \Big(2 g_{u} \lambda_{3}^{2}  -2 g_{u} |Y_2|^4  + 4 g_{d} \Big(- \lambda_4  + \lambda_3\Big)Y_2 Y_3^*  + |Y_2|^2 \Big(3 g_{u} \lambda_4  + 4 g_{d} Y_2 Y_3^* \Big)\Big)\nonumber \\ 
 &-\frac{1}{8} g_{d}^* \Big(g^{4} g_{d} -10 g^{2} g_{d} \lambda_4 +160 g_{d} \lambda_{4}^{2} +12 g_{d} \lambda_4 |Y_3|^2 +16 g_{u} \lambda_3 Y_3 Y_2^* -16 g_{u} \lambda_4 Y_3 Y_2^* \nonumber \\ 
 &+16 g_{u} Y_2 Y_3 Y_{2}^{*\,2} +4 g_{d} |Y_1|^2 \Big(3 \lambda_4  -4 Y_2 Y_2^* \Big)-16 g_{d} Y_3 |Y_2|^2 Y_3^* \nonumber \\ 
 &+2 g_{d} g_{u}^* \Big(3 g_{u} \lambda_4  + 4 g_{d} Y_2 Y_3^*  -4 g_{u} |Y_2|^2 \Big)\Big)\\ 
\beta_{\lambda_3}^{(1)} & =  
+\frac{3}{4} g^{4} +2 \lambda_1 \lambda_2 -3 g^{2} \lambda_3 +4 \lambda_{3}^{2} +8 \lambda_3 \lambda_4 +8 \lambda_3 \lambda_5 +2 \lambda_3 |Y_2|^2 +2 \lambda_3 |Y_3|^2 \nonumber \\ 
 &+g_{d}^* \Big(-2 g_{d} |Y_3|^2  + 2 g_{u} Y_3 Y_2^*  - g_{d} |g_{u}|^2  + g_{d} \lambda_3 \Big)-4 Y_3 |Y_2|^2 Y_3^* \nonumber \\ 
 &+g_{u}^* \Big(2 g_{d} Y_2 Y_3^*  -2 g_{u} |Y_2|^2  + g_{u} \lambda_3 \Big)\\ 
\beta_{\lambda_3}^{(2)} & =  
-\frac{25}{8} g^{6} -4 \lambda_{1}^{2} \lambda_2 -4 \lambda_1 \lambda_{2}^{2} +\frac{43}{8} g^{4} \lambda_3 - \lambda_{1}^{2} \lambda_3 -8 \lambda_1 \lambda_2 \lambda_3 - \lambda_{2}^{2} \lambda_3 +2 g^{2} \lambda_{3}^{2} -10 \lambda_{3}^{3} \nonumber \\ 
 &+5 g^{4} \lambda_4 +16 g^{2} \lambda_3 \lambda_4 -48 \lambda_{3}^{2} \lambda_4 -40 \lambda_3 \lambda_{4}^{2} +5 g^{4} \lambda_5 +16 g^{2} \lambda_3 \lambda_5 -48 \lambda_{3}^{2} \lambda_5 -40 \lambda_3 \lambda_{5}^{2} \nonumber \\ 
 &-4 \lambda_1 \lambda_2 |Y_1|^2 -\frac{1}{4} g^{4} |Y_2|^2 +\frac{5}{4} g^{2} \lambda_3 |Y_2|^2 -4 \lambda_{3}^{2} |Y_2|^2 -16 \lambda_3 \lambda_4 |Y_2|^2 -\frac{1}{4} g^{4} |Y_3|^2 \nonumber \\ 
 &+\frac{5}{4} g^{2} \lambda_3 |Y_3|^2 -4 \lambda_{3}^{2} |Y_3|^2 -16 \lambda_3 \lambda_5 |Y_3|^2 -3 \lambda_3 |Y_2|^4 +10 |Y_3|^2 |Y_2|^4 -3 \lambda_3 |Y_3|^4 \nonumber \\ 
 &+10 |Y_2|^2 |Y_3|^4 -\frac{3}{2} \lambda_3 Y_2 |Y_1|^2 Y_2^* -\frac{1}{4} g_{d}^{2} g_{d}^{*\,2} \Big(3 \lambda_3 -10 |Y_3|^2   -5 |g_{u}|^2 \Big)\nonumber \\ 
 &-3 g_{u} g_{d} g_{d}^{*\,2}Y_3 Y_2^* -\frac{3}{2} \lambda_3 Y_3 |Y_1|^2 Y_3^* +5 \lambda_3 Y_3 |Y_2|^2 Y_3^* +12 Y_2 Y_3 |Y_1|^2 Y_2^* Y_3^* \nonumber \\ 
 &-\frac{1}{4} g_{u} g_{u}^{*\,2} \Big(-10 g_{u} |Y_2|^2  + 12 g_{d} Y_2 Y_3^*  + 3 g_{u} \lambda_3 \Big)\nonumber \\ 
 &+g_{d}^* \Big(-\frac{1}{8} g^{4} g_{d} +\frac{5}{8} g^{2} g_{d} \lambda_3 -2 g_{d} \lambda_{3}^{2} -8 g_{d} \lambda_3 \lambda_4 +\frac{5}{2} g_{d} \lambda_3 |Y_3|^2 +\frac{5}{4} g_{d} |g_{u}|^4 +5 g_{d} |Y_3|^4 \nonumber \\ 
 &+2 g_{u} \lambda_3 Y_3 Y_2^* -8 g_{u} \lambda_4 Y_3 Y_2^* -8 g_{u} \lambda_5 Y_3 Y_2^* -6 g_{u} Y_2 Y_3 Y_{2}^{*\,2} \nonumber \\ 
 &+Y_1^* \Big(2 g_{d} Y_1 |Y_3|^2  -4 g_{u} Y_1 Y_3 Y_2^*  -\frac{3}{4} g_{d} \lambda_3 Y_1 \Big)+6 g_{d} Y_3 |Y_2|^2 Y_3^* -6 g_{u} Y_{3}^{2} Y_2^* Y_3^* \nonumber \\ 
 &+g_{u}^* \Big(3 g_{d} g_{u} |Y_1|^2 - 3 g_{d}^{2} Y_2 Y_3^*  + 3 g_{d} g_{u} |Y_3|^2  + 3 g_{u} \Big(g_{d} Y_2  - g_{u} Y_3 \Big)Y_2^*  + \frac{5}{4} g_{d} g_{u} \lambda_3 \Big)\Big)\nonumber \\ 
 &-\frac{1}{8} g_{u}^* \Big(g^{4} g_{u} -5 g^{2} g_{u} \lambda_3 +16 g_{u} \lambda_{3}^{2} +64 g_{u} \lambda_3 \lambda_5 -40 g_{u} |Y_2|^4 -16 g_{d} \lambda_3 Y_2 Y_3^*  \nonumber \\ 
 &+64 g_{d} (\lambda_4 Y_2 Y_3^* + \lambda_5 Y_2 Y_3^*) +48 g_{d} Y_2 Y_3 Y_{3}^{*\,2} +16 |Y_1|^2 \Big(2 g_{d} Y_2 Y_3^*  + \tfrac{3}{8} g_{u} \lambda_3  \nonumber \\ 
 &-g_{u} Y_2 Y_2^* \Big) +4 |Y_2|^2 \Big(12 \Big(g_{d} Y_2  - g_{u} Y_3 \Big)Y_3^*  -5 g_{u} \lambda_3 \Big)\Big)\\ 
\beta_{\lambda_1}^{(1)} & =  
2 \lambda_1 |Y_2|^2  + 2 \lambda_2 \lambda_3  + 2 |Y_1|^2 \Big(\lambda_1 -2 Y_2 Y_2^* \Big) + 4 \lambda_{1}^{2}  + 8 \lambda_1 \lambda_4  -\frac{3}{2} g^{2} \lambda_1  \nonumber \\
&+ |g_{d}|^2 \Big(\lambda_1 -2 Y_1 Y_1^* \Big)\\ 
\beta_{\lambda_1}^{(2)} & =  
\frac{39}{16} g^{4} \lambda_1 +g^{2} \lambda_{1}^{2} -10 \lambda_{1}^{3} +\frac{5}{4} g^{4} \lambda_2 - \lambda_1 \lambda_{2}^{2} +4 g^{2} \lambda_2 \lambda_3 -8 \lambda_1 \lambda_2 \lambda_3 -4 \lambda_{2}^{2} \lambda_3 - \lambda_1 \lambda_{3}^{2} \nonumber \\ 
 &-4 \lambda_2 \lambda_{3}^{2} +16 g^{2} \lambda_1 \lambda_4 -48 \lambda_{1}^{2} \lambda_4 -40 \lambda_1 \lambda_{4}^{2} -3 g^{4} |Y_1|^2 +\frac{5}{2} g^{2} \lambda_1 |Y_1|^2 -4 \lambda_{1}^{2} |Y_1|^2 \nonumber \\ 
 &+\frac{5}{4} g^{2} \lambda_1 |Y_2|^2 -4 \lambda_{1}^{2} |Y_2|^2 -16 \lambda_1 \lambda_4 |Y_2|^2 -4 \lambda_2 \lambda_3 |Y_3|^2 -\frac{1}{4} \Big(-10 |Y_1|^2  + 3 \lambda_1 \Big)|g_{d}|^4 \nonumber \\ 
 &-3 \lambda_1 |Y_1|^4 +10 |Y_2|^2 |Y_1|^4 -3 \lambda_1 |Y_2|^4 +10 |Y_1|^2 |Y_2|^4 -2 g^{2} Y_2 |Y_1|^2 Y_2^*  \nonumber \\ 
 &+5 \lambda_1 Y_2 |Y_1|^2 Y_2^*-\frac{3}{2} \lambda_1 Y_3 |Y_1|^2 Y_3^* -\frac{3}{2} \lambda_1 Y_3 |Y_2|^2 Y_3^* +12 Y_2 Y_3 |Y_1|^2 Y_2^* Y_3^* \nonumber \\ 
 &-\frac{1}{4} g_{u}^* \Big(8 g_{u} \lambda_2 \lambda_3 +3 g_{u} \lambda_1 |Y_2|^2 -4 g_{d} \lambda_1 Y_2 Y_3^* +8 g_{d} \lambda_2 Y_2 Y_3^* \nonumber \\ 
 &+|Y_1|^2 \Big(24 g_{d} Y_2 Y_3^*  + 3 g_{u} \lambda_1  -8 g_{u} Y_2 Y_2^* \Big)\Big)\nonumber \\ 
 &+\frac{1}{8} g_{d}^* \Big(5 g^{2} g_{d} \lambda_1 -16 g_{d} \lambda_{1}^{2} -64 g_{d} \lambda_1 \lambda_4 -6 g_{d} \lambda_1 |Y_3|^2 +40 g_{d} |Y_1|^4 \nonumber \\ 
 &-3 g_{d} |g_{u}|^2 \Big(\lambda_1-8 Y_1 Y_1^*\Big) +8 g_{u} \lambda_1 Y_3 Y_2^* -16 g_{u} \lambda_2 Y_3 Y_2^*    \nonumber \\ 
 &+4 |Y_1|^2 \Big(12 \Big(g_{d} Y_2  - g_{u} Y_3 \Big)Y_2^* + g_{d} \Big(-2 g^{2}  + 4 Y_3 Y_3^*  + 5 \lambda_1 \Big)\Big)\Big)\\ 
\beta_{\lambda_5}^{(1)} & =  
20 \lambda_{5}^{2}  + 2 \lambda_5 |g_{u}|^2  -2 |Y_3|^4  -3 g^{2} \lambda_5  + 4 \lambda_5 |Y_3|^2  -\frac{1}{2} |g_{u}|^4  + \frac{3}{8} g^{4}  + \lambda_{2}^{2} + \lambda_{3}^{2}\\ 
\beta_{\lambda_5}^{(2)} & =  
-\frac{25}{16} g^{6} -4 \lambda_{2}^{3} +\frac{5}{4} g^{4} \lambda_3 +2 g^{2} \lambda_{3}^{2} -4 \lambda_{3}^{3} +\frac{63}{8} g^{4} \lambda_5 -10 \lambda_{2}^{2} \lambda_5 -10 \lambda_{3}^{2} \lambda_5 +28 g^{2} \lambda_{5}^{2} \nonumber \\ 
 &-240 \lambda_{5}^{3} -2 \lambda_{2}^{2} |Y_1|^2 -2 \lambda_{3}^{2} |Y_2|^2 -\frac{1}{4} g^{4} |Y_3|^2 +\frac{5}{2} g^{2} \lambda_5 |Y_3|^2 -40 \lambda_{5}^{2} |Y_3|^2 +2 \lambda_5 |Y_3|^4 \nonumber \\ 
 &+2 |Y_1|^2 |Y_3|^4 +2 |Y_2|^2 |Y_3|^4 +g_{u}^{3} g_{u}^{*\,3} -3 \lambda_5 Y_3 |Y_1|^2 Y_3^* -3 \lambda_5 Y_3 |Y_2|^2 Y_3^* +8 Y_{3}^{3} Y_{3}^{*\,3} \nonumber \\ 
 &+\frac{1}{2} g_{u} g_{u}^{*\,2} \Big(-2 g_{d} Y_2 Y_3^*  + g_{u} \lambda_5  + g_{u} |Y_1|^2  + g_{u} |Y_2|^2 \Big)\nonumber \\ 
 &-\frac{1}{8} g_{u}^* \Big(g^{4} g_{u} -10 g^{2} g_{u} \lambda_5 +160 g_{u} \lambda_{5}^{2} +16 g_{d} \lambda_3 Y_2 Y_3^* -16 g_{d} \lambda_5 Y_2 Y_3^* +16 g_{d} Y_2 Y_3 Y_{3}^{*\,2} \nonumber \\ 
 &+4 g_{u} |Y_1|^2 \Big(3 \lambda_5  -4 Y_3 Y_3^* \Big)+4 g_{u} |Y_2|^2 \Big(3 \lambda_5  -4 Y_3 Y_3^* \Big)\Big)\nonumber \\ 
 &-\frac{1}{4} g_{d}^* \Big(2 g_{d} \Big(2 \lambda_{3}^{2}  -2 |Y_3|^4  + 3 \lambda_5 |Y_3|^2 \Big) - g_{d} |g_{u}|^4 +8 g_{u} Y_3 \Big(- \lambda_5  + \lambda_3 + |Y_3|^2\Big)Y_2^* \nonumber \\ 
 &+|g_{u}|^2 \Big(3 g_{d} \lambda_5  -4 g_{d} Y_3 Y_3^*  + 4 g_{u} Y_3 Y_2^* \Big)\Big)\\ 
\beta_{\lambda_2}^{(1)} & =  
2 \lambda_1 \lambda_3  + 2 \lambda_2 |Y_3|^2  + 2 |Y_1|^2 \Big(-2 Y_3 Y_3^*  + \lambda_2\Big) + 4 \lambda_{2}^{2}  + 8 \lambda_2 \lambda_5  -\frac{3}{2} g^{2} \lambda_2  \nonumber \\
&+ |g_{u}|^2 \Big(-2 Y_1 Y_1^*  + \lambda_2\Big)\\ 
\beta_{\lambda_2}^{(2)} & =  g^{4} \Big(
\frac{5}{4}  \lambda_1 +\frac{39}{16} \lambda_2 \Big) - \lambda_{1}^{2} \lambda_2 +g^{2} \lambda_{2}^{2} -10 \lambda_{2}^{3} -4 \Big(\lambda_1 \lambda_{3}^{2} +\lambda_{1}^{2} \lambda_3 - g^{2} \lambda_1 \lambda_3\Big) -8 \lambda_1 \lambda_2 \lambda_3  \nonumber \\ 
 &- \lambda_2 \lambda_{3}^{2} +16 g^{2} \lambda_2 \lambda_5 -48 \lambda_{2}^{2} \lambda_5 -40 \lambda_2 \lambda_{5}^{2} -3 g^{4} |Y_1|^2 +\frac{5}{2} g^{2} \lambda_2 |Y_1|^2 -4 \lambda_{2}^{2} |Y_1|^2 \nonumber \\ 
 &-4 \lambda_1 \lambda_3 |Y_2|^2 +\frac{5}{4} g^{2} \lambda_2 |Y_3|^2 -4 \lambda_{2}^{2} |Y_3|^2 -16 \lambda_2 \lambda_5 |Y_3|^2 -\frac{1}{4} \Big(3 \lambda_2 -10 |Y_1|^2  \Big)|g_{u}|^4 \nonumber \\ 
 &-3 \lambda_2 |Y_1|^4 +10 |Y_3|^2 |Y_1|^4 -3 \lambda_2 |Y_3|^4 +10 |Y_1|^2 |Y_3|^4 -\frac{3}{2} \lambda_2 Y_2 |Y_1|^2 Y_2^*  \nonumber \\ 
 &-2 g^{2} Y_3 |Y_1|^2 Y_3^* +5 \lambda_2 Y_3 |Y_1|^2 Y_3^* -\frac{3}{2} \lambda_2 Y_3 |Y_2|^2 Y_3^* +12 Y_2 Y_3 |Y_1|^2 Y_2^* Y_3^* \nonumber \\ 
 &+\frac{1}{8} g_{u}^* \Big(5 g^{2} g_{u} \lambda_2 -16 g_{u} \lambda_{2}^{2} -64 g_{u} \lambda_2 \lambda_5 -6 g_{u} \lambda_2 |Y_2|^2 +40 g_{u} |Y_1|^4 -16 g_{d} \lambda_1 Y_2 Y_3^*  \nonumber \\ 
 &+8 g_{d} \lambda_2 Y_2 Y_3^* +4 |Y_1|^2 \Big(-12 g_{d} Y_2 Y_3^*  + 12 g_{u} Y_3 Y_3^*  -2 g^{2} g_{u}  + 4 g_{u} Y_2 Y_2^*  + 5 g_{u} \lambda_2 \Big)\Big)\nonumber \\ 
 &-\frac{1}{8} g_{d}^* \Big(3 g_{d} |g_{u}|^2 \Big(-8 Y_1 Y_1^*  + \lambda_2\Big)\nonumber \\ 
 &+2 \Big(8 g_{d} \lambda_1 \lambda_3 +3 g_{d} \lambda_2 |Y_3|^2 +8 g_{u} \lambda_1 Y_3 Y_2^* -4 g_{u} \lambda_2 Y_3 Y_2^* \nonumber \\ 
 &+|Y_1|^2 \Big(24 g_{u} Y_3 Y_2^*  + 3 g_{d} \lambda_2  -8 g_{d} Y_3 Y_3^* \Big)\Big)\Big)
\end{align}} 
\subsection{Yukawa couplings}
{\allowdisplaybreaks  \begin{align} 
\beta_{g_{d}}^{(1)} & =  
\frac{1}{4} \Big(2 g_{d} |Y_1|^2  + 2 g_{d} |Y_3|^2  -3 g^{2} g_{d}  + 4 g_{d}^{2} g_{d}^*  + 4 g_{d} |Y_2|^2  -8 g_{u} Y_3 Y_2^*  + g_{d} |g_{u}|^2 \Big)\\ 
\beta_{g_{d}}^{(2)} & =  
\frac{1}{32} \Big(-7 g_{d} |g_{u}|^4 -10 g_{d}^{3} g_{d}^{*\,2} +g_{u}^* \Big(28 g_{d} g_{u} |Y_1|^2  + 2 g_{u} \Big(-7 g_{d} Y_2 + 24 g_{u} Y_3 \Big)Y_2^*  \nonumber \\ 
 &+ g_{d} \Big(16 g_{d} Y_2 Y_3^*  -32 g_{u} \lambda_3  -56 g_{u} |Y_3|^2  -7 g^{2} g_{u} \Big)\Big) - |g_{d}|^2 \Big(2 \Big(12 g_{d} Y_2 Y_2^*  \nonumber \\ 
 & -8 g_{u} Y_3 Y_2^*  -31 g^{2} g_{d}  + 64 g_{d} \lambda_4  + 7 g_{d} Y_1 Y_1^*  + 7 g_{d} Y_3 Y_3^* \Big) + 7 g_{d} g_{u} g_{u}^* \Big)\nonumber \\ 
 &-2 \Big(14 g_{d} |Y_1|^4 -2 \Big(16 g_{u} \lambda_3 Y_3  + \Big(24 g_{u} Y_3 -7 g_{d} Y_2  \Big)|Y_3|^2  + g^{2} \Big(5 g_{d} Y_2 -4 g_{u} Y_3 \Big)\Big)Y_2^* \nonumber \\ 
 &+8 Y_2 \Big(3 g_{d} Y_2  -4 g_{u} Y_3 \Big)Y_{2}^{*\,2} +2 |Y_1|^2 \Big(4 g_{d} \Big(4 \lambda_1  + g^{2}\Big) + \Big(7 g_{d} Y_2  -8 g_{u} Y_3 \Big)Y_2^* \Big)\nonumber \\ 
 &+g_{d} \Big(\Big(-11 g^{2} Y_3  + 32 \lambda_3 Y_3 \Big)Y_3^*  + 14 |Y_3|^4  + 2 \Big(-4 \Big(8 \lambda_{4}^{2}  + \lambda_{1}^{2} + \lambda_{3}^{2}\Big) + g^{4}\Big)\Big)\Big)\Big)\\ 
\beta_{g_{u}}^{(1)} & =  
\frac{1}{4} \Big(2 g_{u} |Y_1|^2  + 2 g_{u} |Y_2|^2  -3 g^{2} g_{u}  + 4 g_{u}^{2} g_{u}^*  + 4 g_{u} |Y_3|^2  -8 g_{d} Y_2 Y_3^*  + g_{u} |g_{d}|^2 \Big)\\ 
\beta_{g_{u}}^{(2)} & =  
\frac{1}{32} \Big(-7 g_{u} |g_{d}|^4 \nonumber \\ 
 &- g_{d}^* \Big(7 g^{2} g_{d} g_{u} +32 g_{d} g_{u} \lambda_3 -28 g_{d} g_{u} |Y_1|^2 +56 g_{d} g_{u} |Y_2|^2 +14 g_{d} g_{u} |Y_3|^2 +7 g_{d} g_{u}^{2} g_{u}^* \nonumber \\ 
 &-16 g_{u}^{2} Y_3 Y_2^* -48 g_{d}^{2} Y_2 Y_3^* \Big)\nonumber \\ 
 &-2 \Big(2 g^{4} g_{u} -8 g_{u} \lambda_{2}^{2} -8 g_{u} \lambda_{3}^{2} -64 g_{u} \lambda_{5}^{2} -11 g^{2} g_{u} |Y_2|^2 +32 g_{u} \lambda_3 |Y_2|^2 -10 g^{2} g_{u} |Y_3|^2 \nonumber \\ 
 &+14 g_{u} |Y_1|^4 +14 g_{u} |Y_2|^4 +24 g_{u} |Y_3|^4 +5 g_{u}^{3} g_{u}^{*\,2} +8 g^{2} g_{d} Y_2 Y_3^* -32 g_{d} \lambda_3 Y_2 Y_3^*  \nonumber \\ 
 &+14 g_{u} Y_3 |Y_2|^2 Y_3^* -48 g_{d} Y_{2}^{2} Y_2^* Y_3^* -32 g_{d} Y_2 Y_3 Y_{3}^{*\,2} \nonumber \\ 
 &+|g_{u}|^2 \Big(12 g_{u} Y_3 Y_3^*  -8 g_{d} Y_2 Y_3^*  -31 g^{2} g_{u}  + 64 g_{u} \lambda_5  + 7 g_{u} Y_1 Y_1^*  + 7 g_{u} Y_2 Y_2^* \Big)\nonumber \\ 
 &+2 |Y_1|^2 \Big(4 g_{u} \Big(4 \lambda_2  + g^{2}\Big) + \Big(7 g_{u} Y_3  -8 g_{d} Y_2 \Big)Y_3^* \Big)\Big)\Big)\\ 
\beta_{Y_3}^{(1)} & =  
\frac{1}{4} \Big(\Big(2 g_{u} Y_3  -4 g_{d} Y_2 \Big)g_{u}^*  + Y_3 \Big(2 |Y_1|^2  + 2 |Y_2|^2  -3 g^{2}  + 8 |Y_3|^2 \Big) + Y_3 |g_{d}|^2 \Big)\\ 
\beta_{Y_3}^{(2)} & =  
\frac{1}{32} \Big(-7 Y_3 |g_{d}|^4 +g_{d}^* \Big(g_{d} \Big(-7 g_{u} Y_3  + 24 g_{d} Y_2 \Big)g_{u}^*  + Y_3 \Big(-8 \Big(-2 g_{u} Y_3  \nonumber \\ 
 &+ 7 g_{d} Y_2 \Big)Y_2^*  + g_{d} \Big(11 g^{2}  -14 |Y_3|^2  -32 \lambda_3 \Big)\Big)\Big) -2 \Big(2 g_{u} \Big(3 g_{u} Y_3  -4 g_{d} Y_2 \Big)g_{u}^{*\,2} \nonumber \\ 
 &+g_{u}^* \Big(4 g^{2} g_{d} Y_2 -16 g_{d} \lambda_3 Y_2 -5 g^{2} g_{u} Y_3 +\Big(7 g_{u} Y_3  -24 g_{d} Y_2 \Big)|Y_2|^2 -8 g_{d} Y_2 |Y_3|^2 \nonumber \\ 
 &+\Big(7 g_{u} Y_1 Y_3  -8 g_{d} Y_1 Y_2 \Big)Y_1^* +12 g_{u} Y_{3}^{2} Y_3^* \Big) +Y_3 \Big(14 |Y_1|^4 +14 |Y_2|^4  \nonumber \\ 
 &+2 |Y_1|^2 \Big(-14 Y_2 Y_2^*  + 4 \Big(4 \lambda_2  + g^{2}\Big) + 7 Y_3 Y_3^* \Big)+|Y_2|^2 \Big(14 Y_3 Y_3^*  + 32 \lambda_3  + 7 g^{2} \Big)\nonumber \\ 
 &+2 \Big(10 |Y_3|^4  + \Big(-31 g^{2} Y_3  + 64 \lambda_5 Y_3 \Big)Y_3^*  -4 \Big(8 \lambda_{5}^{2}  + \lambda_{2}^{2} + \lambda_{3}^{2}\Big) + g^{4}\Big)\Big)\Big)\Big)\\ 
\beta_{Y_2}^{(1)} & =  
\frac{1}{4} \Big(2 \Big(-2 g_{u} Y_3  + g_{d} Y_2 \Big)g_{d}^*  + Y_2 \Big(2 |Y_1|^2  + 2 |Y_3|^2  -3 g^{2}  + 8 |Y_2|^2  + |g_{u}|^2\Big)\Big)\\ 
\beta_{Y_2}^{(2)} & =  
\frac{1}{32} \Big(-4 g_{d} \Big(3 g_{d} Y_2  -4 g_{u} Y_3 \Big)g_{d}^{*\,2} +g_{d}^* \Big(\Big(-7 g_{d} Y_2  + 24 g_{u} Y_3 \Big)|g_{u}|^2 \nonumber \\  
 &+2 \Big(5 g^{2} g_{d} Y_2 -4 g^{2} g_{u} Y_3 +16 g_{u} \lambda_3 Y_3 -12 \Big(-\frac23 g_{u} Y_3  + g_{d} Y_2 \Big)|Y_2|^2 -7 g_{d} Y_2 |Y_3|^2 \nonumber \\ 
 &+\Big(8 g_{u} Y_1 Y_3 -7 g_{d} Y_1 Y_2 \Big)Y_1^* +24 g_{u} Y_{3}^{2} Y_3^* \Big)\Big)\nonumber \\ 
 &- Y_2 \Big(7 |g_{u}|^4 +g_{u}^* \Big(-11 g^{2} g_{u}  + 14 g_{u} |Y_2|^2  + 32 g_{u} \lambda_3  -8 \Big(2 g_{d} Y_2  -7 g_{u} Y_3 \Big)Y_3^* \Big)\nonumber \\ 
 &+2 \Big(2 g^{4} -8 \lambda_{1}^{2} -8 \lambda_{3}^{2} -64 \lambda_{4}^{2} +7 g^{2} |Y_3|^2 +32 \lambda_3 |Y_3|^2 +14 |Y_1|^4 +20 |Y_2|^4 +14 |Y_3|^4 \nonumber \\ 
 &+2 |Y_1|^2 \Big(16 \lambda_1  -14 Y_3 Y_3^* + 4 g^{2}  + 7 Y_2 Y_2^* \Big)+2 |Y_2|^2 \Big(64 \lambda_4 -31 g^{2}  + 7 Y_3 Y_3^* \Big)\Big)\Big)\Big)\\ 
\beta_{Y_1}^{(1)} & =  
\frac{1}{4} Y_1 \Big(2 |Y_2|^2  + 2 |Y_3|^2  -6 g^{2}  + 8 |Y_1|^2  + |g_{d}|^2 + |g_{u}|^2\Big)\\ 
\beta_{Y_1}^{(2)} & =  
-\frac{1}{32} Y_1 \Big(7 |g_{d}|^4 +7 |g_{u}|^4 +g_{d}^* \Big(32 g_{d} \lambda_1 -11 g^{2} g_{d}  -14 g_{d} |g_{u}|^2  + 14 g_{d} |Y_1|^2  \nonumber \\ 
 &-32 g_{u} Y_3 Y_2^*  + 56 g_{d} |Y_2|^2 \Big) +g_{u}^* \Big(14 g_{u} |Y_1|^2 + (32  \lambda_2 -11 g^{2}) g_{u}  -8 \Big(4 g_{d} Y_2  \nonumber \\ 
 &-7 g_{u} Y_3 \Big)Y_3^* \Big) +2 \Big(20 |Y_1|^4 +14 (|Y_2|^4 + |Y_3|^4)  -8 (\lambda_{1}^{2} + \lambda_{2}^{2}) -11 g^{2} (g^{2} + |Y_3|^2 )\nonumber \\ 
 &+ 32 \lambda_2 |Y_3|^2+|Y_2|^2 \Big(32 \lambda_1  -11 g^{2}  -28 Y_3 Y_3^* \Big)+14 |Y_1|^2 \Big( Y_2 Y_2^*  +  Y_3 Y_3^* -\frac{26}{7} g^{2} \Big)\Big)\Big)
\end{align}} 
\subsection{Fermion mass terms}
{\allowdisplaybreaks  \begin{align} 
\beta_{M_1}^{(1)} & =  
\frac{1}{2} M_1 \Big(|g_{d}|^2 + |g_{u}|^2\Big)\\ 
\beta_{M_1}^{(2)} & =  
\frac{1}{16} \Big(M_1 |g_{d}|^4 +g_{u} \Big(16 g_{d} {T}_3 Y_1^*  + g_{u} M_1 g_{u}^{*\,2}  + M_1 \Big(-12 |Y_3|^2  + 17 g^{2}  \nonumber \\ 
 &-2 |Y_1|^2  -2 |Y_2|^2 \Big)g_{u}^* \Big) +M_1 |g_{d}|^2 \Big(-12 Y_2 Y_2^*  + 17 g^{2}  -2 Y_1 Y_1^*  -2 Y_3 Y_3^* \Big)\Big)\\ 
\beta_{M_2}^{(1)} & =  
\frac{1}{4} M_2 \Big(2 |Y_2|^2  + 2 |Y_3|^2  + 4 |Y_1|^2  -6 g^{2}  + |g_{d}|^2 + |g_{u}|^2\Big)\\ 
\beta_{M_2}^{(2)} & =  
\frac{1}{32} \Big(22 g^{4} M_2 +64 g^{2} M_2 |Y_1|^2 +22 g^{2} M_2 |Y_2|^2 +22 g^{2} M_2 |Y_3|^2 -7 M_2 |g_{d}|^4  \nonumber \\ 
 &-7 M_2 |g_{u}|^4  +8 M_2 |Y_1|^4 -28 M_2 |Y_2|^4 -28 M_2 |Y_3|^4 -64 Y_1 |Y_2|^2 T_1^* -64 Y_1 |Y_3|^2 T_2^* \nonumber \\ 
 &  +g_{d}^* \Big(14 g_{d} M_2 |g_{u}|^2  -32 g_{d} Y_1 T_1^*  + M_2 \Big(11 g^{2} g_{d}  -2 g_{d} |Y_1|^2  + 32 g_{u} Y_3 Y_2^*  \nonumber \\ 
 &-56 g_{d} |Y_2|^2 \Big)\Big) -4 M_2 Y_2 |Y_1|^2 Y_2^* -4 M_2 Y_3 |Y_1|^2 Y_3^* +56 M_2 Y_3 |Y_2|^2 Y_3^* \nonumber \\ 
 &+g_{u}^* \Big(-32 g_{u} Y_1 T_2^*  + M_2 \Big(11 g^{2} g_{u}  -2 g_{u} |Y_1|^2  + 8 \Big(4 g_{d} Y_2  -7 g_{u} Y_3 \Big)Y_3^* \Big)\Big)\Big)
\end{align}} 
\subsection{Trilinear scalar couplings}
{\allowdisplaybreaks  \begin{align} 
\beta_{{T}_3}^{(1)} & =  
\frac{1}{2} \Big(g_{d}^* \Big(8 M_1 Y_1 g_{u}^*  + g_{d} {T}_3 \Big) \nonumber \\
&+ {T}_3 \Big(2 |Y_1|^2  + 2 |Y_2|^2  + 2 |Y_3|^2  -3 g^{2}  + 4 \lambda_1  + 4 \lambda_2  + 4 \lambda_3  + |g_{u}|^2\Big)\Big)\\ 
\beta_{{T}_3}^{(2)} & =  
\frac{1}{16} \Big(-2 g_{d}^{*\,2} \Big(3 g_{d}^{2} {T}_3  + 40 g_{d} M_1 Y_1 g_{u}^*  -64 M_1 Y_1 Y_3 Y_2^* \Big)\nonumber \\ 
 &+g_{d}^* \Big(-80 g_{u} M_1 Y_1 g_{u}^{*\,2} +2 g_{u}^* \Big(-32 \lambda_3 M_1 Y_1  \nonumber \\ 
 &-32 M_1 Y_{1}^{2} Y_1^*  -48 M_1 Y_1 |Y_2|^2  -48 M_1 Y_1 |Y_3|^2  + 5 g_{d} g_{u} {T}_3  + 8 g^{2} M_1 Y_1 \Big)\nonumber \\ 
 &+{T}_3 \Big(-12 g_{d} |Y_1|^2  + 16 g_{u} Y_3 Y_2^*  + g_{d} \Big(-12 |Y_3|^2  -16 \Big(\lambda_1 + \lambda_3\Big) + 5 g^{2} \Big)\Big)\Big)\nonumber \\ 
 &+{T}_3 g_{u}^* \Big(-12 g_{u} |Y_1|^2  -12 g_{u} |Y_2|^2  + 16 g_{d} Y_2 Y_3^*  -16 g_{u} \lambda_2  -16 g_{u} \lambda_3  + 5 g^{2} g_{u} \Big)\nonumber \\ 
 &+g_{u}^{*\,2} \Big(128 M_1 Y_1 Y_2 Y_3^*  -6 g_{u}^{2} {T}_3 \Big) +{T}_3 \Big(19 g^{4} +8 g^{2} \lambda_1 \nonumber \\ 
 & -16 \lambda_{1}^{2} +8 g^{2} \lambda_2 -96 \lambda_1 \lambda_2 -16 \lambda_{2}^{2} +64 g^{2} \lambda_3 -96 \lambda_1 \lambda_3 -96 \lambda_2 \lambda_3 -16 \lambda_{3}^{2} \nonumber \\ 
 &-64 (2 \lambda_1 \lambda_4 +2 \lambda_3 \lambda_4 - \lambda_{4}^{2} + 2 \lambda_2 \lambda_5 +2 \lambda_3 \lambda_5 - \lambda_{5}^{2}) +10 g^{2} |Y_3|^2 -32 \lambda_2 |Y_3|^2 \nonumber \\ 
 &-32 \lambda_3 |Y_3|^2 -24 (|Y_1|^4 + |Y_2|^4 + |Y_3|^4) +4 |Y_1|^2 \Big(10 (Y_2 Y_2^*  +  Y_3 Y_3^*)  + 5 g^{2}  \nonumber \\ 
 &-8 (\lambda_1  + \lambda_2) \Big) +2 |Y_2|^2 \Big(-16 \Big(\lambda_1 + \lambda_3\Big) + 20 Y_3 Y_3^*  + 5 g^{2} \Big)\Big)\Big)\\ 
\beta_{T_1}^{(1)} & =  
2 \lambda_3 T_2  + 2 T_1 |Y_2|^2  + 4 \lambda_1 T_1  -4 Y_1 |Y_2|^2 M_2^*  + 8 \lambda_4 T_1  -\frac{3}{2} g^{2} T_1  \nonumber \\
&+ |g_{d}|^2 \Big(-2 Y_1 M_2^*  + T_1\Big) + T_1 |Y_1|^2 \\ 
\beta_{T_1}^{(2)} & =  
\frac{39}{16} g^{4} T_1 +g^{2} \lambda_1 T_1 -\frac{21}{2} \lambda_{1}^{2} T_1 +\frac{1}{2} \lambda_{2}^{2} T_1 -4 \lambda_2 \lambda_3 T_1 - \lambda_{3}^{2} T_1 +16 g^{2} \lambda_4 T_1  \nonumber \\ 
 &-48 \lambda_1 \lambda_4 T_1 -40 \lambda_{4}^{2} T_1 +\frac{5}{4} g^{4} T_2 -2 \lambda_1 \lambda_2 T_2 +4 g^{2} \lambda_3 T_2 -4 \lambda_1 \lambda_3 T_2 -4 \lambda_2 \lambda_3 T_2  \nonumber \\ 
 &-4 \lambda_{3}^{2} T_2 +\frac{5}{4} g^{2} T_1 |Y_1|^2 -4 \lambda_1 T_1 |Y_1|^2 +\frac{5}{4} g^{2} T_1 |Y_2|^2 -4 \lambda_1 T_1 |Y_2|^2 -16 \lambda_4 T_1 |Y_2|^2 \nonumber \\ 
 &-4 \lambda_3 T_2 |Y_3|^2 -\frac{3}{2} T_1 |Y_1|^4 -3 T_1 |Y_2|^4 -3 g^{4} Y_1 M_2^* -2 g^{2} Y_1 |Y_2|^2 M_2^* +4 \lambda_1 Y_1 |Y_2|^2 M_2^* \nonumber \\ 
 &+10 Y_1 |Y_2|^4 M_2^* -\frac{1}{4} |g_{d}|^4 \Big(-10 Y_1 M_2^*  + 3 T_1 \Big)+10 Y_{1}^{2} |Y_2|^2 M_2^* Y_1^* +\frac{7}{4} T_1 Y_2 |Y_1|^2 Y_2^* \nonumber \\ 
 &+\frac{1}{8} g_{d}^* \Big(5 g^{2} g_{d} T_1 -16 g_{d} \lambda_1 T_1 -64 g_{d} \lambda_4 T_1 +7 g_{d} T_1 |Y_1|^2 -6 g_{d} T_1 |Y_3|^2 \nonumber \\ 
 &-3 g_{d} |g_{u}|^2 \Big(-8 Y_1 M_2^*  + T_1\Big) +8 g_{u} T_1 Y_3 Y_2^* -16 g_{u} T_2 Y_3 Y_2^* \nonumber \\ 
 &+8 Y_1 M_2^* \Big(2 g_{d} \lambda_1  + 2 g_{d} |Y_3|^2  + 5 g_{d} |Y_1|^2  + 6 g_{d} |Y_2|^2  -6 g_{u} Y_3 Y_2^*  - g^{2} g_{d} \Big)\Big)\nonumber \\ 
 &-\frac{3}{4} T_1 Y_3 |Y_1|^2 Y_3^* -\frac{3}{2} T_1 Y_3 |Y_2|^2 Y_3^* +12 Y_1 Y_3 |Y_2|^2 M_2^* Y_3^* -\frac{3}{8} g_{u}^*  g_{u} T_1 |Y_1|^2 \nonumber \\ 
 &-\frac{1}{8} g_{u}^* \Big(2 \Big(4 g_{d} Y_2 \Big(2 T_2  +6 Y_1 M_2^*  - T_1\Big)Y_3^*  + 8 g_{u} \lambda_3 T_2  + g_{u} |Y_2|^2 \Big(3 T_1  -8 Y_1 M_2^* \Big)\Big) \Big)\\ 
\beta_{T_2}^{(1)} & =  
2 \lambda_3 T_1  + 2 T_2 |Y_3|^2  + 4 \lambda_2 T_2  -4 Y_1 |Y_3|^2 M_2^*  + 8 \lambda_5 T_2  -\frac{3}{2} g^{2} T_2  \nonumber \\
&+ |g_{u}|^2 \Big(-2 Y_1 M_2^*  + T_2\Big) + T_2 |Y_1|^2 \\ 
\beta_{T_2}^{(2)} & =  
\frac{5}{4} g^{4} T_1 -2 \lambda_1 \lambda_2 T_1 +4 g^{2} \lambda_3 T_1 -4 \lambda_1 \lambda_3 T_1 -4 \lambda_2 \lambda_3 T_1 -4 \lambda_{3}^{2} T_1 +\frac{39}{16} g^{4} T_2 +\frac{1}{2} \lambda_{1}^{2} T_2 \nonumber \\ 
 &+g^{2} \lambda_2 T_2 -\frac{21}{2} \lambda_{2}^{2} T_2 -4 \lambda_1 \lambda_3 T_2 - \lambda_{3}^{2} T_2 +16 g^{2} \lambda_5 T_2 -48 \lambda_2 \lambda_5 T_2 -40 \lambda_{5}^{2} T_2 \nonumber \\ 
 &+\frac{5}{4} g^{2} T_2 |Y_1|^2 -4 \lambda_2 T_2 |Y_1|^2 -4 \lambda_3 T_1 |Y_2|^2 +\frac{5}{4} g^{2} T_2 |Y_3|^2 -4 \lambda_2 T_2 |Y_3|^2 \nonumber \\ 
 &-16 \lambda_5 T_2 |Y_3|^2 -\frac{3}{2} T_2 |Y_1|^4 -3 T_2 |Y_3|^4 -3 g^{4} Y_1 M_2^* -2 g^{2} Y_1 |Y_3|^2 M_2^*  \nonumber \\ 
 &+4 \lambda_2 Y_1 |Y_3|^2 M_2^* +10 Y_1 |Y_3|^4 M_2^* -\frac{1}{4} |g_{u}|^4 \Big(-10 Y_1 M_2^*  + 3 T_2 \Big)+10 Y_{1}^{2} |Y_3|^2 M_2^* Y_1^*  \nonumber \\ 
 &-\frac{3}{4} T_2 Y_2 |Y_1|^2 Y_2^* -\frac{1}{8} g_{d}^* \Big(3 g_{d} T_2 |Y_1|^2 +3 g_{d} |g_{u}|^2 \Big(-8 Y_1 M_2^*  + T_2\Big)\nonumber \\ 
 &+2 \Big(4 g_{u} Y_3 \Big(2 T_1  + 6 Y_1 M_2^*  - T_2 \Big)Y_2^*  + g_{d} \Big(8 \lambda_3 T_1  + |Y_3|^2 \Big(3 T_2  -8 Y_1 M_2^* \Big)\Big)\Big)\Big)\nonumber \\ 
 &+\frac{7}{4} T_2 Y_3 |Y_1|^2 Y_3^* -\frac{3}{2} T_2 Y_3 |Y_2|^2 Y_3^* +12 Y_1 Y_3 |Y_2|^2 M_2^* Y_3^* +\frac{1}{8} g_{u}^* \Big(5 g^{2} g_{u} T_2 \nonumber \\ 
 & -16 g_{u} \lambda_2 T_2 -64 g_{u} \lambda_5 T_2 +7 g_{u} T_2 |Y_1|^2 -6 g_{u} T_2 |Y_2|^2 -16 g_{d} T_1 Y_2 Y_3^* +8 g_{d} T_2 Y_2 Y_3^* \nonumber \\ 
 & +8 Y_1 M_2^* \Big(2 g_{u} \lambda_2  + 2 g_{u} |Y_2|^2  + 5 g_{u} |Y_1|^2  -6 g_{d} Y_2 Y_3^*  + 6 g_{u} |Y_3|^2  - g^{2} g_{u} \Big)\Big)
\end{align}} 
\subsection{Scalar mass terms}
{\allowdisplaybreaks  \begin{align} 
\beta_{{B}}^{(1)} & =  
\frac{1}{2} \Big(2 {B} |Y_2|^2  + 2 {B} |Y_3|^2  -3 {B} g^{2}  + 4 {B} \lambda_3  + 4 {T}_3 T_1^*  + 4 {T}_3 T_2^*  + {B} |g_{u}|^2  \nonumber \\
&+ g_{d}^* \Big(8 M_1 M_2 g_{u}^*  + {B} g_{d} \Big)\Big)\\ 
\beta_{{B}}^{(2)} & =  
+\frac{19}{16} {B} g^{4} +\frac{1}{2} {B} \lambda_{1}^{2} -2 {B} \lambda_1 \lambda_2 +\frac{1}{2} {B} \lambda_{2}^{2} +4 {B} g^{2} \lambda_3 - {B} \lambda_{3}^{2} -8 {B} \lambda_3 \lambda_4 +4 {B} \lambda_{4}^{2} \nonumber \\ 
 &-8 {B} \lambda_3 \lambda_5  +4 {B} \lambda_{5}^{2} +\frac{5}{8} {B} g^{2} |Y_2|^2 -2 {B} \lambda_3 |Y_2|^2 +\frac{5}{8} {B} g^{2} |Y_3|^2 -2 {B} \lambda_3 |Y_3|^2 -\frac{3}{2} {B} |Y_2|^4  \nonumber \\ 
 &-\frac{3}{2} {B} |Y_3|^4 +\frac{1}{2} g^{2} {T}_3 T_1^* -2 \lambda_1 {T}_3 T_1^* -2 \lambda_2 {T}_3 T_1^* -6 \lambda_3 {T}_3 T_1^* -8 \lambda_4 {T}_3 T_1^* -2 {T}_3 |Y_1|^2 T_1^* \nonumber \\ 
 &-2 {T}_3 |Y_2|^2 T_1^* +\frac{1}{2} g^{2} {T}_3 T_2^* -2 \lambda_1 {T}_3 T_2^* -2 \lambda_2 {T}_3 T_2^* -6 \lambda_3 {T}_3 T_2^* -8 \lambda_5 {T}_3 T_2^* \nonumber \\ 
 &-2 {T}_3 |Y_1|^2 T_2^* -2 {T}_3 |Y_3|^2 T_2^* +4 M_2 {T}_3 |Y_2|^2 Y_1^* +4 M_2 {T}_3 |Y_3|^2 Y_1^* -\frac{3}{4} {B} Y_2 |Y_1|^2 Y_2^* \nonumber \\ 
 &+g_{d}^{*\,2} \Big(-5 g_{d} M_1 M_2 g_{u}^*  + 8 M_1 M_2 Y_3 Y_2^*  -\frac{3}{8} {B} g_{d}^{2} \Big)\nonumber \\ 
 &-\frac{1}{16} g_{d}^* \Big(\Big(64 \lambda_3 M_1 M_2  + 64 M_1 M_2 |Y_1|^2  + 96 M_1 M_2 ( |Y_2|^2  +  |Y_3|^2) -10 {B} g_{d} g_{u}   \nonumber \\ 
 &-16 g^{2} M_1 M_2 \Big)g_{u}^* +80 g_{u} M_1 M_2 g_{u}^{*\,2} +16 g_{d} {T}_3 T_1^* +{B} \Big(12 g_{d} |Y_3|^2  + 16 (g_{d} \lambda_3 - g_{u} Y_3 Y_2^*) \nonumber \\ 
 &  -5 g^{2} g_{d}  + 6 g_{d} |Y_1|^2 \Big)\Big)  - {B} Y_3( \frac{3}{4} |Y_1|^2 -\frac{5}{2} |Y_2|^2) Y_3^* +g_{u}^{*\,2} \Big(8 M_1 M_2 Y_2 Y_3^*  -\frac{3}{8} {B} g_{u}^{2} \Big)\nonumber \\ 
 &+g_{u}^* \Big(\frac{1}{16} {B} \Big(16 g_{d} Y_2 Y_3^* -12 g_{u} |Y_2|^2  -16 g_{u} \lambda_3  + 5 g^{2} g_{u}  -6 g_{u} |Y_1|^2 \Big) - g_{u} {T}_3 T_2^* \Big)\\ 
\beta_{m^2_{1}}^{(1)} & =  
-\frac{3}{2} g^{2} m^2_{1} +8 \lambda_4 m^2_{1} +2 \lambda_3 m^2_{2} +2 \lambda_1 m^2_{3} +4 |T_1|^2 +2 |{T}_3|^2 +2 m^2_{1} |Y_2|^2 \nonumber \\ 
 &+|g_{d}|^2 \Big(-2 M_2 M_2^*  -8 M_1 M_1^*  + m^2_{1}\Big)-4 Y_2 |M_2|^2 Y_2^* \\ 
\beta_{m^2_{1}}^{(2)} & =  
\Big(\frac{39}{16} g^{4} - \lambda_{1}^{2}  - \lambda_{3}^{2}  +16 g^{2} \lambda_4  -40 \lambda_{4}^{2}\Big) m^2_{1} +\Big(\frac{5}{4} g^{4}  +4 g^{2} \lambda_3  -4 \lambda_{3}^{2} \Big)m^2_{2} -4 \lambda_{1}^{2} m^2_{3}  \nonumber \\ 
 &-3 g^{4} |M_2|^2 +g^{2} |T_1|^2 -10 \lambda_1 |T_1|^2 -48 \lambda_4 |T_1|^2 -2 \lambda_1 |T_2|^2 -4 \lambda_3 |T_2|^2 \nonumber \\ 
 &+\frac{1}{2} g^{2} |{T}_3|^2 -6 \lambda_1 |{T}_3|^2 -6 \lambda_3 |{T}_3|^2 -8 \lambda_4 |{T}_3|^2 -4 \lambda_1 m^2_{3} |Y_1|^2 +\frac{5}{4} g^{2} m^2_{1} |Y_2|^2 \nonumber \\ 
 &-16 \lambda_4 m^2_{1} |Y_2|^2 -4 \lambda_3 m^2_{2} |Y_3|^2 -\frac{1}{4} \Big(3 m^2_{1} -10 |M_2|^2  -64 |M_1|^2 \Big)|g_{d}|^4 -3 m^2_{1} |Y_2|^4  \nonumber \\ 
 &+10 |M_2|^2 |Y_2|^4 -4 \lambda_3 T_2 T_1^* +4 Y_1 |Y_2|^2 M_2^* T_1^* -4 \lambda_3 T_1 T_2^* -4 Y_1 |T_1|^2 Y_1^* \nonumber \\ 
 &-2 Y_1 |{T}_3|^2 Y_1^*  +4 M_2 T_1 |Y_2|^2 Y_1^* -2 g^{2} Y_2 |M_2|^2 Y_2^* -4 Y_2 |T_1|^2 Y_2^* -\frac{3}{2} m^2_{1} Y_2 |Y_1|^2 Y_2^* \nonumber \\ 
 &+10 Y_1 Y_2 |M_2|^2 Y_1^* Y_2^* -2 Y_3 |{T}_3|^2 Y_3^* -\frac{3}{2} m^2_{1} Y_3 |Y_2|^2 Y_3^* +12 Y_2 Y_3 |M_2|^2 Y_2^* Y_3^* \nonumber \\ 
 &+\frac{1}{8} g_{d}^* \Big(5 g^{2} g_{d} m^2_{1} -64 g_{d} \lambda_4 m^2_{1} -16 g_{d} |T_1|^2 -6 g_{d} m^2_{1} |Y_1|^2 -6 g_{d} m^2_{1} |Y_3|^2 \nonumber \\ 
 &-3 g_{d} |g_{u}|^2 \Big(-16 M_1 M_1^*  -8 M_2 M_2^*  + m^2_{1}\Big)+16 g_{d} M_2 T_1 Y_1^* +32 g_{d} Y_1 |M_1|^2 Y_1^* \nonumber \\ 
 &+8 g_{u} m^2_{1} Y_3 Y_2^*  -16 g_{u} m^2_{2} Y_3 Y_2^* -64 g_{u} Y_3 |M_1|^2 Y_2^* \nonumber \\ 
 &+8 M_2^* \Big(2 g_{d} Y_1 T_1^*  + M_2 \Big(2 g_{d} |Y_3|^2  + 5 g_{d} |Y_1|^2  + 6 g_{d} |Y_2|^2  -6 g_{u} Y_3 Y_2^*  - g^{2} g_{d} \Big)\Big)\nonumber \\ 
 &+32 g_{d} Y_3 |M_1|^2 Y_3^* \Big)-\frac{1}{4} g_{u}^* \Big(4 g_{u} |{T}_3|^2 +g_{u} |Y_2|^2 \Big(3 m^2_{1} -32 M_1 M_1^* -8 M_2 M_2^* \Big) \nonumber \\ 
 &+4 \Big(2 g_{u} \lambda_3 m^2_{2}  + g_{d} Y_2 \Big(2 m^2_{2}  + 6 |M_2|^2  + 8 |M_1|^2  - m^2_{1} \Big)Y_3^* \Big)\Big)\\ 
\beta_{m^2_{2}}^{(1)} & =  
2 \lambda_3 m^2_{1} -\frac{3}{2} g^{2} m^2_{2} +8 \lambda_5 m^2_{2} +2 \lambda_2 m^2_{3} +4 |T_2|^2 +2 |{T}_3|^2 +2 m^2_{2} |Y_3|^2 \nonumber \\ 
 &+|g_{u}|^2 \Big(-2 M_2 M_2^*  -8 M_1 M_1^*  + m^2_{2}\Big)-4 Y_3 |M_2|^2 Y_3^* \\ 
\beta_{m^2_{2}}^{(2)} & =  
\Big(\frac{5}{4} g^{4}  +4 g^{2} \lambda_3  -4 \lambda_{3}^{2}\Big) m^2_{1} +\Big(\frac{39}{16} g^{4}  - \lambda_{2}^{2}  - \lambda_{3}^{2}  +16 g^{2} \lambda_5  -40 \lambda_{5}^{2} \Big) m^2_{2} \nonumber \\ 
 &-4 \lambda_{2}^{2} m^2_{3} -3 g^{4} |M_2|^2 -2 \lambda_2 |T_1|^2 -4 \lambda_3 |T_1|^2 +g^{2} |T_2|^2 -10 \lambda_2 |T_2|^2 -48 \lambda_5 |T_2|^2 \nonumber \\ 
 &+\frac{1}{2} g^{2} |{T}_3|^2 -6 \lambda_2 |{T}_3|^2 -6 \lambda_3 |{T}_3|^2 -8 \lambda_5 |{T}_3|^2 -4 \lambda_2 m^2_{3} |Y_1|^2 -4 \lambda_3 m^2_{1} |Y_2|^2 \nonumber \\ 
 &+\frac{5}{4} g^{2} m^2_{2} |Y_3|^2 -16 \lambda_5 m^2_{2} |Y_3|^2 -\frac{1}{4} \Big(-10 |M_2|^2  + 3 m^2_{2}  -64 |M_1|^2 \Big)|g_{u}|^4   \nonumber \\ 
 &-3 m^2_{2} |Y_3|^4+10 |M_2|^2 |Y_3|^4 -4 \lambda_3 T_2 T_1^* -4 \lambda_3 T_1 T_2^* +4 Y_1 |Y_3|^2 M_2^* T_2^*  \nonumber \\ 
 &-4 Y_1 |T_2|^2 Y_1^* -2 Y_1 |{T}_3|^2 Y_1^* +4 M_2 T_2 |Y_3|^2 Y_1^* -2 Y_2 |{T}_3|^2 Y_2^* \nonumber \\ 
 &-\frac{1}{8} g_{d}^* \Big(3 g_{d} |g_{u}|^2 \Big(m^2_{2} -16 M_1 M_1^*  -8 M_2 M_2^*\Big) +2 \Big(4 g_{d} |{T}_3|^2  \nonumber \\ 
 &+g_{d} \Big(8 \lambda_3 m^2_{1}  + |Y_3|^2 \Big(3 m^2_{2} -32 M_1 M_1^*  -8 M_2 M_2^* \Big)\Big)\nonumber \\ 
 &+4 g_{u} Y_3 \Big(2 m^2_{1}  + 6 |M_2|^2  + 8 |M_1|^2  - m^2_{2} \Big)Y_2^* \Big)\Big) -2 g^{2} Y_3 |M_2|^2 Y_3^* -4 Y_3 |T_2|^2 Y_3^* \nonumber \\ 
 & -\frac{3}{2} m^2_{2} Y_3 |Y_1|^2 Y_3^* -\frac{3}{2} m^2_{2} Y_3 |Y_2|^2 Y_3^* +10 Y_1 Y_3 |M_2|^2 Y_1^* Y_3^*  +12 Y_2 Y_3 |M_2|^2 Y_2^* Y_3^*  \nonumber \\
 &+\frac{1}{8} g_{u}^* \Big(5 g^{2} g_{u} m^2_{2} -g_{u}\Big(64 \lambda_5 m^2_{2} +16  |T_2|^2 +6 m^2_{2} |Y_1|^2 +6  m^2_{2} |Y_2|^2 -16 M_2 T_2 Y_1^*\Big) \nonumber \\ 
 &+32 g_{u} \Big(Y_1 |M_1|^2 Y_1^* +Y_2 |M_1|^2 Y_2^*\Big) -g_{d} \Big(16  m^2_{1} Y_2 Y_3^* -8 m^2_{2} Y_2 Y_3^* +64 Y_2 |M_1|^2 Y_3^* \Big) \nonumber \\ 
 &+8 M_2^* \Big(2 g_{u} Y_1 T_2^*  + M_2 \Big(2 g_{u} |Y_2|^2  + 5 g_{u} |Y_1|^2  -6 g_{d} Y_2 Y_3^*  + 6 g_{u} |Y_3|^2  - g^{2} g_{u} \Big)\Big)\Big)\\ 
\beta_{m^2_{3}}^{(1)} & =  
2 \lambda_1 m^2_{1} +2 \lambda_2 m^2_{2} - T_{1}^{2} - T_{2}^{2} +2 |T_1|^2 +2 |T_2|^2 +2 |{T}_3|^2 +2 m^2_{3} |Y_1|^2 +2 Y_{1}^{2} M_{2}^{*\,2}  \nonumber \\ 
 &- T_{1}^{*\,2} - T_{2}^{*\,2} -8 Y_1 |M_2|^2 Y_1^* +2 M_{2}^{2} Y_{1}^{*\,2} \\ 
\beta_{m^2_{3}}^{(2)} & =  
4 g^{2} \lambda_1 m^2_{1} -4 \lambda_{1}^{2} m^2_{1} +4 g^{2} \lambda_2 m^2_{2} -4 \lambda_{2}^{2} m^2_{2} - \lambda_{1}^{2} m^2_{3} - \lambda_{2}^{2} m^2_{3} -2 g^{2} T_{1}^{2} +4 \lambda_1 T_{1}^{2} \nonumber \\ 
 &-2 g^{2} T_{2}^{2} +4 \lambda_2 T_{2}^{2} +4 g^{2} |T_1|^2 -12 \lambda_1 |T_1|^2 +4 g^{2} |T_2|^2 -12 \lambda_2 |T_2|^2 +4 g^{2} |{T}_3|^2 \nonumber \\ 
 &-6 \lambda_1 |{T}_3|^2 -6 \lambda_2 |{T}_3|^2 +\frac{5}{2} g^{2} m^2_{3} |Y_1|^2 -4 \lambda_1 m^2_{1} |Y_2|^2 +2 T_{1}^{2} |Y_2|^2 -4 \lambda_2 m^2_{2} |Y_3|^2 \nonumber \\ 
 &+2 T_{2}^{2} |Y_3|^2 -3 m^2_{3} |Y_1|^4 +32 |M_2|^2 |Y_1|^4 +2 g^{2} Y_{1}^{2} M_{2}^{*\,2} -2 Y_{1}^{2} \Big(|Y_2|^2 + |Y_3|^2 \Big) M_{2}^{*\,2}  \nonumber \\ 
 &-2 g^{2} T_{1}^{*\,2} +4 \lambda_1 T_{1}^{*\,2} +2 |Y_2|^2 T_{1}^{*\,2} -2 g^{2} T_{2}^{*\,2} +4 \lambda_2 T_{2}^{*\,2} +2 |Y_3|^2 T_{2}^{*\,2} \nonumber \\ 
 &-8 g^{2} Y_1 |M_2|^2 Y_1^* -4 g_{d} g_{u} {T}_3 M_1^* Y_1^* -8 Y_{1}^{3} M_{2}^{*\,2} Y_1^* +2 g^{2} M_{2}^{2} Y_{1}^{*\,2} -2 M_{2}^{2} |Y_2|^2 Y_{1}^{*\,2} \nonumber \\ 
 &-2 M_{2}^{2} |Y_3|^2 Y_{1}^{*\,2} -8 M_{2}^{2} Y_1 Y_{1}^{*\,3} \nonumber \\ 
 &-\frac{1}{4} |g_{u}|^2 \Big(8 \lambda_2 m^2_{2} -4 T_{2}^{2} +4 Y_{1}^{2} M_{2}^{*\,2} +8 T_2 T_2^* -4 T_{2}^{*\,2} +4 {T}_3 {T}_3^* +3 m^2_{3} Y_1 Y_1^* \nonumber \\ 
 &-32 M_1 Y_1 M_1^* Y_1^*  -16 M_2 Y_1 M_2^* Y_1^* +4 M_{2}^{2} Y_{1}^{*\,2} \Big)\nonumber \\ 
 &+g_{d}^* \Big(-2 g_{d} \lambda_1 m^2_{1} +g_{d} T_{1}^{2} -2 g_{d} |T_1|^2 - g_{d} |{T}_3|^2 -\frac{3}{4} g_{d} m^2_{3} |Y_1|^2 - g_{d} Y_{1}^{2} M_{2}^{*\,2} +g_{d} T_{1}^{*\,2} \nonumber \\ 
 &-4 M_1 Y_1 g_{u}^* {T}_3^* +8 g_{d} Y_1 |M_1|^2 Y_1^* +4 g_{d} Y_1 |M_2|^2 Y_1^* - g_{d} M_{2}^{2} Y_{1}^{*\,2} \Big)\nonumber \\ 
 &-4 Y_2 |T_1|^2 Y_2^* -2 Y_2 |{T}_3|^2 Y_2^* -\frac{3}{2} m^2_{3} Y_2 |Y_1|^2 Y_2^* +8 Y_1 Y_2 |M_2|^2 Y_1^* Y_2^* -4 Y_3 |T_2|^2 Y_3^* \nonumber \\ 
 &-2 Y_3 |{T}_3|^2 Y_3^* -\frac{3}{2} m^2_{3} Y_3 |Y_1|^2 Y_3^* +8 Y_1 Y_3 |M_2|^2 Y_1^* Y_3^* 
\end{align}} 

\end{appendix}

\bibliographystyle{JHEP}

\providecommand{\href}[2]{#2}\begingroup\raggedright\endgroup

\end{document}